\documentclass[aps,amsmath,amssymb,floatfix,prb,twocolumn,showpacs,superscriptaddress]{revtex4-1}
\usepackage{bm}
\usepackage{epsfig}
\usepackage[usenames]{color}
\usepackage{graphicx}
\usepackage{amsfonts}
\usepackage[hypertex]{hyperref}

\newcommand{\olcite}[1]{[\onlinecite{#1}]} 

\newcommand{\beq}{\begin{equation}}
\newcommand{\eeq}{\end{equation}}
\newcommand{\beqa}{\begin{eqnarray}}
\newcommand{\eeqa}{\end{eqnarray}}

\newcommand{\sgn}{\mbox{sgn}}

\renewcommand{\Re}{\mbox{Re} }
\renewcommand{\Im}{\mbox{Im} }

\newcommand{\rI}{{\tt I}}

\newcommand{\rIII}{{\tt III}}

\newcommand{\rVII}{{\tt VII}}

\begin{document}
\title{{\it Fluctuoscopy} of Disordered Two-Dimensional Superconductors}


\author{A.~Glatz}
\affiliation{Materials Science Division, Argonne National
Laboratory, 9700 S.Cass Avenue, Argonne, Illinois 60637, USA}

\author{A.~A.~Varlamov}
\affiliation{Materials Science Division, Argonne National
Laboratory, 9700 S.Cass Avenue, Argonne, Illinois 60637, USA}
\affiliation{CNR-SPIN, Viale del Politecnico 1, I-00133 Rome, Italy}

\author{V.~M.~Vinokur}
\affiliation{Materials Science Division, Argonne National
Laboratory, 9700 S.Cass Avenue, Argonne, Illinois 60637, USA}

\date{\today}

\begin{abstract}
We revise the long studied problem of fluctuation conductivity (FC) in
disordered two-dimensional superconductors placed in a perpendicular magnetic field by
finally deriving the complete solution in the temperature-magnetic field
phase diagram. The obtained expressions allow both to perform straightforward
(numerical) calculation of the FC surface $\delta\sigma_{xx}^{\left( \mathrm{%
tot}\right) }(T,H)$ and to get asymptotic expressions in all its
qualitatively different domains. This surface becomes in particular
non-trivial at low temperatures, where it is trough-shaped with $%
\delta\sigma_{xx}^{\left( \mathrm{tot}\right) }(T,H)<0$. In this region,
close to the quantum phase transition, $\delta\sigma_{xx}^{\left(
\mathrm{tot}\right) }(T,H=\mathrm{const})$ is non-monotonic, in agreement
with experimental findings. We reanalyzed and present comparisons to several
experimental measurements. Based on our results we derive a qualitative
picture of superconducting fluctuations close to $H_{\mathrm{c2}}\left(
0\right) $ and $T=0$ where fluctuation Cooper pairs rotate with
cyclotron frequency $\omega_{c}\sim\Delta_{\mathrm{BCS}}^{-1}$ and Larmor
radius $\sim \xi_{\mathrm{BCS}}$, forming some kind of quantum liquid with
 long coherence length $\xi_{\mathrm{QF}}\gg\xi_{\mathrm{BCS}}$ and slow
relaxation ($\tau_{\mathrm{QF}}\gg\hbar\Delta_{\mathrm{BCS}}^{-1}$).
\end{abstract}

\maketitle

\section{Introduction}

The understanding of the mechanisms of superconducting fluctuations (SF),
achieved during the past decades~\cite{LV09} provided a unique tool
obtaining information about the microscopic parameters of superconductors
(SC). SFs are comprised of Cooper pairs with finite lifetime which appear
already above the transition but do not form a stable condensate yet. They
affect thermodynamic and transport properties of the normal state both
directly and through the changes which they cause in the normal
quasi-particle subsystem~\cite{LV09}.

\begin{figure}[htb]
\begin{center}
\includegraphics[width=0.99\columnwidth]{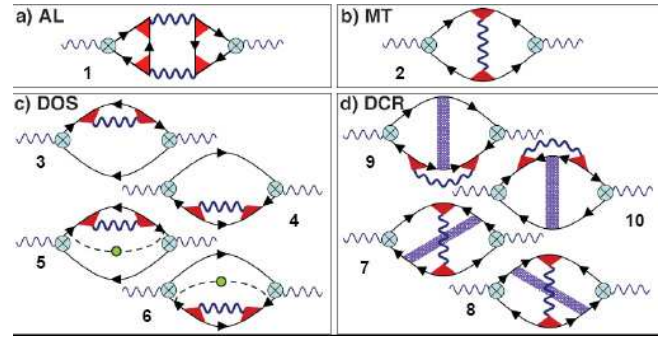}
\end{center}
\caption{(Color online) Feynman diagrams for the leading-order contributions
to the electromagnetic response operator. Wavy lines stand for fluctuation
propagators, solid lines with arrows are impurity-averaged normal state
Green's functions, crossed circles are electric field vertices, dashed lines
with a circle represent additional impurity renormalizations, and triangles
and dotted rectangles are impurity ladders accounting for the electron
scattering at impurities (Cooperons).}
\label{fig.conddia}
\end{figure}

SFs are commonly described in terms of three principal contributions: the
Aslamazov-Larkin (AL) process, corresponding to the opening of a new channel
for the charge transfer~\cite{AL68}, anomalous Maki-Thompson (MT) process,
which describes single-particle quantum interference at impurities in the
presence of SFs\cite{M68,T70}, and the change of the single-particle density
of states (DOS) due to their involvement in fluctuation pairings~\cite%
{ILVY93,BDKLV93}. The first two processes (AL and MT) result in the
appearance of positive and singular contributions to conductivity (diagrams
1 and 2 in Fig.~\ref{fig.conddia}) close to the superconducting critical
temperature $T_{\mathrm{c}0}$ , while the third one (DOS) results in a
decrease of the Drude conductivity due to the lack of single-particle
excitations at the Fermi level (diagrams 3-6 in Fig. \ref{fig.conddia}). The
latter contribution is less singular in temperature than the first two and
can compete with them only if the AL and MT processes are suppressed for
some reasons (for example, c-axis transport in layered superconductors) or
far away from $T_{\mathrm{c}0}$.

The classical results obtained first in the vicinity of $T_{\mathrm{c}0}$
were later generalized to temperatures far from the transition, e.g. in Refs.~\olcite%
{AV80,L80,ARV83}, and to relatively high fields, see Ref.~\olcite{Ab85}. More recently,
quantum fluctuations (QF) came into the focus of investigations. In Ref.~[%
\onlinecite{BE99,BEL00}] it was found, that in granular SC at very low
temperatures and close to $H_{\mathrm{c2}}\left( 0\right) $, the positive AL
contribution to magneto-conductivity decays as $T^{2}$ while the fluctuation
suppression of the DOS results in a temperature independent negative
contribution, logaritmically growing in magnitude for $H\rightarrow H_{%
\mathrm{c2}}\left( 0\right) $. The authors of Ref.~[\onlinecite{SSVG09}]
came to the same conclusion while studying the effect of QFs on the
Nernst-Ettingshausen coefficient in two-dimensional (2D) SCs. For the first
time they attracted the attention~\cite{footnote1} to the special role of
diagrams 9 \& 10 in Fig.~\ref{fig.conddia}.

Special attention should be paid to the paper by Galitski and Larkin, Ref.~[%
\onlinecite{GL01}], where the effects of QFs on magneto-conductivity and
magnetization of 2D superconductors were studied. The authors analyzed all
ten diagrams shown in Fig.~\ref{fig.conddia} in the lowest Landau level
(LLL) approximation, valid at fields close to the critical line $H_{\mathrm{%
c2}}\left( T\right) $. They found an expression for the total fluctuation
contribution to magneto-conductivity in this region and demonstrated that,
analogously to the situation in granular SCs, close to zero temperature and
in the vicinity of $H_{\mathrm{c2}}\left( 0\right) $ this contribution is
negative, i.e. QFs increase resistivity, and not conductivity (in contrast to
the situation close to $T_{\mathrm{c}0}$). Nevertheless, contrary to the
conclusions of Refs.~ [\onlinecite{BE99,BEL00}], in Ref.~[\onlinecite{GL01}]
the logarithmic growth of the magneto-resistance at zero temperature when
the magnetic field approaches $H_{\mathrm{c2}}\left( 0\right) $ from larger
fields, is due \textit{to all} AL, MT and DOS-like contributions.

\begin{figure}[t]
\begin{center}
\includegraphics[width=0.99\columnwidth]{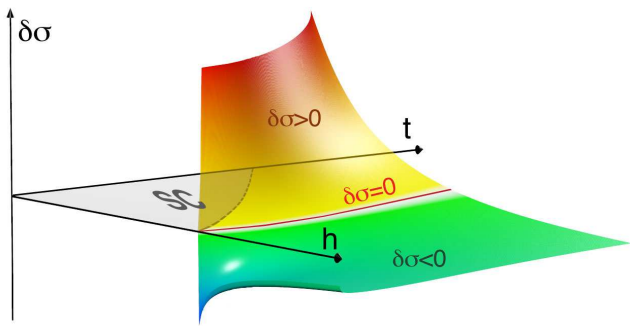}
\end{center}
\caption{ (Color online) Fluctuation correction to conductivity (FC) $%
\protect\delta \protect\sigma =\protect\delta \protect\sigma _{xx}^{\mathrm{%
(tot)}}\left( t,h\right)$ as function of the reduced temperature $%
t=T $/$T_{\mathrm{c0}}$ and magnetic field $h=0.69H/H_{\mathrm{c2}}\left( 0\right) $
plotted as surface. The FC changes its sign along the thick red line ($%
\protect\delta \protect\sigma =0$). The boundary of the superconducting
region is shown by a dashed line. Here $\protect\delta \protect\sigma $ is
plotted for constant $\protect\tau T_{c0}=10^{-2}$ and $\protect\tau _{%
\protect\phi }T_{c0}=10$.}
\label{fig.iceberg}
\end{figure}

Yet, the confidence in the exotic nature of negative fluctuation corrections
and the common believe of fluctuation contributions to conductivity being
positive beyond the narrow domain of the quantum phase transition, has been
persistent and is based on available asymptotic expressions only. The region
near $T=0$ and magnetic fields near $H_{\mathrm{c2} }\left( 0\right) $
remains poorly understood and in addition, an universal picture combining QFs
at high magnetic fields and conventional finite temperature quantum
corrections is still lacking.

This is why we revisit the problem of fluctuation conductivity of a
disordered 2D superconductor placed in a perpendicular magnetic field in
this paper. We present an exact calculation (\textit{without the use of the
LLL approximation}) of all ten diagrams of the first order of fluctuation
theory (see Fig. \ref{fig.conddia}) valid in the whole $H$-$T$ phase diagram
beyond the superconducting region, i.e. for arbitrary fields $H\geq H_{%
\mathrm{c2}}\left( T\right) $ and temperatures $T_{\mathrm{c}}\left(
H\right) \leq T$. The obtained expressions allow both to perform
straightforward (numerical) calculation of the fluctuation conductivity
''surface'' $\delta\sigma_{xx}^{\left( \mathrm{tot}\right) }(T,H)$ and to
get asymptotic expressions in all its qualitatively different domains.

\begin{figure}[t]
\includegraphics[width=0.9\columnwidth]{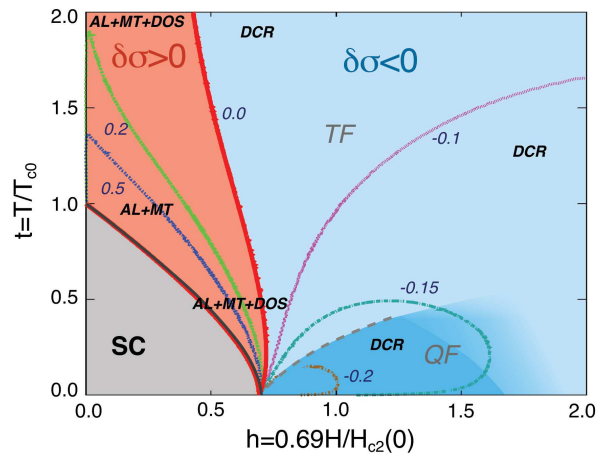}
\caption{(Color online) Contours of constant fluctuation conductivity [$%
\protect\delta \protect\sigma =\protect\delta \protect\sigma _{xx}^{\mathrm{%
(tot)}}\left( t,h\right) $ shown in units of $e^{2}$]. The dominant FC
contributions are indicated by bold-italic labels. The dashed line separates
the domain of quantum fluctuations (QF) [dark area of $\protect\delta 
\protect\sigma <0$] and thermal fluctuations (TF). The contour lines are
obtained from Eq. (\protect\ref{all}) with $T_{c0}\protect\tau =0.01$ and $%
T_{c0}\protect\tau _{\protect\phi }=10$. }
\label{fig.contours}
\end{figure}

A typical example of the surface $\delta\sigma_{xx}^{\left( \mathrm{tot}%
\right) }(T,H)$ is presented in Fig.~\ref{fig.iceberg} and demonstrates that
our revision and completion of the commonly believed understanding of
fluctuation corrections is urgently called for: Its striking feature
consists of the fact, that the FC is positive only in the domain bound by
the separatices $H_{\mathrm{c2}}(T)$ and $\delta\sigma_{xx}^{\left( \mathrm{%
tot}\right) }(T,H)=0$ and is negative throughout all other parts of the
phase diagram (see Fig.~\ref{fig.contours}, in which the domains of
different overall signs of $\delta\sigma_{xx}^{\left( \mathrm{tot}\right)
}(T,H)$ and contours of constant $\delta\sigma_{xx}^{\left( \mathrm{tot}%
\right) }$ in the whole phase diagram are shown). Contrary to the common
assumption, the FC is only positive in the domain of weak fields and
temperatures above $T_{\mathrm{c0}}$, the region of positive corrections
depends on the magnitude of the positive anomalous MT contribution (i.e. on
the value of the phase-breaking time $\tau_{\phi})$. With increasing magnetic
field, the interval of temperatures where $\delta\sigma_{xx}^{\left( \mathrm{%
tot}\right) }(T,H)>0$ shrinks and becomes zero close to $H_{\mathrm{c2}}(0).$
In particular at low temperatures, the behavior of the FC turns out to be
highly non-trivial. In this case, the surface $\delta \sigma_{xx}^{\left( 
\mathrm{tot}\right) }(T,H)$ has a trough-shaped character and the dependence 
$\delta\sigma_{xx}^{\left( \mathrm{tot}\right) }(T,H=\mathrm{const})$ is
non-monotonic. We will see below that this feature is observed
in available experimental results as well.

Our analysis also elucidates the understanding of the hierarchy of the
various contributions to the fluctuation corrections in different domains of
the phase diagram (see Fig.~\ref{fig.contours} in which the dominating
fluctuation contributions to magneto-conductivity are indicated for
different regions of the phase diagram). We reproduce the expression for the
total fluctuation contribution to magneto-conducitivty of 2D disordered
superconductors presented in Ref.~[\onlinecite{GL01}] in the vicinity of the 
$H_{\mathrm{c2}}(T)$ line. Nevertheless, our analysis clearly demonstrates
that the principal fluctuation contributions close to $T_{\mathrm{c0}}$,
paraconductivity (AL), anomalous MT and DOS, in the region of QF become 
\textit{zero} as $\sim $ $T^{2}$ (compare to Refs.~[\onlinecite{BE99,BEL00}%
]) - which is contrary to the picture given in Ref.~[\onlinecite{GL01}]. It
is the fourth, usually ignored, fluctuation contribution, formally
determined by the sum of diagrams 7-10 and the regular part of the MT
diagram, which governs the quantum phase transition (QPT). It can be
identified with the renormalization of the single-particle diffusion
coefficient in the presence of fluctuations (DCR) and it turns out that
this contribution dominates in the periphery of the phase
diagram including the vicinity of the QPT $\left[ t=T/T_{\mathrm{c0}}\ll 
\widetilde{h}=\left( H-H_{\mathrm{c2}}(0)\right) /H_{\mathrm{c2}}(0);H>H_{%
\mathrm{c2}}(0)\right]$.

\begin{figure}[tb]
\includegraphics[width=.9\columnwidth]{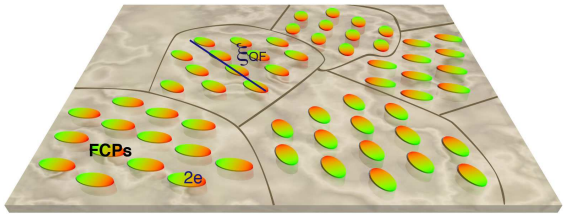}
\caption{(Color online) Illustration of the cluster structure of a FCP (2e)
liquid above the upper critical field. This picture represents a snapshot at
a certain time and would stay that way for time $\protect\tau_{QF}$. The
typical size of a coherent FCP cluster is $\protect\xi_{QF}$.}
\label{fig.abrikosovQC}
\end{figure}

Finally, based on our results, we propose a qualitative picture for QPT,
which drastically differs from the Ginzburg-Landau one, valid close to $T_{%
\mathrm{c0}}$. The latter can be described in terms of a set of
long-wavelength fluctuation modes [with $\lambda\gtrsim\xi_{\mathrm{GL}%
}\left( T\right) \gg\xi_{\mathrm{BCS}}$] of the order parameter, with
characteristic lifetime $\tau_{\mathrm{GL}}=\pi\hbar/8k_{B}\left( T-T_{%
\mathrm{c0}}\right)$. Near the QPT, the order parameter oscillates on much
smaller scales - the fluctuation modes with wave-lengths up $\xi_{\mathrm{BCS%
}}$ are excited. Due to the magnetic field, one can imagine that FCP in this
region rotate with the Larmor radius $\sim\xi_{\mathrm{BCS}}$ and cyclotron
frequency $\omega_{c}\sim\Delta_{\mathrm{BCS}}^{-1}$. We show that close to $%
H_{\mathrm{c2}}\left( 0\right) $ these FCP form some kind of quantum liquid
with long coherence length $\xi_{\mathrm{QF}}\sim\xi_{\mathrm{BCS}}/%
\widetilde{h}^{1/2}$ and slow relaxation $\tau_{\mathrm{QF}}\sim\hbar
\Delta_{\mathrm{BCS}}^{-1}/\widetilde{h}$ (see Fig.~\ref{fig.abrikosovQC}).

In the following sections and in the appendices, we will show the details of
our derivations and calculation and present the general expression for the
fluctuation magneto-conductivity of disordered 2D SC throughout the whole
phase diagram. For the calculation of the complete and various fluctuation
corrections (by numerical integration and summation), we developed an
optimized program which is available at~[\onlinecite{soft}]. It can be used
as the theoretical basis for the \textquotedblleft
fluctuoscopy\textquotedblright\ of superconductors: the study of their
behavior in ultra-high magnetic fields and precise extraction for their
physical parameters like the critical temperature and magnetic field, and
the temperature dependence of the phase-breaking time, and/or e.g. for the
separation of the quantum corrections in studies of the \textquotedblleft
superconductor--insulator\textquotedblright\ transition.

\section{Model}

We consider a disordered 2D superconductor characterized by the diffusion
coefficient $\mathcal{D}$ placed in a perpendicular magnetic field $H$ at
temperatures $T>T_{\mathrm{c}}\left( H\right) $. Temperatures should not be
too close to the critical temperature and remain beyond the region of
critical fluctuations, i.e. $T/T_{\mathrm{c}}\left( H\right) -1\gg \sqrt{%
\mathrm{Gi}_{\left( \mathrm{2}\right) }\left( H\right) }$.
The Ginzburg-Levanyuk number Gi$_{\left( 2\right) }$ for conductivity (see Ref.~[%
\onlinecite{LV09}]) in both extremes of the line $H_{\mathrm{c2}}(T)$%
 (at temperatures close to $T_{\mathrm{c0}}$ and at zero
temperature) is on the order of $\left( p_{F}^{2}ld\right) ^{-1}$,
where $d$ is the SC film thickness, and it can reach values of up to $%
10^{-2}$. 
We assume the temperature $T\ll \min \left\{ \tau ^{-1},\omega
_{D}\right\} $ in order to remain in the diffusive regime of electron scattering
and in the frameworks of the BCS model ($\tau $\ is the electron elastic
scattering time on impurities, $\omega _{\mathrm{D}}$ is the Debey
frequency).  The restrictions on magnetic field are dictated by the
requirements to be below the regime of Shubnikov-de Haas oscillations [$%
\omega _{\mathrm{c}}\tau \lesssim 1\Longleftrightarrow H\lesssim \left( T_{%
\mathrm{c0}}\tau \right) ^{-1}H_{\mathrm{c2}}\left( 0\right) ,$ where $%
\omega _{\mathrm{c}}=$ $4\mathcal{D}eH$ is the fluctuation Cooper pair
cyclotron frequency] and to be below the Clogston limit: $H\lesssim \left(
\varepsilon _{\mathrm{F}}\tau \right) H_{\mathrm{c2}}\left( 0\right) ,$ i.e. 
$H/H_{\mathrm{c2}}\left( 0\right) \ll \min \left\{ \left( T_{\mathrm{c0}%
}\tau \right) ^{-1},\varepsilon _{\mathrm{F}}\tau \right\} $.

Under these rather non-restrictive assumptions the DC fluctuation
conductivity 
\begin{equation}
\delta \sigma ^{\left( \mathrm{fl}\right) }\left( T,H\right) =-\lim_{\omega
\rightarrow 0}\frac{\Im Q^{\left( \mathrm{fl}\right) }\left( \omega
,T,H\right) }{\omega }  \label{quom}
\end{equation}%
is determined by the imaginary part of the fluctuation contribution $%
Q^{\left( \mathrm{fl}\right) }\left( \omega ,T,H\right) $ to the
electromagnetic response operator\cite{LV09}. The latter is described
graphically by the ten standard diagrams shown in Fig. \ref{fig.conddia}.
The solid lines denote the one-electron Green function%
\begin{equation*}
G\left( x,x^{\prime },p_{y},p_{z},\varepsilon _{l}\right) =\sum_{k}\frac{%
\varphi _{k}\left( x-l_{H}^{2}p_{y}\right) \varphi _{k}^{\ast }\left(
x^{\prime }-l_{H}^{2}p_{y}\right) }{i\widetilde{\varepsilon }_{l}-\xi \left(
k,p_{z}\right) },
\end{equation*}%
wavy lines correspond to the fluctuation propagator 
\begin{align}
& L_{n}^{-1}(\Omega _{k})=   \label{propagator}\\
& -\nu _{0}\left[ \ln \frac{T}{T_{\mathrm{c0}}}+\psi \left( \frac{1}{2}+%
\frac{|\Omega _{k}|+\omega _{\mathrm{c}}(n+\frac{1}{2})}{4\pi T}\right)
-\psi \left( \frac{1}{2}\right) \right]   \notag
\end{align}%
and shaded three- and four-leg blocks indicate the results of the average
over elastic impurity scattering of electrons (Cooperons):%
\begin{equation}
\lambda _{n}(\varepsilon _{1},\varepsilon _{2})=\frac{\tau ^{-1}\theta
(-\varepsilon _{1}\varepsilon _{2})}{|\varepsilon _{1}-\varepsilon
_{2}|+\omega _{\mathrm{c}}(n+1/2)+\tau _{\varphi }^{-1}},  \notag
\end{equation}%
\begin{equation*}
C_{n}(\varepsilon _{1},\varepsilon _{2})=\frac{1}{2\pi \nu _{0}\tau }\frac{%
\tau ^{-1}\theta (-\varepsilon _{1}\varepsilon _{2})}{|\varepsilon
_{1}-\varepsilon _{2}|+\omega _{\mathrm{c}}(n+1/2)+\tau _{\varphi }^{-1}}.
\end{equation*}%
Here $\nu _{0}$\ is the one-electron density of states, $n,m$ are the quantum
numbers of the Cooper pair Landau states, $\Omega _{k}=2\pi kT$ , $%
\varepsilon_{l}=2\pi T\left( l+1/2\right) $ are the bosonic and fermionic
Matsubara frequencies. An important characteristic of these expressions is
that they are valid even far from the critical temperature [for temperatures 
$T\ll \min \{\tau ^{-1},\omega _{D}\}$] and for $|\Omega _{k}|\ll \omega _{D}
$ and $n\ll \left( T_{\mathrm{c0}}\tau \right) ^{-1}.$

In the Appendices we present the details of the calculation of all ten
diagrams performed under the above general assumptions. In the following
sections of the main text, we restrict ourselves to the discussion and
analysis of the main result: the complete expression of the fluctuations
corrections and the individual contributions from AL, MT, DOS, and DCR
processes.

\section{Results}

The complete expression for the total fluctuation correction to conductivity 
$\delta\sigma_{xx}^{\mathrm{(tot)}}\left( T,H\right) $ of a disordered 2D SC
in a perpendicular magnetic field that holds in the complete $T$-$H$\ phase
diagram above the line $H_{c2}(T)$ is given by the sum of Eqs.\thinspace (%
\ref{ALgener}), (\ref{MTgener}), (\ref{sigdos}), and (\ref{DCRf}):

\begin{widetext}%
\begin{align}
\delta\sigma_{xx}^{\left(  \mathrm{tot}\right)  }\left(  t,h\right)    &
=\underbrace{\frac{e^{2}}{\pi}\sum_{m=0}^{\infty}(m+1)\int_{-\infty}^{\infty
}\frac{dx}{\sinh^{2}\pi x}\left\{  \frac{\operatorname{Im}^{2}\mathcal{E}_{m}%
}{\left\vert \mathcal{E}_{m}\right\vert ^{2}}+\frac{\operatorname{Im}%
^{2}\mathcal{E}_{m+1}}{\left\vert \mathcal{E}_{m+1}\right\vert ^{2}}%
+\frac{\operatorname{Im}^{2}\mathcal{E}_{m+1}-\operatorname{Im}^{2}%
\mathcal{E}_{m}}{\left\vert \mathcal{E}_{m}\right\vert ^{2}\left\vert
\mathcal{E}_{m+1}\right\vert ^{2}}\operatorname{Re}\left[  \mathcal{E}%
_{m}\mathcal{E}_{m+1}\right]  \right\}  }_{\delta\sigma_{xx}^{\mathrm{AL}}%
}\nonumber\\
& +\underbrace{\frac{e^{2}}{\pi}\left(  \frac{h}{t}\right)  \sum_{m=0}%
^{M}\frac{{1}}{\gamma_{\phi}+\frac{2h}{t}\left(  m+1/2\right)  }\int_{-\infty
}^{\infty}\frac{dx}{\sinh^{2}\pi x}\frac{\operatorname{Im}^{2}\mathcal{E}_{m}%
}{\left\vert \mathcal{E}_{m}\right\vert ^{2}}}_{\delta\sigma_{xx}%
^{\mathrm{MT(an)}}+\delta\sigma_{xx}^{\mathrm{MT(reg2)}}}+\underbrace
{\frac{e^{2}}{\pi^{4}}\left(  \frac{h}{t}\right)  \sum_{m=0}^{M}%
\sum_{k=-\infty}^{\infty}\frac{4\mathcal{E}_{m}^{\prime\prime}\left(
t,h,|k|\right)  }{\mathcal{E}_{m}\left(  t,h,|k|\right)  }}_{\delta\sigma
_{xx}^{\mathrm{MT(reg1)}}}\;\nonumber\\
& +\underbrace{\frac{4e^{2}}{\pi^{3}}\left(  \frac{h}{t}\right)  \sum
_{m=0}^{M}\int_{-\infty}^{\infty}\frac{dx}{\sinh^{2}\pi x}\frac
{\operatorname{Im}\mathcal{E}_{m}\operatorname{Im}\mathcal{E}_{m}^{\prime}%
}{\left\vert \mathcal{E}_{m}\right\vert ^{2}}}_{\delta\sigma_{xx}%
^{\mathrm{DOS}}}\text{ }\quad+\underbrace{\frac{4e^{2}}{3\pi^{6}}\left(
\frac{h}{t}\right)  ^{2}\sum_{m=0}^{M}(m+\frac{1}{2})\sum_{k=-\infty}^{\infty
}\frac{8\mathcal{E}_{m}^{\prime\prime\prime}\left(  t,h,|k|\right)
}{\mathcal{E}_{m}\left(  t,h,|k|\right)  }}_{\delta\sigma_{xx}^{\mathrm{7-10}}%
}.
\label{all}
\end{align}
\end{widetext}

Here $t=T/T_{\mathrm{c0}}$, 
\begin{equation*}
h=\frac{\pi ^{2}}{8\gamma _{E}}\frac{H}{H_{\mathrm{c2}}\left( 0\right) }=0.69%
\frac{H}{H_{\mathrm{c2}}\left( 0\right) },
\end{equation*}%
$\gamma _{E}=e^{\gamma _{e}}$ ($\gamma _{e}$ is the Euler constant), $%
M=\left( tT_{\mathrm{c0}}\tau \right) ^{-1},\gamma _{\phi }=\pi /\left( 8T_{%
\mathrm{c0}}\tau _{\phi }\right) $, $\tau _{\phi }$ is the phase-breaking
time, 
\begin{equation*}
\mathcal{E}_{m}\equiv \mathcal{E}_{m}\left( t,h,z\right) =\ln t+\psi \left[ 
\frac{1+z}{2}+\frac{2h}{t}\frac{\left( 2m+1\right) }{\pi ^{2}}\right] -\psi
\left( \frac{1}{2}\right) 
\end{equation*}%
and its derivatives $\mathcal{E}_{m}^{(p)}\left( t,h,z\right) \equiv
\partial _{z}^{p}\mathcal{E}_{m}\left( t,h,z\right) $. Apart from the
detailed derivation of the result, Eq. (\ref{all}), one can also do a careful
study of the asymptotic expressions for different fluctuation contributions
throughout the $h$-$t$ phase diagram, presented in the Appendices. All of them, side by
side with the asymptotic expressions for $\delta \sigma _{xx}^{\mathrm{(tot)}%
}$ are summarized in table~\ref{tab.asym}.

\begin{table*}[tbp]
{\tiny 
\begin{tabular}{|c|l|l|l|l|l|}
\hline
& $\delta \sigma _{xx}^{\mathrm{AL}}$ & $\delta \sigma _{xx}^{\mathrm{MT}}$
& $\delta \sigma _{xx}^{\mathrm{DOS}}$ & $\delta \sigma _{xx}^{\mathrm{7-10}}
$ & $\delta \sigma _{xx}^{\mathrm{tot}}$ \\ \hline
\texttt{I} & $%
\begin{array}{c}
\frac{e^{2}}{16\epsilon } \\ 
-\frac{7\zeta \left( 3\right) e^{2}}{8\pi ^{4}}\ln \frac{1}{\epsilon }%
\end{array}%
$ & $\frac{e^{2}}{8\left( \epsilon -\gamma _{\phi }\right) }\ln \frac{%
\epsilon }{\gamma _{\phi }}-\frac{14\zeta \left( 3\right) e^{2}}{\pi ^{4}}%
\ln \frac{1}{\epsilon }$ & $-\frac{14\zeta \left( 3\right) e^{2}}{\pi ^{4}}%
\ln \frac{1}{\epsilon }$ & $%
\begin{array}{c}
\frac{e^{2}}{3\pi ^{2}}\ln \ln \frac{1}{T_{\mathrm{c0}}\tau } \\ 
+O\left( \epsilon \right) 
\end{array}%
$ & $%
\begin{array}{c}
\frac{e^{2}}{16\epsilon }+\frac{e^{2}}{8\left( \epsilon -\gamma _{\phi
}\right) }\ln \frac{\epsilon }{\gamma _{\phi }} \\ 
+\frac{e^{2}}{3\pi ^{2}}\ln \ln \frac{1}{T_{\mathrm{c0}}\tau }%
\end{array}%
$ \\ \hline
$%
\begin{array}{c}
\mathtt{I}- \\ 
\mathtt{III}%
\end{array}%
$ & $%
\begin{array}{c}
\frac{e^{2}}{2\epsilon }\left( \frac{\epsilon }{2h}\right) ^{2}\left[ \psi
\left( \frac{1}{2}+\frac{\epsilon }{2h}\right) \right.  \\ 
\left. -\psi \left( \frac{\epsilon }{2h}\right) -\frac{h}{\epsilon }\right] 
\end{array}%
$ & $%
\begin{array}{c}
\frac{e^{2}}{8}\frac{1}{\epsilon -\gamma _{\phi }}\left[ \psi \left( \frac{1%
}{2}+\frac{t\epsilon }{2h}\right) -\psi \left( \frac{1}{2}+\frac{t\gamma
_{\phi }}{2h}\right) \right]  \\ 
-\frac{14\zeta \left( 3\right) e^{2}}{\pi ^{4}}\left[ \ln \left( \frac{t}{2h}%
\right) -\psi \left( \frac{1}{2}+\frac{t\epsilon }{2h}\right) \right] 
\end{array}%
$ & $%
\begin{array}{c}
-\frac{14\zeta \left( 3\right) e^{2}}{\pi ^{4}}\left[ \ln \left( \frac{t}{2h}%
\right) \right.  \\ 
\left. -\psi \left( \frac{1}{2}+\frac{t\epsilon }{2h}\right) \right] 
\end{array}%
$ & $%
\begin{array}{c}
\frac{e^{2}}{3\pi ^{2}}\ln \ln \frac{1}{T_{\mathrm{c0}}\tau } \\ 
+O\left( \max \left[ \epsilon ,h^{2}\right] \right) 
\end{array}%
$ &  \\ \hline
\texttt{IV} & $\frac{4e^{2}\gamma _{E}^{2}t^{2}}{3\pi ^{2}\widetilde{h}^{2}}$
& $-\frac{2e^{2}}{\pi ^{2}}\ln \frac{1}{\widetilde{h}}-\frac{2\gamma
_{E}e^{2}}{\pi ^{2}}\left( \frac{t}{\widetilde{h}}\right) $ & $-\frac{%
4e^{2}\gamma _{E}^{2}t^{2}}{3\pi ^{2}\widetilde{h}^{2}}$ & $\frac{4e^{2}}{%
3\pi ^{2}}\ln \frac{1}{\widetilde{h}}$ & $-\frac{2e^{2}}{3\pi ^{2}}\left(
\ln \frac{1}{\widetilde{h}}+\frac{3t}{\widetilde{h}}\right) $ \\ \hline
\texttt{V} & $\frac{2\gamma _{E}e^{2}}{\pi ^{2}}\left( \frac{t}{\widetilde{h}%
}\right) $ & $-\frac{2e^{2}}{3\pi ^{2}}\ln \frac{1}{4\gamma _{E}t}$ & $-%
\frac{2\gamma _{E}e^{2}}{\pi ^{2}}\left( \frac{t}{\widetilde{h}}\right) $ & $%
\frac{4e^{2}}{3\pi ^{2}}\ln \frac{1}{4\gamma _{E}t}$ & $-\frac{2e^{2}}{3\pi
^{2}}\ln \frac{1}{4\gamma _{E}t}$ \\ \hline
$%
\begin{array}{c}
\mathtt{VI}- \\ 
\mathtt{VII}%
\end{array}%
$ & $\frac{e^{2}}{4}\frac{t}{h-h_{c2}\left( t\right) }$ & $-\frac{2e^{2}}{%
3\pi ^{2}}\ln \frac{2h}{\pi ^{2}t}$ & $-\frac{e^{2}}{4}\frac{t}{%
h-h_{c2}\left( t\right) }$ & $\frac{4e^{2}}{3\pi ^{2}}\ln \frac{2h}{\pi ^{2}t%
}$ & $-\frac{2e^{2}}{3\pi ^{2}}\ln \frac{h_{c2}\left( t\right) }{%
h-h_{c2}\left( t\right) }$ \\ \hline
\texttt{VIII} & $\frac{e^{2}}{6\pi ^{2}}\frac{C_{1}}{\ln ^{3}t}$ & $-\frac{%
e^{2}}{\pi ^{2}}\ln \frac{\ln \frac{1}{T_{\mathrm{c0}}\tau }}{\ln t}+\frac{%
\pi ^{2}e^{2}}{192}\frac{\ln \frac{\pi ^{2}}{2\gamma _{\phi }}}{\ln ^{2}t}$
& $-\frac{\pi ^{2}e^{2}}{192}\frac{1}{\ln ^{2}t}$ & $\frac{e^{2}}{3\pi ^{2}}%
\ln \frac{\ln \frac{1}{T_{\mathrm{c0}}\tau }}{\ln t}$ & $-\frac{2e^{2}}{3\pi
^{2}}\ln \frac{\ln \frac{1}{T_{\mathrm{c0}}\tau }}{\ln t}$ \\ \hline
\texttt{IX} & $\frac{\pi ^{2}e^{2}}{192}\left( \frac{t}{h}\right) ^{2}\frac{%
C_{2}}{\ln ^{3}\frac{2h}{\pi ^{2}}}$ & $-\frac{e^{2}}{\pi ^{2}}\ln \frac{\ln 
\frac{1}{T_{\mathrm{c0}}\tau }}{\ln \frac{2h}{\pi ^{2}}}+\frac{7\zeta \left(
3\right) \pi ^{2}e^{2}}{768}\left( \frac{t}{h}\right) ^{2}\frac{1}{\ln ^{2}%
\frac{2h}{\pi ^{2}}}$ & $-\frac{7\zeta \left( 3\right) \pi ^{2}e^{2}}{384}%
\left( \frac{t}{h}\right) ^{2}\frac{1}{\ln ^{2}\frac{2h}{\pi ^{2}}}$ & $%
\frac{e^{2}}{3\pi ^{2}}\ln \frac{\ln \frac{1}{T_{\mathrm{c0}}\tau }}{\ln 
\frac{2h}{\pi ^{2}}}$ & $%
\begin{array}{c}
-\frac{2e^{2}}{3\pi ^{2}}\ln \frac{\ln \frac{1}{T_{\mathrm{c0}}\tau }}{\ln 
\frac{2h}{\pi ^{2}}} \\ 
-\frac{7\zeta \left( 3\right) \pi ^{2}e^{2}}{768}\left( \frac{t}{h}\right)
^{2}\frac{1}{\ln ^{2}\frac{2h}{\pi ^{2}}}%
\end{array}%
$ \\ \hline
\end{tabular}
}
\caption{Asymptotic expressions in different domains, shown in Fig.~\protect
\ref{fig.domains}. The first column gives the domain according to that
figure and is determined by the $t$ \& $h$ regions given in table~\protect\ref{tab.domains}.}
\label{tab.asym}
\end{table*}

\begin{figure}[htb]
\begin{center}
\includegraphics[width=0.95\columnwidth]{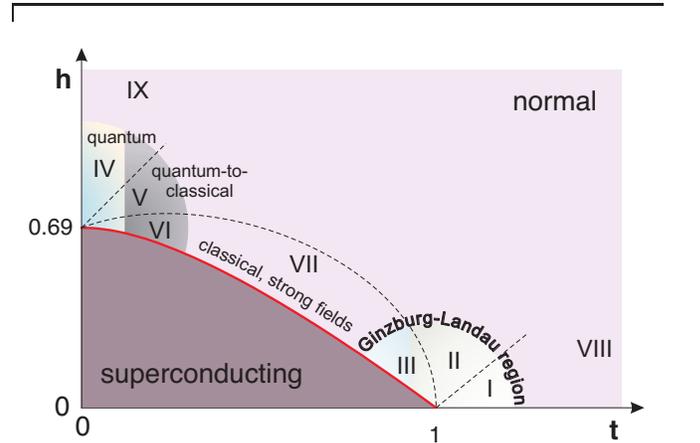}
\end{center}
\caption{(Color online) Schematic representation of the regions of different
behavior of fluctuation conductivity in the $h$-$t$ diagram. See table~%
\protect\ref{tab.domains} for more explanations on the domains.}
\label{fig.domains}
\end{figure}

\begin{table}[tbh]
\begin{tabular}{c|l|l}
domain & $t$ and $h$ range & description \\ \hline
\texttt{I} & $h=0$, $\epsilon \ll 1$ & zero field, near $T_{c0}$ \\ 
\texttt{II} & $h-h_{c2}\sim $ $\epsilon \ll 1$ & near $T_{c0}$-reflected $%
h_{c2}$-line \\ 
\texttt{III} & $h-h_{c2}(t)\ll 1$, $\epsilon \ll 1$ & near $h_{c2}$-line \\ 
\texttt{I} - \texttt{III} & $h\ll 1$, $\epsilon \ll 1$ & full GL region
\\ 
\texttt{IV} & $t\ll h-h_{c2}(t)$ & region of QFs \\ 
\texttt{V} & $\widetilde{h}\sim t\ll 1$ & quantum-to-classical \\ 
\texttt{VI} & $\widetilde{h}\lesssim t\ll 1$ & classical, near $%
h_{c2}(t\rightarrow 0)$ \\ 
\texttt{VII} & $h-h_{c2}(t)\lesssim t\ll h_{c2}(t)$ & classical, strong
fields \\ 
\texttt{VIII} & $\ln t\gtrsim 1,h\ll t$ & high temperatures \\ 
\texttt{IX} & $h\gg \max \{1,t\}$ & high magnetic fields%
\end{tabular}%
\caption{Explanation of the different domains with $t$ \& $h$ ranges. Here $\protect\epsilon =\ln t$%
. }
\label{tab.domains}
\end{table}

We start the discussion of table~\ref{tab.asym} for domains \texttt{I} -%
\texttt{III}, corresponding to the Ginzburg-Landau region of fluctuations
close to $T_{\mathrm{c0}}$ and in zero magnetic field (domain \texttt{I}).
One can see, that our general expression Eq. (\ref{all}) naturally
reproduces the well known AL, MT and DOS contributions. The only new result
here is the explicitly written contribution $\delta \sigma^{\left( 
\mathrm{DCR}\right) }$ (diagrams 7-10), which was usually ignored in view of
the lack of its divergence close to $T_{\mathrm{c0}}.$ Nevertheless, one can
see that its constant contribution $\sim\ln\ln\left( T_{\mathrm{c0}%
}\tau\right) ^{-1}$ is necessary for matching the GL results with the
neighboring  domains \texttt{VIII} \& \texttt{IX}. The domains \texttt{II} \& \texttt{%
III} \ are still described by the GL theory in weak magnetic fields and Eq. (%
\ref{all}) reproduces all available asymptotic expressions found in literature.

\begin{figure*}[tb]
\begin{center}
\includegraphics[width=0.8\textwidth]{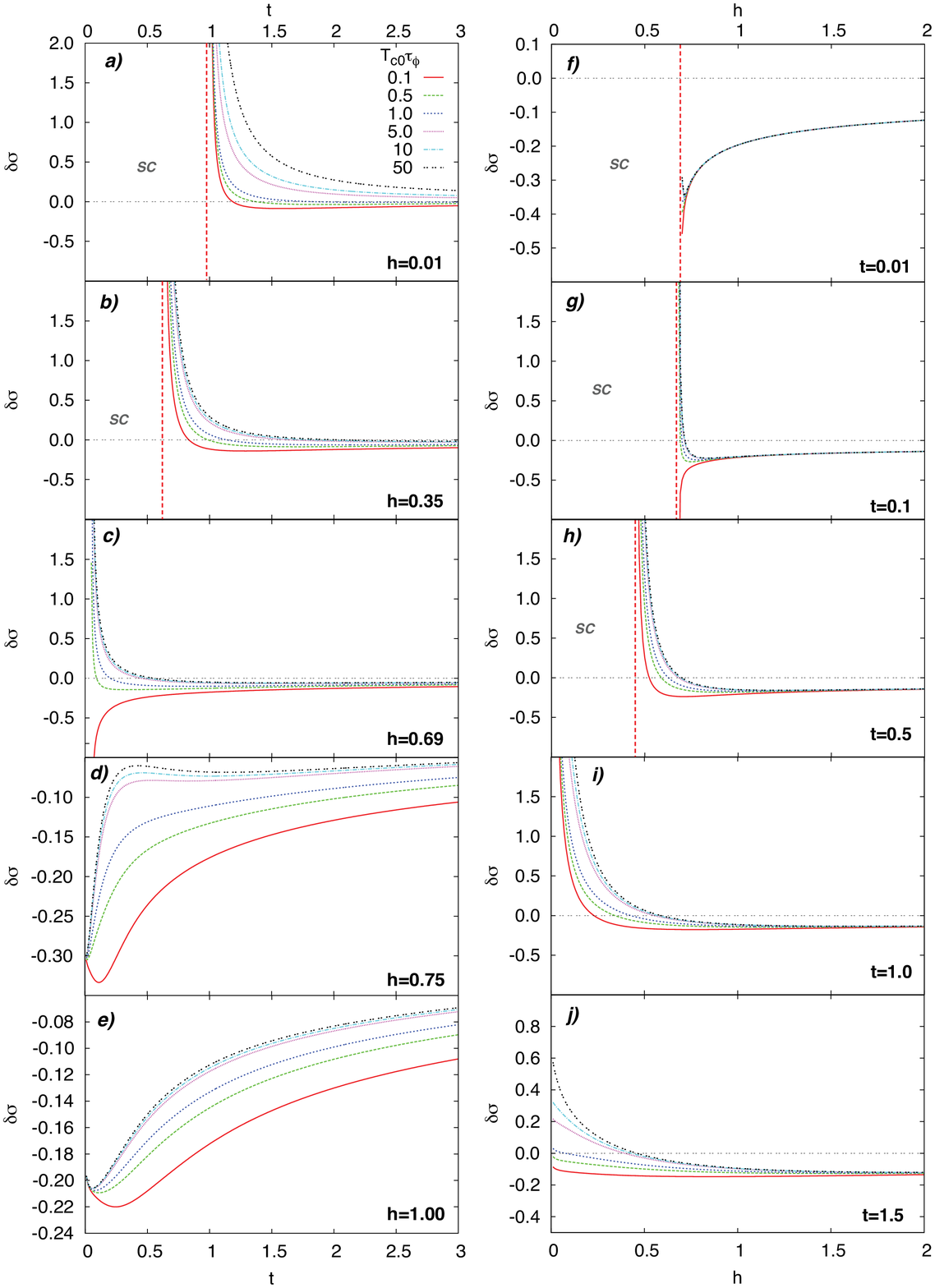}
\end{center}
\caption{(Color online) Total fluctuation conductivity for different $T_{c0}%
\protect\tau_\protect\phi$ for several constant temperatures $t$ and
magnetic fields $h$. 
a) to e) show $\protect\delta\protect\sigma(t)$ for
different magnetic fields below ($h=0.01,0.35$, superconducting region
marked by ''\textit{SC}''), at ($h=0.69$), and above ($h=0.75,1.0$) the zero
temperature critical field $H_{c2}(0)$. The legend key of a) applies to all
panels. Note, that the value of $T_{c0}\protect\tau_\protect\phi$ near the
transition at $t=0$ can determine the sign of the fluctuation conductivity -
however at very low temperatures the FC becomes independent of the
phase-breaking time and all lines coalesce which is not resolved in that plot. f) to j) show $\protect\delta%
\protect\sigma(h)$ for constant temperatures, below ($t=0.01,0.1,0.5$), at ($%
t=1.0$), and above ($t=1.5$) the transition temperature $T_{c0}$. All plots
are calculated for $T_{c0}\protect\tau=10^{-3}$. If $\protect\delta\protect%
\sigma=0$ is within plot range, it is marked by a horizontal dashed line,
and the critical magnetic fields $H_{c2}(T)$ are shown as (red) vertical
dashed lines. See detailed discussion in the text.}
\label{fig.cuts}
\end{figure*}

The most surprising result in table~\ref{tab.asym} is the domain \texttt{IV%
}, the region of quantum fluctuations (see Fig.~\ref{fig.contours}): one
sees that the positive AL (the anomalous MT contribution is equal to the AL
one in that domain) decays with decreasing temperature as $T^{2}$. Moreover,
it is exactly cancelled by the negative contribution of the four DOS-like
diagrams 3-6: 
\begin{equation}
\delta\sigma_{xx}^{\mathrm{AL}}=\delta\sigma_{xx}^{\mathrm{MT(an)}%
}=-\delta\sigma_{xx}^{\mathrm{DOS}}=\frac{4e^{2}\gamma_{E}^{2}t^{2}}{3\pi
^{2}\widetilde{h}^{2}}.  \label{ALlow}
\end{equation}
The total fluctuation contribution to conductivity $\delta\sigma_{xx}^{%
\left( \mathrm{tot}\right) }$ in this important region ($t\ll\widetilde{h})$
is \textit{completely determined by the renormalization of the diffusion
coefficient} (the regular part of the MT contribution and diagrams 7-10). It
turns out to be negative and at zero temperature diverges logarithmically
when the magnetic field approaches $H_{c2}\left( 0\right)$. The non-trivial
fact following from Eq.~(\ref{all}) is that an increase of temperature at a
fixed value of the magnetic field in this domain mainly results in a
further decrease of conductivity 
\begin{equation}
\delta\sigma_{xx}^{\mathrm{(tot)}}=-\frac{2e^{2}}{3\pi^{2}}\ln\frac {1}{%
\widetilde{h}}-\frac{6\gamma_{E}e^{2}}{\pi^{2}}\frac{t}{\widetilde{h}}+O%
\left[ \left( \frac{t}{\widetilde{h}}\right) ^{2}\right],  \label{tot}
\end{equation}
and only at the boundary with domain \texttt{V}, when $t\sim\widetilde{h},$
the total fluctuation contribution $\delta\sigma_{xx}^{\mathrm{(tot)}}$
passes through a minimum and starts to grow. Such non-monotonic behavior of
the conductivity close to $H_{c2}\left( 0\right) $ was multiple times observed in
experiments~\cite{Bridgitte,Batur} (see Fig.~\ref{fig.sigmahconst}).

The domain \texttt{V} describes the transition regime between quantum and
classical fluctuations, while in the domains \texttt{VI}-\texttt{VII},
extended along the line $H_{c2}\left( T\right)$, superconducting
fluctuations have already classical (but non-Ginzburg-Landau) character. In
all these three regions one observes exactly the same cancellation of the AL
and DOS contributions as in domain \texttt{IV} and $\delta\sigma_{xx}^{%
\mathrm{(tot)}}$ is determined by the negative DCR contribution.

\begin{figure}[htb]
\begin{center}
\includegraphics[width=0.9\columnwidth]{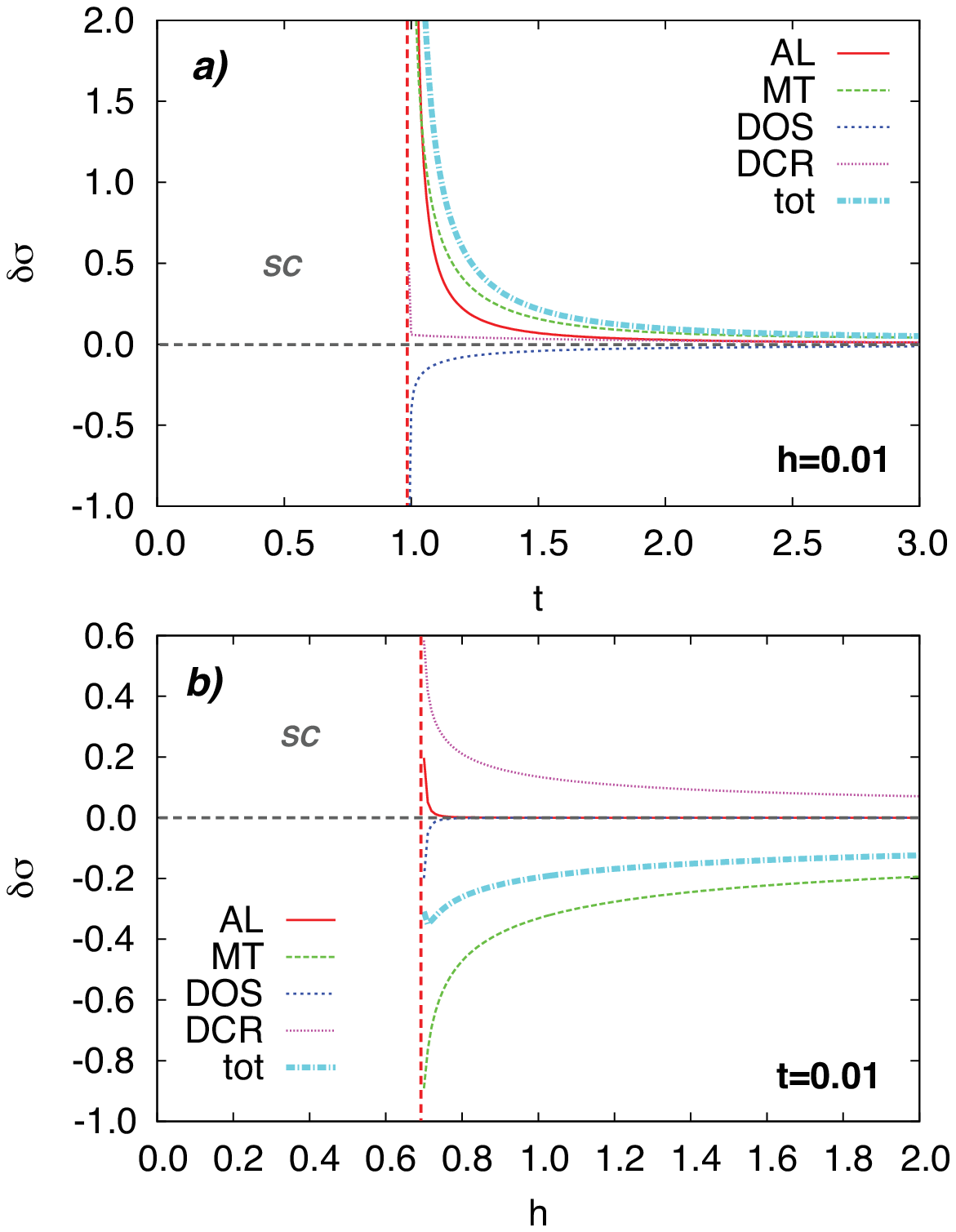}
\end{center}
\caption{(Color online) Fluctuation conductivity contributions: AL, MT, DOS,
DCR, and total (tot) for $T_{c0}\protect\tau=10^{-3}$ and $T_{c0}\protect\tau%
_{\protect\phi}=5$. a) shows the temperature dependence at low field $h=0.01$%
, and b) the field dependence at low temperature $t=0.01$.}
\label{fig.low_ht}
\end{figure}

Finally, in the peripheral domains \texttt{VIII}-\texttt{IX}, the direct
positive contribution of fluctuation Cooper pairs (AL) to conductivity
decays faster than all the other: $\sim \ln ^{-3}\left( T/T_{\mathrm{c0}%
}\right)$. We stress, that this exact result differs from the evaluation of
the AL paraconductivity far from the transition of Ref.~[\onlinecite{AV80}],
but is in complete agreement with the high temperature asymptotic expression
for the paraconductivity of a clean 2D superconductor, see Ref.~[%
\onlinecite{RVV91}]. This agreement seems natural: fluctuation Cooper pair
transport is insensitive to impurity scattering. The anomalous MT
contribution, in complete accordance with Refs.~[\onlinecite{AV80,L80}], decays
as $\sim \ln \gamma _{\phi }^{-1}/\ln ^{-2}\left( T/T_{\mathrm{c0}}\right)$.
The contribution of diagrams 3-6 also decays as $\ln ^{-2}\left( T/T_{%
\mathrm{c0}}\right)$, but without the large factor $\ln \gamma _{\phi }^{-1}$%
. Finally, the regular MT contribution together with the ones from diagrams 7-10 decay
extremely slow, in fact double logarithmically:

\begin{equation}
\delta \sigma _{xx}^{\mathrm{(DCR)}}=-\frac{2e^{2}}{3\pi ^{2}}\left( \ln \ln 
\frac{1}{T_{\mathrm{c0}}\tau }-\ln \ln \frac{T}{T_{\mathrm{c0}}}\right) .
\end{equation}
Up to the numerical prefactor this expression coincides with the results of
Ref.~[\onlinecite{AV80,ARV83}].

Eq. (\ref{all}) provides the basis for a \textquotedblleft
fluctuoscope\textquotedblright for superconductors, i.e. the extraction of
its microscopic parameters from the analysis of fluctuation corrections.
Indeed one can see that $\delta \sigma _{xx}^{\mathrm{(tot)}}$ depends on
two superconducting parameters: $T_{\mathrm{c0}},H_{c2}\left( 0\right) $,
the elastic scattering time $\tau $, and (temperature dependent)
phase-breaking time $\tau _{\phi }\left( T\right) $. The elastic scattering
time can be obtained from the normal state properties of the superconductor,
while the Eq. (\ref{all}) can become the instrument for the precise
determination of the critical temperature $T_{\mathrm{c0}}$ (instead of the
often used rule \textquotedblleft half width of transition\textquotedblright ) and $%
H_{c2}\left( 0\right) $. Moreover, it can be an invaluable tool for the
study of the temperature dependence of the phase-breaking time $\tau _{\phi
}\left( T\right) $. 

The exemplary surface of $\delta \sigma _{xx}^{\mathrm{(tot)}}\left(
T,H\right) $ presented in Fig. \ref{fig.iceberg} for $T_{\mathrm{c0}}\tau
=10^{-2}$ and $T_{\mathrm{c0}}\tau _{\phi }=10$ shows that the value of $%
\tau _{\phi }$ determines the behavior of fluctuation corrections only in
the region of low fields. It is convenient to analyze Fig.~\ref{fig.iceberg}
side-by-side with Fig.~\ref{fig.contours} where lines $\delta \sigma
_{xx}^{\left( \mathrm{tot}\right) }(T,H)=\mathrm{const}$ through the phase
diagram are shown. It is interesting to note that the numerical analysis of
Eq.~(\ref{all}) shows that the logarithmic asymptotic Eq.~(\ref{tot}) is
valid only within an extremely narrow field range $\widetilde{h}\lesssim
10^{-6}$.

In order to get a broader overview of the richness of our main result, we
complied several magneto-conductivity single-parameter dependencies (cuts
through the ''surface'' at constant $t$ or $h$) in Fig.~\ref{fig.cuts}. Each
individual panel a) through j) of this figure shows the total FC for six
different values of $T_{c0}\tau_\phi$ between $0.1$ and $50$ and fixed $%
T_{c0}\tau=10^{-3}$. If $\delta\sigma=0$ is within plot range, it is marked
by a horizontal dashed line, and the critical magnetic fields $H_{c2}(T)$
are shown as (red) vertical dashed lines. a) to e) show $\delta\sigma(t)$
for different magnetic fields below ($h=0.01,0.35$, superconducting region
marked by ''\textit{SC}''), at ($h=0.69$), and above ($h=0.75,1.0$) the zero
temperature critical field $H_{c2}(0)$. The legend key of a) applies to all
panels. The behavior is as expected from the above discussion, but it is
very educative to take a closer look at the behavior near the QPT [panel
c)]: As mentioned above, the asymptotic expression for the quantum regime is
only valid at extremely small temperatures, which cannot be resolved in this
plot. Therefore one sees in particular for the smallest $T_{c0}\tau_\phi$%
-value a sharp dip in the FC at low temperatures, which will eventually
coalesce with all other curves at even lower temperature (not visible) and
become independent of the phase-breaking time, as can be seen at larger $h$
in panel d) and e).

f) to j) show $\delta\sigma(h)$ for constant temperatures, below ($%
t=0.01,0.1,0.5$), at ($t=1.0$), and above ($t=1.5$) the transition
temperature $T_{c0}$.
Here again, it is seen in panel f) that in the quantum regime the FC is
mostly independent of the phase-breaking time and only close to the QPT a
separation of the curves becomes visible since even the small temperature $%
t=0.01$ becomes of order $\widetilde{h}$ or even larger and the asymptotic
expression does not hold anymore.

In Fig.~\ref{fig.low_ht} we plotted two particular curves of Fig.~\ref%
{fig.cuts} in more detail, showing the different contributions from
the diagram groups a)-d) of Fig.~\ref{fig.conddia}. These are the curves for $%
T_{c0}\tau _{\phi }=5$ at lowest magnetic field $h=0.01$ [in a)] and
temperature $t=0.01$ [in b)]. Comparing these curves to the
asymptotics of table~\ref{tab.asym}, one sees, that the behavior near 
$T_{c0}$ is as expected [see a)], and in particular the contribution from diagrams 7-10 is
negligible. However, in the quantum regime it becomes the dominating
contribution, rendering the total FC negative and only close to the QPT is
cancelled by the MT contribution.

Despite Eq. (\ref{all}) being a closed expression, its specific
evaluation in the most general case requires sophisticated numerical
summation and integration. While being straight-forward, one might encounter
technical difficulties in the evaluation of the complex Poly-Gamma functions 
$\psi ^{(n)}(z)$.  
Moreover,  the summation cut-off parameter $M$%
 can reach   extremely large values at low temperatures [
experimental values $\left( T_{c0}\tau \right) _{\exp }^{-1}$ for materials near the
 superconductor-insulator transition can be on the order $10^{6}$],
which slows down the numerical procedure significantly. 
The latter difficulty
can be partially overcome by evaluation of the slowly divergent tails of the 
$m$-sums in Eq. (\ref{all}) as integrals. Here, we should also note
that for fitting purposes one does not need to choose the real, often
extremelly small, experimental values $\left( T_{c0}\tau \right) _{\exp }$%
. To save CPU time, one can assume the value $\left( T_{c0}\tau
\right) _{\mathrm{num}}$ of this parameter to be much larger than $%
\left( T_{c0}\tau \right) _{\exp }$ (but still much less than $%
T_{c0}\tau _{\phi }$) and only at the very end to shift the final
expression by $\ln \ln \frac{\left( T_{c0}\tau \right) _{\mathrm{num}}}{%
\left( T_{c0}\tau \right) _{\exp }}$.  Nevertheless, the numerics of
the problem remains challenging: for the surface plot in Fig.~\ref%
{fig.iceberg} we evaluated $10^{6}$ values for $\delta\sigma$
with the modest assumption $\left( T_{c0}\tau\right) _{\mathrm{%
num}}=0.01$, yet it still took three month of single
CPU time for its calculation. Our optimized tool for the evaluation of Eq. (\ref{all}) can be
found at~[\onlinecite{soft}].

\section{Comparison with Experimental Results}

A main aspect of this work is, that the complete expression, Eq.~\ref{all}, can be
used to extract experimental parameters of thin superconducting films from
measured data (''fluctuoscopy''). In particular the critical temperature $%
T_{c0}$, the critical magnetic field $H_{c2}(0)$, and the phase-breaking
time $\tau_\phi$.

\begin{figure}[htb]
\begin{center}
\includegraphics[width=0.95\columnwidth]{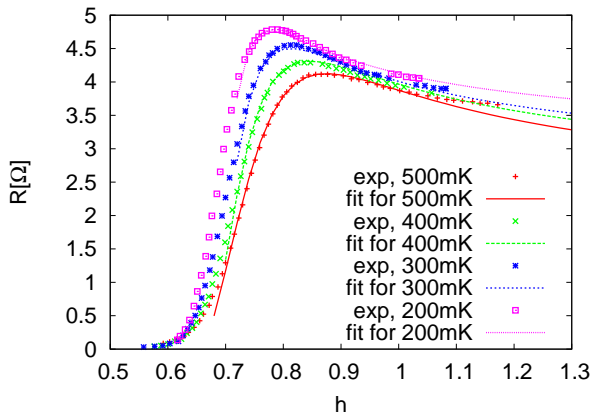}
\end{center}
\caption{(Color online) Comparison to resistivity measurements in thin
indium oxide films, published in Ref.~[\onlinecite{SK05}]. Here we present
the data taken from Fig. 4a of~[\onlinecite{SK05}] for the ''Weak'' sample
with thickness $30$nm, $T_{c0}=3.35$K, and $B_{c2}(0)=13$T. We fitted the
resistivity $R$ for temperatures $0.2$, $0.3$, $0.4$, and $0.5$K using our
full expression for $\protect\delta\protect\sigma$ with the experimentally
found $T_{c0}$. For $B_{c2}(0)$ we fitted a slightly larger value of $13.7$T
and $T_{c0}\protect\tau_{\protect\phi}=5\pm 1$.}
\label{fig.SKfit}
\end{figure}

As an example of the practical use for our results, we fitted two different
sets of experimental data. First we compared our general Eq.~(\ref{all}) to
resistivity measurements in thin disordered indium oxide films, presented in
Ref.~[\onlinecite{SK05}]. Figure~\ref{fig.SKfit} shows the low temperature
data for one sample (referred to as \textquotedblright Weak\textquotedblright in Ref.~\olcite{SK05}) of a
film with thickness $30$nm, transition temperature $T_{c0}=3.35$K and
critical magnetic field $B_{c2}(0)=13$T. The resistivity was measured,
depending on magnetic field, for low temperature values $T=200,300,400,500$%
mK. We plotted the theoretical expression for $\delta \sigma _{xx}^{\mathrm{%
(tot)}}$ using the fitting parameter values $B_{c2}(0)=13.7 $T, $T_{c0}\tau
_{\phi }=5\pm 1$, and the experimentally found value of $T_{c0}=3.35$K.
Overall the fitted FC curves show good agreement with the results of the
measurements.

\begin{figure}[tbh]
\begin{center}
\includegraphics[width=0.9\columnwidth]{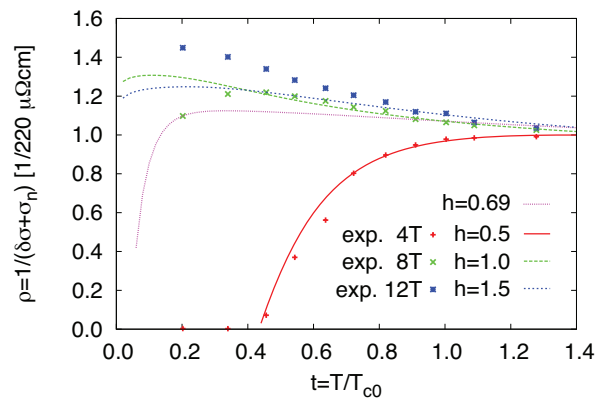}
\end{center}
\caption{(Color online) Temperature dependence of the FC at different fields
close to $H_{c2}(0)$ and comparison to experimental data for a thin film of La$_{2-x}$Ce$_{x}$CuO$_{4}$ ($x=0.09$) with $T_{c0}\approx 22.5$K and  fitted$%
B_{c2}(0)\approx 6$T (see Fig. 1a of Ref.~[\onlinecite{Bridgitte}]a). 
For the theoretical curves a fixed $T_{c0}\protect\tau _{\protect\phi }=0.5$ is used
which nevertheless, captured the overall behavior. All curves
are numerically calculated with $T_{c0}\protect\tau =10^{-3}$. The highest field curve could be possible fitted better, using a smaller $\tau_\phi$. However,
as described in the text, $\tau_\phi$ is in general temperature dependent and the available data does not allow to extract this values.}
\label{fig.sigmahconst}
\end{figure}

Second we re-analyzed data for thin films of the compound La$_{2-x}$Ce$%
_{x}$CuO$_{4}$, with $x=0.09$, in high magnetic fields, see Ref.~\olcite{Bridgitte}a, with thickness $100$nm. 
In contrast
to the indium oxide film, we fitted the temperature dependence of the FC
at different constant high magnetic fields here. This sample has $T_{c0}\approx
22.5$K and we fitted $B_{c2}(0)\approx 6$T, which the $4$T data suggests. In Fig.~\ref{fig.sigmahconst} we plotted
the temperature dependence of Eq. (\ref{all}) for different magnetic fields for
fixed $T_{c0}\tau _{\phi }=0.5$ and $T_{c0}\tau =10^{-3}$. 

At this point it is important to remark, that $\tau _{\phi }$ depends  on
temperature in general, such that for a better fit one needs first to analyze FC data
at constant temperatures to extract $\tau _{\phi }(T)$ and then fit
temperature dependent data. This way one can obtain precise values for the
otherwise difficult to determine experimental parameters $T_{c0}$, $%
H_{c2}(0) $, and $\tau_\phi(T)$. However, the data of Ref.~[\onlinecite{Bridgitte}]a 
does not allow to extract this information for a better fit in Fig.~\ref{fig.sigmahconst} .

\section{Quantum liquid of FCP in the vicinity of $H_{\mathrm{c2}}(0).$}

An analysis of the obtained results allows us to offer a qualitative
picture of the quantum phase transition (QPT) occurring in the vicinity of $%
H_{\mathrm{c2}}(0)$ at very low temperatures.
Above we presented the complete microscopic calculation.
However, it is instructive to start
our discussion of the QFs by describing and refreshing the qualitative 
picture of SFs in the vicinity of $T_{\mathrm{c0}}$, in the
Ginzburg-Landau region\cite{LV09}, for further comparison.
In domains \rI - \rIII, the lifetime of fluctuation-induced
Cooper pairs $\tau_{\mathrm{GL}}$ can be obtained in the simplest way by using
the uncertainty principle.
 Indeed, $\tau_{\mathrm{GL}}\sim\hbar/\Delta E$,
where $\Delta E$ is the energy difference $k_{B}(T-T_{\mathrm{c0}})$ ensuring that $%
\tau_{\mathrm{GL}}$ should become infinite at the transition point. This
yields the standard Ginzburg-Landau time%
\begin{equation}  \label{eq.tauGL}
\tau_{\mathrm{GL}}\sim\hbar/k_{B}(T-T_{\mathrm{c0}})\sim\hbar/\left( k_{B}T_{%
\mathrm{c0}}\epsilon\right) ,
\end{equation}
where $\epsilon=\left( T-T_{\mathrm{c0}}\right) /T_{\mathrm{c0}}\ll1$ is the
reduced temperature. In its turn the coherence length $\xi_{\mathrm{GL}%
}\left( T\right) $ can be estimated as the distance, which two electrons move
apart during the GL time:%
\begin{equation*}
\xi_{\mathrm{GL}}\left( \epsilon\right) =\left( \mathcal{D}\tau _{\mathrm{GL}%
}\right) ^{1/2}\sim\xi_{\mathrm{BCS}}/\sqrt{\epsilon}. 
\end{equation*}
Here $\xi_{\mathrm{BCS}}\sim\sqrt{\mathcal{D}/T_{\mathrm{c0}}}$ is the BCS
coherence length, $\mathcal{D}$\ is the diffusion coefficient. The
fluctuating order parameter $\Delta^{\mathrm{(fl)}}\left( \mathbf{r}%
,t\right) $ varies close to $T_{\mathrm{c0}}$ on a larger scale $\xi_{%
\mathrm{GL}}\left( \epsilon\right) \gg\xi_{\mathrm{BCS}}$. The ratio of the
FCP concentration to the corresponding effective mass with logarithmic
accuracy can be estimated as $n_{\mathrm{c.p.}}/m_{\mathrm{c.p.}}\sim\xi_{%
\mathrm{GL}}^{2-D}\left( \epsilon\right) $ and in the 2D case assume as
constant (which is the case we will discuss in the following)~\cite{LV09}.

The two principal fluctuation contributions to
conductivity close to $T_{\mathrm{c0}}$,  are positive and originate from a direct FCP charge transfer
(AL contribution)%
\begin{equation}
\delta\sigma_{xx}^{\mathrm{AL}}\sim\left( n_{\mathrm{c.p.}}/m_{\mathrm{c.p.}%
}\right) e^{2}\tau_{\mathrm{GL}}\sim e^{2}/\hbar\epsilon  \label{AL}
\end{equation}
and from the specific quantum process of  one-electron charge transfer
related to coherent scattering of electrons on elastic
impurities, which leads to the formation of FCPs  (anomalous MT contribution)%
\begin{equation*}
\delta\sigma_{xx}^{\mathrm{MT(an)}}\sim\frac{e^{2}}{\hbar\epsilon}\ln\left(
\epsilon/\gamma_{\phi}\right) . 
\end{equation*}

However, these two contributions do not capture the complete effect of
fluctuations on conductivity. The involvement of quasi-particles in the
fluctuation pairing results in their absence at the Fermi level, i.e., in the
opening of a pseudo-gap in the one-electron spectrum and consequently
decrease the one-particle Drude-like conductivity. Such an indirect
effect of the FCP formation is usually referred as the DOS contribution. Being
proportional to the concentration of the FCPs $n_{\mathrm{c.p.}}$, the DOS
contribution formally appears by integration of the Fourier-component $%
\left\langle \left\vert \Delta^{\mathrm{\left( fl\right) }}\left( \mathbf{q}%
,\omega\right) \right\vert ^{2}\right\rangle $  of the order parameter  over all
long-wave-length fluctuation modes ($q\lesssim\xi_{\mathrm{BCS}}^{-1}\sqrt{%
\epsilon}$); in the static approximation ($\omega\rightarrow0$) given by: 
\begin{equation}
\delta\sigma_{xx}^{\mathrm{DOS}}\sim-\frac{2n_{\mathrm{c.p.}}e^{2}\tau }{m_{%
\mathrm{e}}}\sim-e^{2}\int\frac{\xi_{\mathrm{BCS}}^{2}d^{2}\mathbf{q}}{%
\epsilon+\xi_{\mathrm{BCS}}^{2}q^{2}}\sim-\frac{e^{2}}{\hbar}\ln\frac {1}{%
\epsilon}.  \label{DOS}
\end{equation}
One sees that the DOS contribution has the opposite sign with respect to the AL
and MT contributions, but close to $T_{\mathrm{c0}}$ does not compete with
those, since it turns out to be less singular as a function of temperature.

Finally, the one-electron diffusion coefficient is renormalized
in the presence of fluctuation pairing (DCR). 
Close to $T_{\mathrm{c0}%
} $ this contribution is not singular in $\epsilon$ (see table~\ref{tab.asym}) and was
usually ignored in literature, but as was mentioned before, it becomes of primary importance
relatively far from $T_{\mathrm{c0}}$, and at very low temperatures. It is
due to $\delta\sigma_{xx}^{\mathrm{DCR}}$\ that the sign
of the total contribution of fluctuations to conductivity $%
\delta\sigma_{xx}^{\mathrm{(tot)}}$ changes in a wide domain of the phase diagram
and in particular close to $T=0,$ in the region of quantum fluctuations (see
Fig.~\ref{fig.contours}, where the regions with dominating fluctuation
contributions to magneto-conductivity are shown).

At zero temperature and fields above $H_{\mathrm{c2}}\left( 0\right) $, the
systematics of the fluctuation contributions to the conductivity
changes considerably  with respect to that close to $T_{\mathrm{c0}}$. Due to the
collision-less rotation of FCPs (they do not \textquotedblright feel" the
presence of elastic impurities, all information concerning electron
scattering is already included in the effective mass of the Cooper pairs)
they do not contribute directly to the longitudinal (along
the applied electric field) electric transport (analogously to the
suppression of the one-electron conductivity in strong magnetic fields $\left(
\omega_{c}\tau\gg1\right) $: $\delta \sigma_{xx}^{\left( \mathrm{e}\right)
}\sim$ $\left( \omega_{c}\tau\right) ^{-2}$, see Ref.~[\onlinecite{A88}])
and the AL contribution to $\delta\sigma _{xx}^{\left( \mathrm{tot}\right) }$
becomes zero. The anomalous MT and DOS contributions tend to zero as well but
because of different reasons. Namely, the former vanishes since magnetic fields
as large as $H_{\mathrm{c2}}\left( 0\right) $ completely destroy the phase
coherence, whereas the latter disappears since magnetic field suppresses the
fluctuation gap in the one-electron spectrum. Therefore the effect of
fluctuations on the conductivity at zero temperature is reduced to the
renormalization of the one-electron diffusion coefficient. FCPs in the quantum region occupy
the lowest Landau level, but all dynamic fluctuations in the 
frequency interval from $0$ to $\Delta_{\mathrm{BCS}}$ have to be taken into
account. The corresponding fluctuation propagator at zero temperature close
to $H_{\mathrm{c2}}\left( 0\right) $ has the form (see Eq. (\ref{emas})) 
\begin{equation*}
L_{0}\left( \omega\right) =-\nu_{0}^{-1}\frac{1}{\widetilde{h}+\omega
/\Delta_{\mathrm{BCS}}} 
\end{equation*}
and 
\begin{equation}
\delta\sigma_{xx}^{\mathrm{DCR}}\sim-\frac{e^{2}}{\Delta_{\mathrm{BCS}}}%
\int_{0}^{\Delta_{\mathrm{BCS}}}\frac{d\omega}{\widetilde{h}+\frac{\omega }{%
\Delta_{\mathrm{BCS}}}}\sim-\frac{e^{2}}{\hbar}\ln\frac{1}{\widetilde{h}}.
\label{DCR}
\end{equation}
The parameter $\widetilde{h}=\left[ H-H_{\mathrm{c2}}\left( 0\right) \right]
/H_{\mathrm{c2}}\left( 0\right) $ plays the same role as the reduced
temperature $\epsilon$ in the case of the classical transition; $%
\Delta_{\mathrm{BCS}}$ is the BCS value of the gap at zero temperature in
zero field.

While the denominator of the integrand in Eq. (\ref{DOS}) defines the
characteristic wavelength $\xi_{\mathrm{GL}}\left( T\right) $ of the
fluctuation modes close to $T_{\mathrm{c0}}$, the one in Eq. \ (\ref{DCR})
defines the characteristic coherence time $\tau_{\mathrm{QF}}\left( 
\widetilde{h}\right) $ of QFs near $H_{\mathrm{c2}}\left( 0\right) $ (where $%
t\ll\widetilde{h}$). The value of the integral is determined by its lower
cut-off $\omega_{\mathrm{QF}}\sim\Delta_{\mathrm{BCS}}\widetilde{h}$, and
the corresponding time scale is 
\begin{equation}
\tau_{\mathrm{QF}}\sim\hbar\left( \Delta_{\mathrm{BCS}}\widetilde{h}\right)
^{-1}.  \label{tauqf}
\end{equation}
One sees that the functional form of $\tau_{\mathrm{QF}}$ is completely
analogous to that of $\tau_{\mathrm{GL}}$: $\Delta_{\mathrm{BCS}}\triangleq T_{%
\mathrm{c0}}$ and the reduced field $\widetilde{h}$ plays the role of
reduced temperature $\epsilon$. 
Eq. (\ref{tauqf}) can also be obtained from the uncertainty principle. Indeed, the energy, characterizing the
proximity to the quantum phase transition is $\Delta E=\hbar\omega_{c}\left(
H\right) -\hbar\omega_{c}\left( H_{\mathrm{c2}}\left( 0\right) \right)
\sim\Delta_{\mathrm{BCS}}\widetilde{h}$ and namely this value should be used
in the Heisenberg relation instead of $k_{B}(T-T_{\mathrm{c0}})$, as was
done in the vicinity of $T_{\mathrm{c0}}$. The spatial coherence scale $\xi_{%
\mathrm{QF}}\left( \widetilde{h}\right) $ can be estimated from the value of 
$\tau_{\mathrm{QF}}$ analogously to the consideration near $T_{\mathrm{c0}}$.
Namely, two electrons with coherent phase starting from the same point get separated by the distance 
\begin{equation*}
\xi_{\mathrm{QF}}\left( \widetilde{h}\right) \sim\left( \emph{D}\tau_{%
\mathrm{QF}}\right) ^{1/2}\sim\xi_{\mathrm{BCS}}/\sqrt{\widetilde{h}},
\end{equation*}
after time $\tau_{\mathrm{QF}}$.

To clarify the physical meaning of $\tau_{\mathrm{QF}}$ and $\xi_{\mathrm{QF}%
}$, note that near the quantum phase transition at zero temperature, where $%
H\rightarrow H_{\mathrm{c2}}\left( 0\right) $, the fluctuations of the order
parameter $\Delta^{\mathrm{(fl)}}\left( \mathbf{r},t\right) $ become highly
inhomogeneous, contrary to the situation near $T_{\mathrm{c0}}$. Indeed,
below $H_{\mathrm{c2}}\left( 0\right) $, the spatial distribution of the
order parameter at finite magnetic field reflects the appearance of Abrikosov
vortices with average spacing [close to $H_{\mathrm{c2}}\left( 0\right) $
but in the region where the notion of vortices is still adequate] equal to%
\begin{equation*}
a\left( H\right) =\xi_{\mathrm{BCS}}/\sqrt{H/H_{\mathrm{c2}}\left( 0\right) }%
\rightarrow\xi_{\mathrm{BCS}}. 
\end{equation*}
Therefore, one expects that close to and above $H_{\mathrm{c2}}\left(
0\right) $ the fluctuation order parameter $\Delta^{\mathrm{(fl)}}\left( 
\mathbf{r},t\right) $ also has a \textquotedblright vortex-like" spatial
structure and varies over the scale  $\xi_{\mathrm{BCS}}$ and being preserved
over time $\tau_{\mathrm{QF}}$. In the language of FCPs, one
describes this situation in the following way: A FCP at zero temperature
and in magnetic field close to $H_{\mathrm{c2}}\left( 0\right) $ rotates
with Larmor radius $r_{L}\sim v_{\mathrm{F}}/\omega_{\mathrm{c}}\left(
H_{\mathrm{c2}}\left( 0\right) \right) \sim v_{\mathrm{F}}/\Delta _{\mathrm{%
BCS}}\sim\xi_{\mathrm{BCS}}$, which represents its effective size. During
 time $\tau_{\mathrm{QF}}$ two initially selected electrons participate
in multiple fluctuating Cooper pairings maintaining their coherence.
The coherence length $\xi_{\mathrm{QF}}\left( \widetilde{h}\right) \gg\xi_{%
\mathrm{BCS}}$ is thus a characteristic size of a cluster of such coherently
rotating FCP, and $\tau_{\mathrm{QF}}$ estimates the lifetime of such a
{\it flickering} cluster. One can view the whole system as an ensemble of
flickering domains of coherently rotating FCP, precursors of vortices (see
Fig.~\ref{fig.abrikosovQC}).

In view of the qualitative picture of SFs in the regime of the QPT, let us
continue with the scenario of Abrikosov lattice defragmentation:
 Approaching $H_{%
\mathrm{c2}}\left( 0\right) $ from below, puddles of fluctuating
vortices are formed, which are nothing else as FCPs rotating in a magnetic field. 
Their characteristic size is $\xi_{\mathrm{QF}}\left( |\widetilde {h}%
|\right) ,$ and they flicker in the characteristic time $\tau _{\mathrm{QF}%
}\left( |\widetilde{h}|\right) $. 
In this situation, the supercurrent  can still
flow through the sample until these puddles do not break the last
percolating superconductive channel. The corresponding field determines the
value of the by QFs renormalized second critical field: $H_{\mathrm{c2}%
}^{\ast}\left( 0\right) =H_{\mathrm{c2}}\left( 0\right) \left[ 1-2\mathrm{Gi}%
\ln\left( 1/\mathrm{Gi}\right) \right] $ (see Ref.~[\onlinecite{LV09}]).
Above this field no supercurrent can flow through the sample anymore, i.e., the system
is in the normal state. Nevertheless, as demonstrated by the above estimates, its
properties are strongly affected by the QF. Fragments of the Abrikosov
lattice can be still observed in this region by the following Gedanken experiment: The
clusters of rotating FCP (''ex-vortices'') of size $\xi _{\mathrm{QF}}$ with
some kind of the superconducting order should be found in the background of the
normal state, if one takes a picture with exposure time shorter than $%
\tau_{\mathrm{QF}}$. For exposure times longer than $\tau_{%
\mathrm{QF}}$, the picture is smeared out and no traces of the Abrikosov vortex
state can be found. However, the detailed nature of the order which exists there is still
unclear. 
It would be attractive to identify these clusters with fragments of the Abrikosov lattice, 
but most probable this is some kind of
quantum FCP liquid. 
Indeed, the presence of  structural disorder can result
in the formation  of a
hexatic phase close to $H_{\mathrm{c2}}^{\ast}\left( 0\right) $,
 where the translational invariance no longer exists, while at the same time 
 conserving the orientational order
 or the vortices.

\section{Discussion}

In terms of the introduced QF characteristics $\tau _{\mathrm{QF}}$
and $\xi _{\mathrm{QF}}$, one can understand the meaning of already found
microscopic QF contributions to different physical values in the vicinity of 
$H_{\mathrm{c2}}\left( 0\right) $ and derive others which are related.

\subsection{In-plane conductivity}

For example, the physical meaning of Eq. (\ref{ALlow}) can be understood as
follows: one could estimate the FCP conductivity by merely replacing $\tau _{%
\mathrm{GL}}\rightarrow \tau _{\mathrm{QF}}$ in the classical AL expression (%
\ref{AL}), which would give $\delta \widetilde{\sigma }^{AL}\sim e^{2}\tau _{%
\mathrm{QF}}.$ Nevertheless, as we already noticed, a FCP at zero
temperature cannot drift along the electric field but only rotates around a
fixed center. As temperature deviates from zero,  FCPs can change their
state due to the interaction with the thermal bath, i.e. their hopping to an
adjacent rotation trajectory along the applied electric field becomes
possible. This means that FCP  can participate in longitudinal charge
transfer now. This process can be mapped onto the paraconductivity of a granular
superconductor~\cite{LVV08} at temperatures above $T_{\mathrm{c0}}$, where
the FCP tunneling between grains occurs in two steps: first one electron
jumps, then the second follows. The probability of each hopping event is
proportional to the inter-grain tunneling rate $\Gamma .$ To conserve the
superconducting coherence between both events, the latter should occur
during the FCP lifetime $\tau _{\mathrm{GL}}.$\ The probability of FCPs
tunneling between two grains is determined by the conditional probability of
two one-electron hopping events and is proportional to $W_{\Gamma }=\Gamma
^{2}$\ $\tau _{\mathrm{GL}}.$ \ Coming back to the situation of FCPs above $%
H_{\mathrm{c2}}\left( 0\right) $, one can identify the tunneling rate with
temperature $T$ while $\tau _{\mathrm{GL}}$ corresponds to $\tau _{\mathrm{QF%
}}.$ Therefore, in order to obtain a final expression, $\delta \widetilde{%
\sigma }^{\mathrm{AL}}$ should be multiplied by the probability factor $W_{%
\mathrm{QF}}=t^{2}\tau _{\mathrm{QF}}$ of the FCP hopping to the neighboring
trajectory: 
\begin{equation*}
\delta \sigma _{xx}^{\mathrm{AL}}\sim \delta \widetilde{\sigma }^{\mathrm{AL}%
}W_{\mathrm{QF}}\sim e^{2}t^{2}/\widetilde{h}^{2}, 
\end{equation*}
which corresponds to the asymptotic Eq. (\ref{ALlow}).

\subsection{Magnetic susceptibility}

In order to estimate the contribution of QFs to the fluctuation induced magnetic
susceptibility of the SC in the vicinity of $H_{\mathrm{c2}}\left( 0\right)$,
one can apply the Langevin formula to a coherent cluster of FCPs and
identify its average size by the rotator radius. One finds
\begin{equation*}
\chi ^{\mathrm{AL}}=\frac{e^{2}n_{\mathrm{c.p}.}}{m_{\mathrm{c.p}.}c}%
\left\langle \xi _{\mathrm{QF}}^{2}\left( \widetilde{h}\right) \right\rangle
\sim \xi _{\mathrm{BCS}}^{2}/c\widetilde{h} 
\end{equation*}
in complete agreement with the result of Ref.~[\onlinecite{GL01}].

\subsection{Nernst coefficient}

One further reproduces the contribution of QFs to the Nernst coefficient.
Close to $H_{\mathrm{c2}}\left( 0\right) $ the chemical potential of FCPs \
can be identified as $\mu _{\mathrm{FCP}}=\hbar \omega _{\mathrm{c}}\left(
H_{\mathrm{c2}}\left( 0\right) \right) -\hbar \omega _{\mathrm{c}}\left(
H\right) $ [as in Ref.~[\onlinecite{SSVG09}], close to $T_{\mathrm{c0}}$, $%
\mu _{\mathrm{FCP}}=k_{\mathrm{B}}\left( T_{\mathrm{c0}}-T\right) $].
The corresponding derivative is $d\mu _{\mathrm{FCP}}/dT\sim dH_{\mathrm{c2}}\left(
T\right) /dT\sim -T/\Delta _{\mathrm{BCS}}.$ Using the relation between the
latter and the Nernst coefficient, it is possible to reproduce one of the
results of Ref.~[\onlinecite{SSVG09}]:

\begin{equation*}
\nu^{\mathrm{AL}}\sim\left[ \tau_{\mathrm{QF}}/m_{\mathrm{c.p}.}\right]
d\mu_{\mathrm{FCP}}/dT\sim\xi_{\mathrm{BCS}}^{2}t/\widetilde{h}. 
\end{equation*}

\subsection{Transversal magneto-resistance above $H_{\mathrm{c2}}\left(
0\right) $}

The proposed qualitative approach can also explain the non-monotonic
behavior of the transversal magneto-resistance observed in the {\it
layered} organic superconductor $\kappa -(BEDT-TTF)_{2}Cu(NCS)_{2}$ above $%
H_{\mathrm{c2}}\left( 0\right) $ at low temperatures~\cite{MK10}. Indeed,
the motion of FCPs along the z-axis in such a system has hopping character and
the quasi-particle spectrum can be assumed to have the form of a corrugated
cylinder. Close to $T_{\mathrm{c0}}$ the fluctuation magneto-conductivity
tensor in this model was already studied in details in Ref.~[%
\onlinecite{BDKLV93}]. There it was demonstrated that the transverse
paraconductivity in that case is suppressed by the square of the small
anisotropy parameter $\left( \xi _{z}/\xi _{x}\right) ^{2}$, while the
dependence on the reduced temperature $\epsilon $ is even more singular
than in plane. In terms of the Ginzburg-Landau FCP life-time (\ref{eq.tauGL}), 
it can be written as
\begin{equation}
\delta \sigma _{zz}^{\mathrm{AL}}\left( \epsilon \right) =\frac{4e^{2}\xi
_{z}^{4}}{\pi ^{2}\xi _{xy}^{2}s^{3}}T_{\mathrm{c0}}^{2}\tau _{\mathrm{GL}%
}^{2}\left( \epsilon \right) ,  \label{ALperp}
\end{equation}%
where $s$ is the interlayer distance. In principle this
result could be obtained, even from the Drude formula applied to the FCP
charge transfer [see above, how  Eq. (\ref{AL}) for $\delta
\sigma _{xx}^{\mathrm{AL}}\left( \epsilon \right) $ was obtained] combined with the above
speculations regarding the hopping of FCPs along z-axis \cite{LVV08}. This
general approach, which does not involve the GL scheme, allows us to map Eq. (\ref{ALperp}) on the case of the QPT by just
replacing $\tau _{\mathrm{GL}}\left( \epsilon \right) \rightarrow \tau _{%
\mathrm{QF}}\left( \widetilde{h}\right):$ 
\begin{equation*}
\delta \sigma _{zz}^{\mathrm{AL}}\left( \widetilde{h}\right) =\frac{%
4e^{2}\xi _{z}^{4}}{\xi _{x}^{2}s^{3}}T_{\mathrm{c0}}^{2}\tau _{\mathrm{QF}%
}^{2}\left( \widetilde{h}\right) =\frac{4e^{2}\xi _{z}^{4}}{\xi _{x}^{2}s^{3}%
}\left( \frac{\gamma _{E}}{\pi }\right) ^{2}\frac{1}{\widetilde{h}^{2}}\,.
\end{equation*}%

The negative contribution appearing from the diffusion coefficient
renormalization competes with the positive $\delta \sigma _{zz}^{\mathrm{AL}%
}\left( \widetilde{h}\right) $.
The only difference between the in-plane [see Eqs. (\ref{tot}) \& (\ref{DCR})]
and z-axis components of this one-particle contribution consists in the
anysotropy factor $\left\langle v_{z} ^{2}\right\rangle/v_{x}^{2}=\xi
_{z}^{2}/\xi _{x}^{2}.$ As a result one gets:
\begin{equation*}
\delta \sigma _{zz}^{\mathrm{(DCR)}}=-\frac{2e^{2}}{3\pi ^{2}s}\frac{\xi
_{z}^{2}}{\xi _{x}^{2}}\ln \frac{1}{\widetilde{h}} 
\end{equation*}
and the total fluctuation correction to the $z$-axis magnetoconductivity at
zero temperature above $H_{\mathrm{c2}}\left( 0\right) $ can be written as 
\begin{equation}  \label{eq.sigmazz}
\delta \sigma _{zz}^{\mathrm{(tot)}}=\frac{2e^{2}\xi _{z}^{2}}{3\pi ^{2}\xi
_{x}^{2}s}\left[ 1.94\left( \frac{\xi _{z}}{s}\right) ^{2}\frac{1}{%
\widetilde{h}^{2}}-\ln \frac{1}{\widetilde{h}}\right] .
\end{equation}

\begin{figure}[tbh]
\begin{center}
\includegraphics[width=0.95\columnwidth]{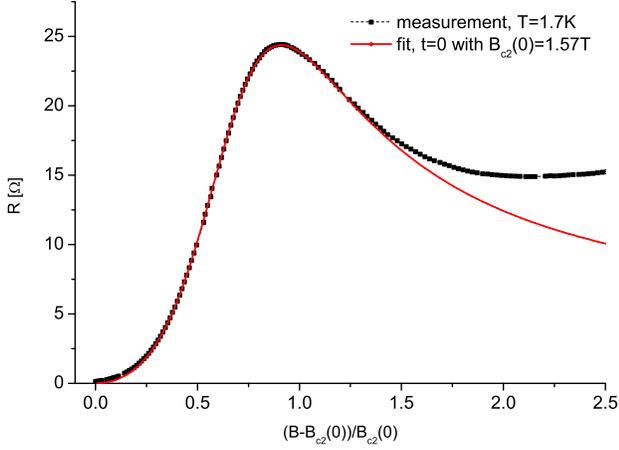}
\end{center}
\caption{(Color online) Comparison to resistivity measurements of the
layered organic superconductor $\protect\kappa -(BEDT-TTF)_{2}Cu(NCS)_{2}$~[\onlinecite{MK10}]. The material has a transition temperature of $%
T_{c0}\approx 9.5$K, $B_{c2}(0)\approx 1.57$T, and $\protect\tau =1.7$ps.
This experimental curve is taken at $T=1.7$K and fitted by expression in Eq.~(\protect
\ref{eq.sigmazz}), which is in perfect agreement with the experiment.
Specifics are given in the text.}
\label{fig.MKfit}
\end{figure}

We used Eq. (\ref{eq.sigmazz}) for the analysis of unpublished data by M.
Kartzovnik~\cite{MK10} on the magneto-resistance of the layered organic
superconductor $\kappa -(BEDT-TTF)_{2}Cu(NCS)_{2}$ at low temperatures and
magnetic fields above $H_{\mathrm{c2}}\left( 0\right) $. The measurement was
taken at $T=1.7$K with a $T_{c0}\approx 9.5$K and $B_{c2}(0)\approx 1.57$T
and this curve was fitted by $0.23\left(0.18/\widetilde{h}^2+\ln \widetilde{h%
}\right)$, see Fig.~\ref{fig.MKfit}. 
For the material parameters of this compound, the author reports $\tau
=1.7$ps, $\xi _{z}=0.3-0.4$ nm, and $s=1$ nm. The fitting shown in Fig.~\ref%
{fig.MKfit} corresponds to the ratio $\xi _{z}/s=0.32$ and looks rather
convincing.

The discrepancy appearing between the theoretical and experimental curves in
the high field region, M. Kartsovnik attributes to the large normal-state
magneto resistance, reflecting the specifics of the cyclotron orbits on the
multi-connected Fermi surface of the compound (due to the low crystal
symmetry it is quite difficult to fit).

\begin{acknowledgments}
We thank T. \thinspace Baturina, Yu.\thinspace Galperin, M.\thinspace Kartsovnik,  A.\thinspace Koshelev,
B.\thinspace Leridon, and M.\thinspace Norman  for useful discussions. The work was
supported by the U.S. Department of Energy Office of Science under the
Contract No. DE-AC02-06CH11357. A.A.V. acknowledges support of the MIUR under
the project PRIN 2008 and the European Community FP7-IRSES programs: ''ROBOCON'' and ''SIMTECH''.
\end{acknowledgments}

\appendix

\section{Aslamazov-Larkin contribution}

\subsection{General expression}

Let us start with the discussion of the AL contribution (diagram 1 in Fig. %
\ref{fig.conddia}). The corresponding analytic expression is%
\begin{align}
Q_{xx}^{\mathrm{AL}}(\omega_{\nu})= &
-4e^{2}T\sum_{\Omega_{k}}\sum_{\{n,m\}=0}^{\infty}\mathbf{B}_{nm}^{\left(
x\right) }(\Omega_{k+\nu},\Omega _{k})L_{m}(\Omega_{k})\label{sigfield}  \\
& \times B_{mn}^{\left( x\right)
}(\Omega_{k},\Omega_{k+\nu})L_{n}(\Omega_{k+\nu}).   \notag
\end{align}
The block of three Green functions $\mathbf{B}_{nm}$ with velocity operator
(originating from the current vertex) and two Cooperons is given by 
\begin{align}
\mathbf{B}_{nm}(\Omega_{k+\nu},\Omega_{k}) & =T\sum_{\varepsilon_{i}}\mathrm{%
Tr}\left\{ G\left( \varepsilon_{i}\right) \widehat{\mathbf{v}}G\left(
\varepsilon_{i+\nu}\right) \times\right.  \notag \\
& \left. \widehat{\lambda}_{n}(\varepsilon_{i+\nu},\Omega_{k-i})G\left(
\Omega_{k-i}\right) \widehat{\lambda}_{m}(\Omega_{k-i},\varepsilon
_{i})\right\} .  \label{b}
\end{align}
The trace operator $\mathrm{Tr}$ denotes the integration over all electron
quantum numbers. The corresponding block was calculated in~[\onlinecite{GL01}%
] exactly for fields with $\omega_{\mathrm{c}}\tau\ll1$, i.e. for the
case of our interest. Under this condition the Landau quantization affects
the motion of Cooper pairs, while the Green functions in the block Eq. (\ref%
{b}) can be used in $\tau-$approximation. As the result, using the
properties of the velocity operator in Landau representation, one finds

\begin{align}
B_{mn}^{\left( x\right) }(\Omega_{k+\nu},\Omega_{k}) & =-2\nu _{0}\mathcal{D}%
\left[ \sqrt{eH(n+1)}\delta_{m,n+1}\right.  \notag \\
& \left. +\sqrt{eHn}\delta_{m,n-1}\right] \Xi_{nm}(\Omega_{k+\nu},%
\Omega_{k}),  \label{Bsi}
\end{align}
with%
\begin{align}
\Xi_{nm}(\Omega_{k},\Omega_{k+\nu}) & =2\pi T\sum_{\varepsilon_{i}}\frac{%
\Theta\left( -\varepsilon_{i+\nu}\Omega_{k-i}\right) }{|2\varepsilon
_{i}+\omega_{\nu}-\Omega_{k}|+\omega_{\mathrm{c}}(n+1/2)}  \notag \\
& \cdot\frac{\Theta\left( -\varepsilon_{i}\Omega_{k-i}\right) }{%
|2\varepsilon_{i}-\Omega_{k}|+\omega_{\mathrm{c}}(m+1/2)}.  \label{thesi}
\end{align}
Substituting Eq. (\ref{Bsi}) in Eq. (\ref{sigfield}) and further
summation over Landau levels\ in Eq. (\ref{sigfield}), results in the cancellation
of the terms containing the products $\delta_{m,n+1}\delta_{n,m+1}$ and $%
\delta_{m,n-1}\delta_{n,m-1}.$ The analysis of the theta-functions in Eq. (\ref%
{thesi}) results in the possibility of separation of different domains of
analyticity in the plane of bosonic frequencies $\Omega_{k}:$

\begin{figure}[htb]
\begin{center}
\includegraphics[width=0.6\columnwidth]{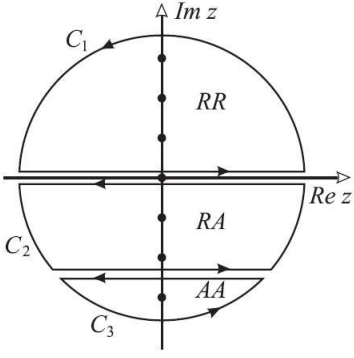}
\end{center}
\caption{The integration contour in the plane of complex frequencies.}
\label{fig.AL}
\end{figure}


\begin{widetext}
\begin{align}
\Xi_{mn}(\Omega_{k},\Omega_{k}+\omega_{\nu})= &  2\pi T\left[
\Theta\left( \Omega_{k}\right)  \sum_{i=k}^{\infty}+\Theta\left(
-\Omega_{k}\right) \sum_{i=0}^{\infty}+\Theta\left(
-\Omega_{k}-\omega_{\nu}\right) \sum_{i=-\infty}^{k-1}+\Theta\left(
\Omega_{k}+\omega_{\nu}\right)
\sum_{i=-\infty}^{\nu-1}\right]  \cdot\nonumber\\
&  \frac{1}{|2\varepsilon_{i}+\omega_{\nu}-\Omega_{k}|+\omega_{\mathrm{c}%
}(n+1/2)}\frac{1}{|2\varepsilon_{i}-\Omega_{k}|+\omega_{\mathrm{c}}%
(m+1/2)}.\label{sigma}%
\end{align}
Summation over fermionic frequency in this expression can  already be
performed in
terms of $\psi-$functions:%
\begin{align}
\Xi_{mn}(\Omega_{k},\Omega_{k}+\omega_{\nu})=  &  \frac{1}{2\omega
_{\mathrm{c}}\left(  n-m\right)  }\left[  \psi\left(
\frac{1}{2}+\frac
{\omega_{\nu}+|\Omega_{k}|+\omega_{\mathrm{c}}(n+1/2)}{4\pi
T}\right) -\psi\left(
\frac{1}{2}+\frac{|\Omega_{k}|+\omega_{\mathrm{c}}(m+1/2)}{4\pi
T}\right)  \right. \nonumber\\
&  +\left.  \psi\left(  \frac{1}{2}+\frac{|\Omega_{k+\nu}|+\omega_{\mathrm{c}%
}(n+1/2)}{4\pi T}\right)  -\psi\left(  \frac{1}{2}+\frac{\omega_{\nu}%
+|\Omega_{k+\nu}|+\omega_{\mathrm{c}}(m+1/2)}{4\pi T}\right)
\right]  .
\label{gransig}%
\end{align}
\end{widetext}
Being interested in the d.c. fluctuation conductivity, i.e.
taking into account the limit $\omega_{\nu}%
\rightarrow-i\omega\rightarrow0$ after analytical continuation,  in Eq. (\ref{gransig}) we neglected the
frequency $\omega_{\nu}$ in comparison with $\omega_{\mathrm{c}}\left(
n-m\right) $ in denominator since the diagonal term ($m=n$) disappears in the process of
summation over Landau levels in Eq. (\ref{sigfield})  as
follows from Eq. (\ref{Bsi}). One notices the useful fact that the
permutation $\Omega _{k}\ \Leftrightarrow\Omega_{k}+\omega_{\nu}$
simultaneously with $m\Leftrightarrow n$ in Eq. (\ref{gransig}) does not
change the function $\Xi_{mn}(\Omega_{k},\Omega_{k}+\omega_{\nu}):$%
\begin{equation}
\Xi_{mn}(\Omega_{k},\Omega_{k}+\omega_{\nu})\equiv\Xi_{nm}(\Omega_{k}+%
\omega_{\nu},\Omega_{k}).  \label{permut}
\end{equation}

Let us return to the general expression for paraconductivity Eq. (\ref%
{sigfield}). One can transform the sum over the bosonic frequencies $%
\Omega_{k} $ to the contour integral $I^{\mathrm{AL}}$ in the plane of
complex frequency $\Omega_{k}\rightarrow -iz$:%
\begin{equation}
Q_{xx}^{\mathrm{AL}}(\omega_{\nu})=-16e^{2}\nu_{0}^{2}\mathcal{D}%
^{2}eH\sum_{n,m}^{\infty}C_{mn}I_{nm}^{\mathrm{AL}}\left(
\omega_{\nu}\right) ,  \label{qan}
\end{equation}%
\begin{align}
I_{nm}^{\mathrm{AL}}\left( \omega_{\nu}\right) & =\frac{1}{4\pi i}\oint
\coth\left( \frac{z}{2T}\right) dz\Xi_{nm}(-iz+\omega_{\nu},-iz)\times
\label{eliashx} \\
& \Xi_{mn}(-iz,-iz+\omega_{\nu})L_{m}(-iz)L_{n}(-iz+\omega_{\nu}),  \notag
\end{align}
where the contour integral encloses all frequencies $\Omega_{k}$ [in the
plane of frequency $z$ these are poles of $\coth\left( z/2T\right) $, see
Fig. \ref{fig.AL}]. The coefficients%
\begin{equation}
C_{mn}=\left( \delta_{m,n+1}\delta_{n,m-1}+\delta_{n,m+1}\delta
_{m,n-1}\right) \sqrt{n}\sqrt{(n+1)}  \label{c}
\end{equation}
control the summation over Landau levels.

Let us stress that both functions $\Xi$ in Eq. (\ref{eliashx}) have breaks
of their analyticity along the lines $\Im z=0$ and $\Im z z=-\omega_{\nu}$,
the same as the product of the propagators. As a result, one gets three domains
where the integrand function is analytical:\ above the line $\Im z=0$,
between the lines $\Im z=0$ and $\Im z z=-\omega_{\nu} $ and below $\Im
z=-\omega_{\nu}.$ For the analytical continuation of function (\ref{gransig})
to the whole complex plane from Matsubara frequencies, three different
functions: $\Xi_{nm}^{RR},\Xi_{nm}^{RA}$, and $\Xi_{nm}^{AA},$should be introduced, which are analytical in their
corresponding domains. They differ by the combinations
of the signs of the explicit absolute values appearing in Eq. (\ref{gransig}). Due to observation (\ref%
{permut}) one can write the useful identities%
\begin{equation*}
\Xi_{nm}^{RR}(-iz+\omega_{\nu},-iz)=\Xi_{mn}^{RR}(-iz,-iz+\omega_{\nu}) 
\end{equation*}%
\begin{equation*}
\Xi_{nm}^{AA}(-iz,-iz-\omega_{\nu})=\Xi_{mn}^{AA}(-iz-\omega_{\nu},-iz) 
\end{equation*}%
\begin{equation*}
\Xi_{nm}^{RA}(-iz+\omega_{\nu},-iz)=\Xi_{mn}^{RA}(-iz,-iz+\omega_{\nu}) 
\end{equation*}
and get for the contour integral in Eq. (\ref{eliashx}) : 
\begin{widetext}
\begin{align*} 4\pi iI_{nm}^{\mathrm{AL}}\left(  \omega_{\nu}\right)
=\int_{-\infty}^{\infty
}\coth\left(  \frac{z}{2T}\right)  dz\left\{  \left[  \Xi_{nm}^{RR}%
(-iz+\omega_{\nu},-iz)\right]  ^{2}L_{m}^{R}(-iz)
-\left[  \Xi_{nm}^{RA}(-iz+\omega_{\nu},-iz)\right]  ^{2}L_{m}%
^{A}(-iz)\right\}  L_{n}^{R}(-iz+\omega_{\nu})\\
+\int_{-\infty-i\omega_{\nu}}^{\infty-i\omega_{\nu}}\coth\left(
\frac {z}{2T}\right)  dz\left\{  \left[
\Xi_{nm}^{RA}(-iz+\omega_{\nu},-iz)\right]
^{2}L_{n}^{R}(-iz+\omega_{\nu})
- \left[  \Xi_{nm}^{AA}(-iz+\omega_{\nu},-iz)\right]  ^{2}L_{n}%
^{A}(-iz+\omega_{\nu})\right\}  L_{m}^{A}(-iz).
\end{align*}
The last integration can be reduced to that along the real
axis by means of shifting the variable $-iz+\omega_{\nu}\rightarrow
-iz^{\prime}.$ The resulting expression (\ref{eliashx}) \ for the
electromagnetic response operator -- still defined on Matsubara frequencies $%
\omega_{\nu}$ -- takes the form:

\begin{align}
Q_{xx}^{\mathrm{AL}}(\omega_{\nu})=4ie^{2}\nu_0^{2}\mathcal{D}^{2}\frac{eH}{\pi}%
\sum_{n,m}^{\infty}C_{mn}\int_{-\infty}^{\infty}\coth\left(  \frac{z}%
{2T}\right)  \Phi_{mn}\left(  z,\omega_{\nu}\right)  dz, \label{dobavit}
\end{align}
where
\begin{align}
\Phi_{mn}\left(  z,\omega_{\nu}\right)   &  =\left\{  \left[  \Xi_{nm}%
^{RR}(-iz+\omega_{\nu},-iz)\right]  ^{2}L_{m}^{R}(-iz)-\left[  \Xi_{nm}%
^{RA}(-iz+\omega_{\nu},-iz)\right]  ^{2}L_{m}^{A}(-iz)\right\}  L_{n}%
^{R}(-iz+\omega_{\nu})\nonumber\\
&  +\left\{  \left[  \Xi_{mn}^{RA}(-iz-\omega_{\nu},-iz)\right]  ^{2}L_{n}%
^{R}(-iz)-\left[  \Xi_{nm}^{AA}(-iz,-iz-\omega_{\nu})\right]  ^{2}L_{n}%
^{A}(-iz)\right\}  L_{m}^{A}(-iz-\omega_{\nu}).\label{fgran}%
\end{align}
The rules for performing the analytical continuations of \
the function\ $\Xi_{mn}(\Omega_{k},\Omega_{k}+\omega_{\nu})$ in Eq. (\ref%
{fgran}) are simple: the sign of the explicitly written absolute values of the corresponding frequency in Eq. (%
\ref{gransig}) is chosen as "$+$" in the case of retarded continuation
(superscript R) and it is chosen as "$-$" in the case of the advanced one
(superscript A). For instance

\begin{align}
\Xi_{mn}^{RA}(\Omega_{k},\Omega_{k}+\omega_{\nu})= &
\frac{1}{2\omega _{\mathrm{c}}\left(  n-m\right)  }\left[
\psi\left(  \frac{1}{2}+\frac
{\omega_{\nu}-\Omega_{k}+\omega_{\mathrm{c}}(n+1/2)}{4\pi T}\right)
-\psi\left(
\frac{1}{2}+\frac{-\Omega_{k}+\omega_{\mathrm{c}}(m+1/2)}{4\pi
T}\right)  \right.  \nonumber\\
&  +\left.  \psi\left(  \frac{1}{2}+\frac{\omega_{\nu}+\Omega_{k}%
+\omega_{\mathrm{c}}(n+1/2)}{4\pi T}\right)  -\psi\left(  \frac{1}{2}%
+\frac{2\omega_{\nu}+\Omega_{k}+\omega_{\mathrm{c}}(m+1/2)}{4\pi T}\right)
\right]  \nonumber
\end{align}
\end{widetext}
and analogously for $\Xi_{nm}^{RR}$ and $\Xi_{nm}^{AA}.$

Now one can perform  the last analytical continuation $%
\omega_{\nu}\rightarrow-i\omega$ in Eq. (\ref{fgran}) and obtain $\Phi_{mn}^{\left( R\right)
}\left( z,\omega\right) $ as an analytic function of the real external
frequency $\omega.$ Since we are interested in the d.c. limit of the FC, i.e. $\omega\rightarrow0$, the function $\Phi_{mn}^{\left( R\right) }\left(
z,\omega\right) $ can be presented in the form of its Taylor expansion: 
\begin{equation*}
\Phi_{mn}^{\left( R\right) }\left( z,\omega\right) =\Phi_{mn}^{\left(
R\right) }\left( z,0\right) -\frac{i\omega}{\omega_{\mathrm{c}}^{2}\left(
n-m\right) ^{2}}\digamma_{nm}(-iz). 
\end{equation*}
The first term is not of interest here: all frequency independent
contributions which form $Q^{\left( \mathrm{fl}\right) }\left( 0,T,H\right) $
are cancelled out: this is a necessary requirement of the absence of the
diamagnetic response in the normal phase of superconductors. 
Actually, in
order to find the FC we need to know only $\Im Q^{\left( \mathrm{fl}\right)
}\left( \omega,T,H\right) ,$ i.e. we are interested only the imaginary part
of $\digamma_{nm}(-iz)$. \ It can be obtained by expansion of all functions $%
\Xi_{nm}^{\alpha\beta}$ ($\alpha ,\beta=R,A$) and propagators in $%
\Phi_{mn}^{\left( R\right) }\left( z,\omega\right) $ over $\omega.$
Introducing the function%
\begin{align*}
\Psi_{nm}(iz) & =\psi\left( \frac{1}{2}+\frac{iz+\omega_{\mathrm{c}}(n+1/2)}{%
4\pi T}\right) \\
& -\psi\left( \frac{1}{2}+\frac{iz+\omega_{\mathrm{c}}(m+1/2)}{4\pi T}\right)
\end{align*}
one can find the analytically continued expressions for the products of Eq. (%
\ref{fgran}):%
\begin{align*}
& \left[ \Xi_{nm}^{RR}\right] ^{2}=\frac{\left[ \Psi_{nm}^{2}(-iz)-\frac{%
i\omega}{2\pi T}\Psi_{nm}(-iz)\Psi_{nm}^{\prime}(-iz)\right] }{\omega_{%
\mathrm{c}}^{2}\left( n-m\right) ^{2}}, \\
& \left[ \Xi_{mn}^{RA}(-iz\pm i\omega,-iz)\right] ^{2} \\
& =\frac{\Re \Psi_{nm}(-iz)}{\omega_{\mathrm{c}}^{2}\left( n-m\right) ^{2}}%
\left[ \Re \Psi_{nm}(-iz)\pm\frac{i\omega}{4\pi T}\Psi_{nm}^{\prime}(\pm iz)%
\right] , \\
& \left[ \Xi_{nm}^{AA}\right] ^{2}=\frac{1}{\omega_{\mathrm{c}}^{2}\left(
n-m\right) ^{2}}\left[ \Psi_{nm}^{2}\left( iz\right) +O\left( \omega
^{2}\right) \right]
\end{align*}
which leads to%
\begin{align*}
\Im \digamma_{nm}(-iz) & =-\frac{\partial}{\partial z}\left\{ 2\Re
\Psi_{nm}^{2}\Im L_{m}^{R}\Im L_{n}^{R}\right. \\
+ & \Im \Psi_{nm}^{2}\left. \left[ \Im L_{n}^{R}\Re L_{m}^{R}+\Im
L_{m}^{R}\Re L_{n}^{R}\right] \right\} .
\end{align*}
One can see that this function is symmetric with respect to subscripts
permutation: $\Im\digamma_{nm}(-iz)=\Im \digamma_{mn}(-iz)$. Let us stress
that we have could present the linear in $\omega$ part of the
function $\Phi_{mn}^{\left( R\right) }\left( z,\omega\right) $ in the form
of full derivative with respect to $z$. The same situation was found in the original
paper of Aslamazov and Larkin~\cite{AL68} for the simple case when the Green
functions block could be assumed to be constant. As a consequence of this
important property of $\Phi_{mn}^{\left( R\right) }\left( z,\omega\right)$,
the integration over $z$ in Eq. (\ref{dobavit}) can be performed by parts.
After summation over $m$, the Eqs. (\ref{quom}) and (\ref{qan}) read as 
\begin{align}
& \delta\sigma_{xx}^{\mathrm{AL}}(T,H)=\frac{e^{2}}{2\pi T}%
\nu_{0}^{2}\sum_{n=0}^{\infty}(n+1)\int_{-\infty}^{\infty}\frac{dz}{%
\sinh^{2}\left( z/2T\right) }  \notag \\
\times & \left\{ 2\Re \Psi_{n,n+1}^{2}(-iz)\Im L_{n}^{R}(-iz)\Im
L_{n+1}^{R}(-iz)\right.  \notag \\
& +\Im \Psi_{n,n+1}^{2}(-iz)\left[ \Im L_{n}^{R}(-iz)\Re
L_{n+1}^{R}(-iz\right. )  \notag \\
& +\left. \left. \Im L_{n+1}^{R}(-iz)\Re L_{n}^{R}(-iz)\right] \right\}.
\label{conduc0}
\end{align}
Let us attract the attention to the fact that due to the integration by parts $\coth
z/2T$ disappeared from the integral Eq. (\ref{dobavit}) being replaced in
Eq. (\ref{conduc0}) by its derivative $\sinh^{-2}\left( z/2T\right).$ 
This fact makes our answer different from the one of Ref.~[\onlinecite{GL01}]
and physically means, as we will see below, that at low temperatures the
paraconducting contribution tend to zero: fluctuation Cooper pairs above $H_{%
\mathrm{c2}}\left( 0\right) $ exist but do not move and do not
participate directly in the charge transfer~\cite{BE99,LV09}.

It is convenient to introduce the dimensionless variable: $x=z/\left( 2\pi
T\right) ,$ parameters $t=T/T_{\mathrm{c0}}$ and $h=2e\xi^{2}H,$ where $%
\xi^{2}=\pi\mathcal{D}/\left( 8T\right) ,$ and the function%
\begin{widetext}
\begin{equation}
\mathcal{E}_{m}\left(  t,h,ix\right)  =\ln t+\psi\left[  \frac{1+ix}{2}%
+\frac{2}{\pi^{2}}\left(  \frac{h}{t}\right)  \left(  2m+1\right)  \right]
-\psi\left(  \frac{1}{2}\right)  .\label{e}%
\end{equation}
In this representation, Eq. (\ref{conduc0}) takes the form ($\mathcal{E}%
_{k}\left(  t,h,ix\right)  \equiv\mathcal{E}_{k}$)%
\begin{align}
\delta\sigma_{xx}^{\mathrm{AL}}\left(  t,h \right)  & =&\frac{e^{2}}{\pi}\sum_{m=0}^{\infty}%
(m+1)\int_{-\infty}^{\infty}\frac{dx}{\sinh^{2}\pi x}\left\{  \frac
{\operatorname{Im}^{2}\mathcal{E}_{m}}{\left\vert \mathcal{E}_{m}\right\vert
^{2}}+\frac{\operatorname{Im}^{2}\mathcal{E}_{m+1}}{\left\vert \mathcal{E}%
_{m+1}\right\vert ^{2}}+\frac{\operatorname{Im}^{2}\mathcal{E}_{m+1}%
-\operatorname{Im}^{2}\mathcal{E}_{m}}{\left\vert \mathcal{E}_{m}\right\vert
^{2}\left\vert \mathcal{E}_{m+1}\right\vert ^{2}}\operatorname{Re}\left[
\mathcal{E}_{m}\mathcal{E}_{m+1}\right]  \right\} .\label{ALgener}
\end{align}
\end{widetext}
This is the general expression for fluctuation
paraconductivity valid in all domains of temperatures and fields under
consideration.

We will see that all diagrams presented in Fig.\ref{fig.conddia} are
relevant in different regions of the phase diagram, depicted in Fig.~\ref%
{fig.domains}. Nine regions of different asymptotic behavior can be
distinguished and below we will analyze all contributions in each
domain.

\subsection{Asymptotic behavior}

\subsubsection{Vicinity of $T_{\mathrm{c0}},$ fields $h\ll1(H\ll H_{\mathrm{%
c2}}\left( 0\right) )$}

In this case $\ln t=\epsilon\ll1$ and the $\psi$-function in Eq.(\ref{e})
can be expanded. In first approximation:%
\begin{equation}
\mathcal{E}_{m}^{\left( 1\right) }\left( t,h,ix\right) =\epsilon +\frac{%
i\pi^{2}x}{4}+\left( \frac{2h}{t}\right) \left( m+\frac{1}{2}\right) .
\label{e1}
\end{equation}
The integral in Eq. (\ref{ALgener}) can be easily carried our: only the
first fraction in the parenthesis should be taken into account. Further
summation over Landau levels can be performed exactly in terms of the $\psi$%
-function:%
\begin{equation}
\delta\sigma_{xx}^{\mathrm{AL}}(\epsilon,h\ll1)=\frac{e^{2}}{2\epsilon}%
\left( \frac{\epsilon}{2h}\right) ^{2}\left[ \psi\left( \frac12+\frac{%
\epsilon }{2h}\right) -\psi\left( \frac{\epsilon}{2h}\right) -\frac{h}{%
\epsilon }\right] ,  \label{AL0}
\end{equation}
which coincides with the known expression for the Cooper pairs contribution
to the magneto-conductivity in the Ginzburg-Landau region~\cite{LV09}.

The general Eq. (\ref{ALgener}) allows to obtain the next order correction in $%
\epsilon$ with respect to the AL result. In order to do this, one should take
into account both terms and expand up to the second order:%
\begin{align}
& \mathcal{E}_{m}^{\left( 2\right) }\left( t,h,ix\right) =\mathcal{E}%
_{m}^{\left( 1\right) }\left( t,h,ix\right) -\frac{14\zeta\left( 3\right) ix%
}{\pi^{2}}\left( \frac{2h}{t}\right) \left( m+\frac{1}{2}\right)  \notag \\
& +7\zeta\left( 3\right) \frac{x^{2}}{4}-\frac{28\zeta\left( 3\right) }{%
\pi^{4}}\left( \frac{2h}{t}\right) ^{2}\left( m+\frac{1}{2}\right) ^{2}
\label{e2}
\end{align}
After some simple but cumbersome calculation in the limit of small fields, one
finds%
\begin{equation}
\delta\sigma_{xx}^{\mathrm{AL}}(\epsilon\ll1)=\frac{e^{2}}{16\epsilon}-\frac{%
7\zeta\left( 3\right) e^{2}}{8\pi^{4}}\ln\frac{1}{\epsilon},  \label{AL1}
\end{equation}
which demonstrates that the next order correction to the well known AL result
in the vicinity of $T_{\mathrm{c0}}$ is smaller than the DOS~\cite{ILVY93}
and the regular MT~\cite{LV09} contributions (see below)  by only numerical
pre-factor.

\subsubsection{High temperatures $T\gg T_{\mathrm{c0}},$ weak fields $h\ll t$}

Let us move to the discussion of the high-temperature asymptotic. We
will assume $\ln t\gg1$ in Eq. (\ref{e}) and get: 
\begin{equation}
\Im \mathcal{E}_{m}\left( t,h,ix\right) =\frac{x}{2}\psi^{\prime}\left[ 
\frac{1}{2}+\frac{4}{\pi^{2}}\left( \frac{h}{t}\right) \left( m+1/2\right) %
\right] .  \label{ehigh}
\end{equation}
The sum in Eq. (\ref{ALgener}) converges at $n_{\max}\sim t/h\gg1$ and
can be replaced by an integral. The integration over $x$ involves only the
region $x\sim1$ and can be performed first. As a result one gets%
\begin{equation*}
\delta\sigma_{xx}^{\mathrm{AL}}(t\gg1,h\ll t)=\frac{e^{2}}{6\pi^{2}}\frac{%
C_{1}}{\ln^{3}t} 
\end{equation*}
with $C_{1}=\frac{1}{3}\int_{0}^{\infty}\left[ \psi^{\prime}\left(
1/2+x\right) \right] ^{3}dx=6.97.$ Let us stress that this asymptotic expression
coincides with the high temperature behavior of the AL contribution obtained
in clean case~\cite{RVV91} which emphasises the statement that the 2D
paraconductivity is an universal function of $\ln t$ throughout the complete
temperature range.

\subsubsection{Fields close to the line $H_{\mathrm{c2}}\left( T\right) $}

The line separating normal and superconductiving phases $H_{\mathrm{c2}}\left(
T\right) $ (in our dimensionless units the line of critical fields $h_{%
\mathrm{c2}}\left( t\right) )$ is determined by the requirement that the
propagator (\ref{propagator}) has a pole when $\Omega_{k}=0$ and $m=0$:%
\begin{equation*}
\ln t+\psi\left( \frac{1}{2}+\frac{2}{\pi^{2}}\frac{h_{\mathrm{c2}}\left(
t\right) }{t}\right) -\psi\left( \frac{1}{2}\right) =0. 
\end{equation*}
At low temperatures $T\ll T_{\mathrm{c0}},$ close to the point $T=0$ and $H=H_{%
\mathrm{c2}}(0)$, the critical field is $h_{\mathrm{c2}}\left( t\right)
=2\xi^{2}H_{\mathrm{c2}}(0)/e\sim1$.
Then one can substitute the $\psi $-function
by its asymptotic expression $\psi(x)=\ln x-1/(2x)$ and take into account that $%
\psi\left( 1/2\right) =-\ln4\gamma_{E}$ ($\gamma_{E}=1.781..$ is the Euler's
constant) which results in
\begin{equation}
h_{\mathrm{c2}}\left( t\rightarrow0\right) =\frac{\pi^{2}}{8\gamma_{E}}.
\label{hc2}
\end{equation}

In order to find the paraconducting contribution to FC above the curve $H_{%
\mathrm{c2}}\left( T\right) $ in Fig.~\ref{fig.domains}, let us rewrite
Eq. (\ref{e}) in terms of the reduced field 
\begin{equation*}
\widetilde{h}\left( t\right) =\frac{h-h_{\mathrm{c2}}\left( t\right) }{h_{%
\mathrm{c2}}\left( t\right) }\ll 1 
\end{equation*}%
Below we will see that the  Cooper
pair contribution to FC, which is singular in $\widetilde{h}^{-1}$, originates in Eq. (\ref{ALgener}) only from the
term with $m=0,$ i.e. we can restrict ourselves to the Lowest Landau Level
(LLL) approximation. Hence we will need the explicit expression for $%
\mathcal{E}_{m}\left( t,\widetilde{h},ix\right) $ only for $m=0,1$ and $%
\widetilde{h}\ll t\ll h_{\mathrm{c2}}\left( t\right) .$ In order to get this,
one can use in Eq~ (\ref{e}) a parametrization in terms of $\widetilde{h}$
and expand it $\widetilde{h}\ll 1$ and $h-h_{\mathrm{c2}}\left(
t\right) =\widetilde{h}\cdot h_{\mathrm{c2}}\left( t\right) \ll t.$ This
gives 
\begin{equation}
\mathcal{E}_{0}\left( t,\widetilde{h},ix\right) =\widetilde{h}+\frac{i\pi
^{2}xt}{4h_{\mathrm{c2}}\left( t\right) }.  \label{emas}
\end{equation}%
The substitution of Eq. (\ref{emas}) \ to Eq. (\ref{ALgener}) results in 
\begin{equation}
\delta \sigma _{xx}^{\mathrm{AL}}(t,h)=\frac{e^{2}}{\pi ^{2}}J_{GL}\left( 
\frac{4h_{\mathrm{c2}}\left( t\right) \widetilde{h}}{\pi ^{2}t}\right) .
\label{sigmaline1}
\end{equation}%
with%
\begin{equation}
J_{GL}\left( r\right) =\int_{-\infty }^{\infty }\frac{dx}{\sinh ^{2}x}\frac{%
x^{2}}{x^{2}+\pi ^{2}r^{2}}=2r\psi ^{\prime }\left( r\right) -\frac{1}{r}-2
\label{JGL}
\end{equation}%
first calculated in Ref.~[\onlinecite{GL01}]. This formula is valid along
all the line $h_{\mathrm{c2}}\left( t\right) $ until $t\sim h_{\mathrm{c2}%
}\left( t\right) $. Taking into account the asymptotic expressions 
\begin{equation}
\psi ^{\prime }\left( r\rightarrow \infty \right) =\frac{1}{r}+\frac{1}{%
2r^{2}}+\frac{1}{6r^{3}};\;\psi ^{\prime }\left( r\rightarrow 0\right)
=1/r^{2}.  \label{asspsi}
\end{equation}%
one finds that in this domain 
\begin{equation}
\delta \sigma _{xx}^{\mathrm{AL}}(t,h)=\left\{ 
\begin{array}{c}
\frac{4e^{2}\gamma _{E}^{2}t^{2}}{3\pi ^{2}\widetilde{h}^{2}},t\ll 
\widetilde{h}, \\ 
\frac{e^{2}t}{4h_{\mathrm{c2}}\left( t\right) \widetilde{h}},h_{\mathrm{c2}%
}\left( t\right) \widetilde{h}\ll t.%
\end{array}%
\right. .  \label{salgen}
\end{equation}%
The first line of Eq. (\ref{salgen}) corresponds to the \emph{quantum
fluctuations} which are realized in the limit of lowest temperatures \ $t\ll 
\widetilde{h}$ close to $H_{\mathrm{c2}}\left( 0\right).$ One sees that the
paraconductivity decays here as $T^{2}$. In the temperature range $h_{%
\mathrm{c2}}\left( t\right) \widetilde{h}\ll t\ll h_{\mathrm{c2}}\left(
t\right) $ the paraconductivity is determined by the first line of Eq. (\ref%
{salgen}). Close to $H_{\mathrm{c2}}\left( 0\right) $ but for relatively
high temperatures $t\sim \widetilde{h}$ the corresponding expression can be
rewritten using the explicit expression for $h_{\mathrm{c2}}\left( 0\right) $
Eq. (\ref{hc2}):
\begin{equation}
\delta \sigma _{xx}^{\mathrm{AL}}(t,h)=\frac{2\gamma _{E}e^{2}}{\pi ^{2}}%
\left( \frac{t}{\widetilde{h}}\right) ,  \label{salmid}
\end{equation}%
which perfectly matches to the first line of the Eq. (\ref{salgen}). Here the
transition from quantum to classical fluctuations takes place. At higher
temperatures along the line $h_{\mathrm{c2}}\left( t\right) $ one should
take into account the temperature dependence of $h_{\mathrm{c2}}\left(
t\right)$: 
\begin{equation}
\delta \sigma _{xx}^{\mathrm{AL}}(t,h)=\frac{e^{2}}{4}\frac{t}{h-h_{\mathrm{%
c2}}\left( t\right) }.  \label{salhigh}
\end{equation}%
This expression is valid along the line $h_{\mathrm{c2}}\left( t\right)$
until $t\ll h_{\mathrm{c2}}\left( t\right) $ where Eq. (\ref{salhigh})
matches to Eq. (\ref{AL0}).

\subsubsection{High fields ($H\gg H_{\mathrm{c2}}\left( 0\right) $),
temperatures $t\ll h$}

In this domain we are far from the transition line $H_{\mathrm{c2}}\left(
T\right)$, $\left[ h\gg h_{\mathrm{c2}}\left( t\right) \right)]$ and the LLL
approximation is not applicable. Nevertheless, one can substitute the summation
over Landau levels by an integration. Replacing the \ $\psi-$function in Eq. (%
\ref{e}) by a logarithm, one finds that%
\begin{equation}
\mathcal{E}_{m}\left( t,h,ix\right) =\ln\frac{4h}{\pi^{2}}\left( m+\frac {1}{%
2}\right) -\psi\left( \frac{1}{2}\right) +\frac{i\pi^{2}xt}{4h\left(
2m+1\right) }.  \label{en}
\end{equation}
One can see that this expression reproduces Eq. (\ref{emas}) when $%
h\rightarrow h_{\mathrm{c2}}\left( t\right) $ and $m=0$. Let us substitute
Eq. (\ref{en}) to Eq. (\ref{ALgener}). As we will see below, the sum
converges at $m\sim 1.$ It is why the main contribution comes from the second term
of Eq. (\ref{ALgener}), where the sum $\Im \mathcal{E}_{m}\Re \mathcal{E}%
_{m+1}+\Im \mathcal{E}_{m+1}\Re \mathcal{E}_{m}\sim\ln\left( \frac{4h}{%
\pi^{2}}\right) .$ As a result we get%
\begin{equation*}
\delta\sigma_{xx}^{\mathrm{AL}}(t,h)=\frac{\pi^{2}e^{2}}{192}\left( \frac {t%
}{h}\right) ^{2}\frac{C_{2}}{\ln^{3}\frac{2h}{\pi^{2}}} 
\end{equation*}
with $C_{2}=0.545.$ Let us recall that this expression is valid for arbitrary
temperatures small in comparison to reduced field $t\ll h$.

\section{Maki-Thompson contribution}

\subsection{General expression}

Below we will calculate the fluctuation renormalization of the one-electron
contributions to conductivity. It is technically convenient to
start with the usual expressions for the Maki-Thompson and other diagrams from
Fig.~\ref{fig.conddia}, written in momentum representation. Only at the
very end one should quantize the motion of Cooper pairs in a magnetic field in
accordance with the rule 
\begin{equation*}
\frac{\mathcal{D}}{8T}\int\frac{d^{2}q}{(2\pi)^{2}}f\left[ \mathcal{D}q^{2}%
\right] =\frac{h}{2\pi^{2}t}\sum_{m=0}^{M}f\left[ \omega_{\mathrm{c}}(n+1/2)%
\right] . 
\end{equation*}

Diagram 2 from Fig. \ref{fig.conddia} can be written as
\begin{equation}
Q_{xx}^{\mathrm{MT}}(\omega_{\nu})=2e^{2}T\sum_{\Omega_{k}}\int{\frac{{d^{2}}%
\mathbf{q}}{{(2\pi)^{2}}}}L(\mathbf{q},\Omega_{k})\Sigma_{xx}^{\mathrm{MT}}(%
\mathbf{q},\Omega_{k},\omega_{\nu}),  \label{d21}
\end{equation}
where 
\begin{align}
\Sigma_{xx}^{\mathrm{MT}}(\mathbf{q},\Omega_{k},\omega_{\nu}) &
=T\sum_{\varepsilon_{n}}\lambda(\mathbf{q},\varepsilon_{n+\nu},\Omega
_{k-n-\nu})  \notag \\
& \times\lambda(\mathbf{q},\varepsilon_{n},\Omega_{k-n})I_{xx}^{\mathrm{MT}}(%
\mathbf{q},\varepsilon_{n},\Omega_{k},\omega_{\nu})  \label{d22}
\end{align}
and%
\begin{align*}
I_{xx}^{\mathrm{MT}} & =\int{\frac{{d^{3}}\mathbf{p}}{{(2\pi)^{3}}}}v_{x}(%
\mathbf{p})v_{x}(\mathbf{q-p})G(\mathbf{p},\varepsilon_{n+\nu}) \\
& \times G(\mathbf{p},\varepsilon_{n})G(\mathbf{q-p},\Omega_{k-n-\nu })G(%
\mathbf{q-p},\Omega_{k-n}).
\end{align*}
The main $q$-dependence in (\ref{d21}) arises from the propagator and
vertices $\lambda$. That is why we can assume $q=0$ in the Green functions
and calculate the electron momentum integral by changing, as usual, to a $\xi (%
\mathbf{p})$ integration:%
\begin{align}
I_{xx}^{\mathrm{MT}} & =-\mathcal{D}\tau^{-1}\nu_{0}\int_{-\infty}^{\infty }%
\frac{d\xi}{\xi-i\widetilde{\varepsilon}_{n}}\frac{1}{\xi-i\widetilde {%
\varepsilon}_{n+\nu}}  \notag \\
& \times\frac{1}{\xi-i\widetilde{\Omega}_{k-n}}\frac{1}{\xi-i\widetilde {%
\Omega}_{k-n-\nu}}.  \label{integ}
\end{align}

\begin{figure}[htb]
\begin{center}
\includegraphics[width=0.9\columnwidth]{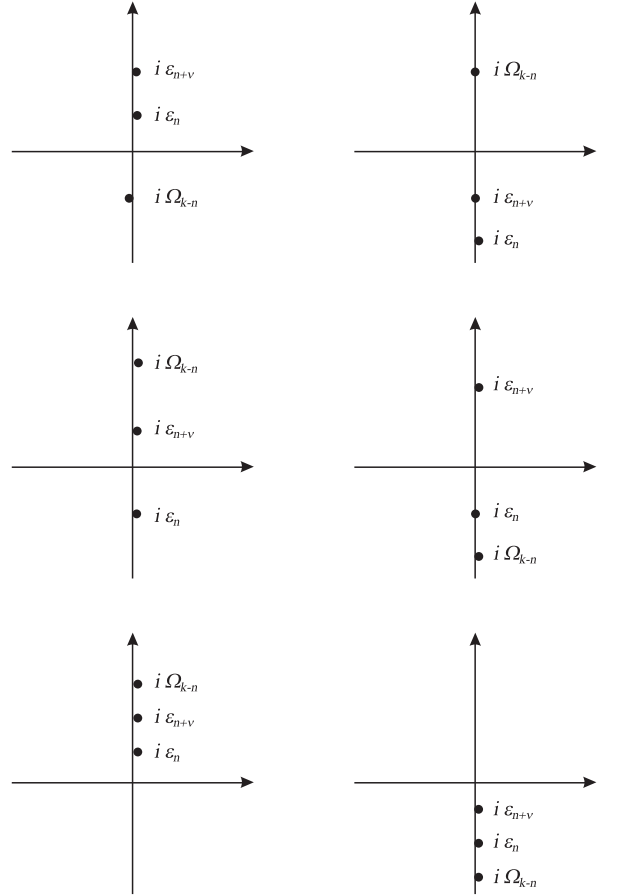}
\end{center}
\caption{$\protect\xi$-integration in Eq. (\protect\ref{integ}): all six
possible positions of the poles in the complex plane of $\protect\xi$ are shown.}
\label{fig.poles}
\end{figure}


This integral, Eq.~(\ref{integ}), can be calculated using the Cauchy theorem.
Closing the contour in upper or lower half-plane by the large semicircle and
noticing, that, due to fast decrease of the integrand the function in Eq. (\ref%
{integ}), the integral over the semicircle becomes zero, one can express $%
J_{xx}$ in terms of the sum of the corresponding residues. There are six different
combinations of the pole positions with respect to the real axis in the
complex plane of $\xi$, leading to non-zero results (see Fig. \ref{fig.poles}%
): two realization corresponding to $\theta\left(
-\varepsilon_{n}\varepsilon_{n+\nu}\right) \theta\left(
\Omega_{k-n}\Omega_{k-n-\nu}\right) \neq0,$ one realization corresponding to 
$\ \theta\left( -\varepsilon_{n}\varepsilon_{n+\nu}\right) \theta\left(
-\Omega_{k-n}\Omega_{k-n-\nu}\right) \neq0,$ two realization corresponding
to $\theta\left( \varepsilon_{n}\varepsilon_{n+\nu}\right) \theta\left(
\Omega_{k-n}\Omega_{k-n-\nu}\right) \neq0,$ and the realization
corresponding to $\theta\left( \varepsilon_{n}\varepsilon_{n+\nu}\right)
\theta\left( -\Omega_{k-n}\Omega_{k-n-\nu}\right) \neq0.$ Calculating the
residues for each situation and assuming that $\widetilde{\varepsilon}%
_{n}=\left( 2\tau\right) ^{-1}\sgn\varepsilon_{n}$ (let us recall that we
consider the disordered limit $T\ll\tau^{-1})$ one finds:%
\begin{align}
I_{xx}^{\mathrm{MT}} & =2\pi\mathcal{D}\nu_{0}\tau^{2}\left\{ \theta\left(
-\varepsilon_{n}\varepsilon_{n+\nu}\right) \theta\left(
\Omega_{k-n}\Omega_{k-n-\nu}\right) \right.  \notag \\
& +\theta\left( \varepsilon_{n}\varepsilon_{n+\nu}\right) \theta\left(
-\Omega_{k-n}\Omega_{k-n-\nu}\right)  \notag \\
& -2\theta\left( -\varepsilon_{n}\varepsilon_{n+\nu}\right) \theta\left(
-\Omega_{k-n}\Omega_{k-n-\nu}\right)  \notag \\
& \left. -2\theta\left( \varepsilon_{n}\varepsilon_{n+\nu}\right)
\theta\left( \Omega_{k-n}\Omega_{k-n-\nu}\right) \right\} .  \notag
\end{align}
Now one should substitute this expression to Eq. (\ref{d22}) and perform the
summation over the fermionic frequencies. This is a cumbersome exercise,
which, nevertheless, can be followed through analytically. Here we mention some
useful transformations which are important to perform the summations: One can see, that the simultaneous
permutations $n\rightarrow-n$ and $k\rightarrow-k$ allows to simplify the
sums:%
\begin{align*}
& \Sigma_{xx}^{\mathrm{MT}} =\left( \Sigma_{xx}^{\mathrm{MT}\left( an\right)
}+\Sigma_{xx}^{\mathrm{MT(reg2)}}\right) +\Sigma_{xx}^{\mathrm{MT(reg1)}%
}=-2\pi\nu_{0}\mathcal{D}T \\
& \times\left\{ \sum_{n=-\nu}^{-1}\frac{2\theta\left(
-\Omega_{k-n}\Omega_{k-n-\nu}\right) -\theta\left(
\Omega_{k-n}\Omega_{k-n-\nu}\right) }{\left(
|\varepsilon_{n+\nu}-\Omega_{k-n-\nu}|+\mathcal{D}q^{2}\right) \left(
|\varepsilon_{n}-\Omega_{k-n}|+\mathcal{D}q^{2}\right) }\right. \\
& +\left. 2\sum_{n=0}^{\infty}\frac{2\theta\left(
\Omega_{k-n}\Omega_{k-n-\nu}\right) -\theta\left(
-\Omega_{k-n}\Omega_{k-n-\nu}\right) }{\left(
|\varepsilon_{n+\nu}-\Omega_{k-n-\nu}|+\mathcal{D}q^{2}\right) \left(
|\varepsilon_{n}-\Omega_{k-n}|+\mathcal{D}q^{2}\right) }\right\} .
\end{align*}
The rules writing the absolute values explicitly in the sum using the first \ $\theta$%
-function is evident. In the second sum, containing $\theta\left(
\Omega_{k-n}\Omega_{k-n-\nu}\right) ,$ one should make a shift $%
n^{\prime}=n+\nu$ . After rewriting the absolute values for the Cooperons, the sums
can be expressed in terms of $\psi$-functions. Using the identity%
\begin{equation*}
\psi\left( 1/2+iz\right) -\psi\left( 1/2-iz\right) =\pi i\tanh\pi z 
\end{equation*}
allows to write the final expression for the first sum as 
\begin{align}
& \Sigma_{xx}^{\mathrm{MT(an)}}+\Sigma_{xx}^{\mathrm{MT(reg2)}}=-\frac{%
\mathcal{D}\nu_{0}\theta\left( \omega_{\nu-1}-|\Omega_{k}|\right) }{2\left(
\omega_{\nu}+\mathcal{D}q^{2}\right) }  \notag \\
& \left[ \psi\left( \frac{1}{2}+\frac{2\omega_{\nu}-|\Omega_{k}|+\mathcal{D}%
q^{2}}{4\pi T}\right) -\psi\left( \frac{1}{2}+\frac{|\Omega _{k}|+\mathcal{D}%
q^{2}}{4\pi T}\right) \right] .  \label{Ian}
\end{align}
Looking at the denominator of this expression one can recognize that $%
\Sigma_{xx}^{\mathrm{MT(an)}}$ is responsible for the anomalous
Maki-Thompson term.

Next we consider the remaining second sum in $\Sigma_{xx}^{\mathrm{MT}}$. One
can see that in the first term both arguments of the absolute values are positive. In the second term
we can replace $k\rightarrow-k,$ with an additional change of the order of the
summation over bosonic frequencies. \ The sum with $\theta\left( \Omega
_{k-n}\Omega_{k-n-\nu}\right) $ can be calculated in the spirit of Eq. (\ref%
{Ian}). Regarding the last sum, containing $\theta\left(
-\Omega_{k-n}\Omega_{k-n-\nu}\right) ,$ one can find that it is exactly
equals to zero for any $\Omega_{k}.$ Finally%
\begin{align}
\Sigma_{xx}^{\mathrm{MT(reg1)}} & =-\frac{\mathcal{D}\nu_{0}}{\omega_{\nu}}%
\left[ \psi\left( \frac{1}{2}+\frac{2\omega_{\nu}+|\Omega_{k}|+\mathcal{D}%
q^{2}}{4\pi T}\right) \right.  \notag \\
& \left. -\psi\left( \frac{1}{2}+\frac{|\Omega_{k}|+\mathcal{D}q^{2}}{4\pi T}%
\right) \right] .  \label{Ireg}
\end{align}
Using the explicit Eqs. (\ref{Ian})-(\ref{Ireg}) we can perform the
final summation over bosonic frequencies in Eq. (\ref{d21}) and the analytical
continuation of $Q_{xx}^{\mathrm{MT}}\left( \omega_{\nu}\right) $ to \ the
axis of real frequencies. The analytical continuation of $Q_{xx}^{\mathrm{%
MT(reg1)}}$ is trivial since Eq. (\ref{Ireg}) is the analytical function of $%
\omega_{\nu}$. As a result we get
\begin{align}
Q_{xx}^{\mathrm{MT(reg1)R}}(\omega) & =i\omega e^{2}\frac{\mathcal{D}\nu _{0}%
}{4\pi^{2}T}\int{\frac{{d^{2}}\mathbf{q}}{{(2\pi)^{2}}}}  \notag \\
& \times\sum_{k=-\infty}^{\infty}L(\mathbf{q},\Omega_{k})\psi^{\prime\prime
}\left( \frac{1}{2}+\frac{|\Omega_{k}|+\mathcal{D}q^{2}}{4\pi T}\right)\,.
\notag
\end{align}
Next we go over from the integration over the momentum of the Cooper pair
center of mass $\mathbf{q}$ to the summation over Landau levels. Recalling that
the density of states at the Landau level is $H/\Phi_{0}$, one finds 
\begin{equation}
\delta\sigma_{xx}^{\mathrm{MT(reg1)}}=\frac{e^{2}}{4\pi^{2}T}\frac {\mathcal{%
D}H}{\Phi_{0}}\sum_{m}\sum_{k=-\infty}^{\infty}\frac{4\mathcal{E}%
_{m}^{\prime\prime}\left( t,h,|k|\right) }{\mathcal{E}_{m}\left(
t,h,|k|\right) },  \label{MTreg1}
\end{equation}
where $\mathcal{E}_{m}^{(p)}\left( t,h,z\right) \equiv\partial_{z}^{p}%
\mathcal{E}_{m}\left( t,h,z\right) $, i.e%
\begin{equation*}
\mathcal{E}_{m}^{\prime\prime}\left( t,h,|k|\right) =\psi^{\prime\prime }%
\left[ \frac{1+|k|}{2}+\frac{2h}{\pi^{2}t}\left( 2m+1\right) \right] . 
\end{equation*}

\begin{figure}[htb]
\begin{center}
\includegraphics[width=0.6\columnwidth]{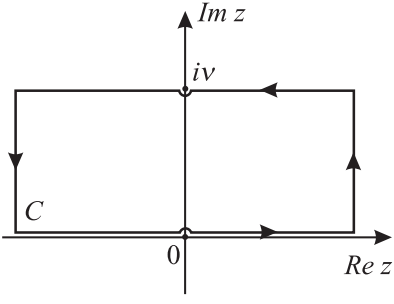}
\end{center}
\caption{Integration contour used in the analytic continuation of the MT
contribution.}
\label{fig.MTcontour}
\end{figure}

In the part of the electromagnetic operator related to Eq. (\ref{Ian}%
), the external frequency $\omega_{\nu}$ appears in the upper limit of
the bosonic sum: 
\begin{align*}
& Q_{xx}^{\mathrm{MT(an)}}+Q_{xx}^{\mathrm{MT(reg2)}} \\
& =-2e^{2}T\mathcal{D}\nu_{0}\int{\frac{{d^{2}}\mathbf{q}}{{(2\pi)^{2}}}}%
\frac{1}{\omega_{\nu}+\mathcal{D}q^{2}}\sum_{|k|=0}^{\nu-1}L(\mathbf{q}%
,\Omega_{k}) \\
& \left[ \psi\left( \frac{1}{2}+\frac{2\omega_{\nu}-|\Omega_{k}|+\mathcal{D}%
q^{2}}{4\pi T}\right) -\psi\left( \frac{1}{2}+\frac{|\Omega _{k}|+\mathcal{D}%
q^{2}}{4\pi T}\right) \right]
\end{align*}
and the procedure of analytical continuation is more sophisticated. First of
all one can easily see that the contributions of positive and negative $%
k $ are equal. The method to continue such a sum the real frequencies was
developed in Ref.~[\onlinecite{AV80}] and consists in an Eliashberg
transformation of the sum over $\Omega_{k}$ to an integral over the contour ${\cal C}$
(see Fig. \ref{fig.MTcontour} and the detailed description of this
procedure in Ref.~[\onlinecite{LV09}]). One finds

\begin{align*}
& Q_{xx}^{\mathrm{MT(an)}}+Q_{xx}^{\mathrm{MT(reg2)R}}=-4e^{2}T\mathcal{D}%
\nu_{0}\int{\frac{{d^{2}}\mathbf{q}}{{(2\pi)^{2}}}}\frac{1}{\omega_{\nu }+%
\mathcal{D}q^{2}} \\
& \left\{ \frac{1}{2}L(\mathbf{q},0)\left[ \psi\left( \frac{1}{2}+\frac{%
2\omega_{\nu}+\mathcal{D}q^{2}}{4\pi T}\right) -\psi\left( \frac {1}{2}+%
\frac{\mathcal{D}q^{2}}{4\pi T}\right) \right] \right. \\
& +\frac{1}{2i}\oint \limits_{C_{2}}dz\coth\left( \pi z\right) L(\mathbf{q}%
,-iz)\left[ \psi\left( \frac{1}{2}+\frac{2\omega_{\nu}+\mathcal{D}q^{2}}{%
4\pi T}+\frac{iz}{2}\right) \right. \\
& \left. \left. -\psi\left( \frac{1}{2}+\frac{iz}{2}+\frac{\mathcal{D}q^{2}}{%
4\pi T}\right) \right] \right\}\,.
\end{align*}
The residue at the point $z=i\nu$ is equal to zero. Shifting the variables
in the integral over the upper line $\Im z=\nu$ as $%
z_{1}=z^{\prime}+i\nu$, one can present $Q_{xx}^{\mathrm{MT}}$ as an
analytical function of $\omega_{\nu}$ and analytically continue it in the standard way 
$i\omega_{\nu}\rightarrow\omega$ $\rightarrow0.$\ Expanding it in small $\omega$
and integrating by parts, one gets 
\begin{widetext}
\begin{align*}
\delta\sigma_{xx}^{\mathrm{MT(an)}}+\delta\sigma_{xx}^{\mathrm{MT(reg2)}} &  =\\
&  =\frac{e^{2}}{2}\mathcal{D}\int{\frac{{d^{2}}\mathbf{q}}{{(2\pi)^{2}}}%
}\frac{1}{\tau_{\phi}^{-1}+\mathcal{D}q^{2}}\int_{-\infty}^{\infty}\frac
{dz}{\sinh^{2}\pi z}\frac{\psi\left(  \frac{1-iz}{2}+\frac{\mathcal{D}q^{2}%
}{4\pi T}\right)  -\psi\left(
\frac{1+iz}{2}+\frac{\mathcal{D}q^{2}}{4\pi T}\right)
}{\ln\frac{T}{T_{c}}+\psi\left(  \frac{1-iz}{2}+\frac
{\mathcal{D}q^{2}}{4\pi T}\right)  -\psi\left(  \frac{1}{2}\right)  }%
\end{align*}
This expression is then transformed to the summations over Landau levels
and written in the dimensionless variables, in the same way as it was done before. \
Adding the regular part Eq. (\ref{MTreg1}), we finally write the most general
expression for the MT contribution valid in all domains of temperatures and
magnetic field under consideration:%

\begin{align}
\delta\sigma^{\mathrm{MT}}  &  =\delta\sigma_{xx}^{\mathrm{MT(reg1)}}+\left(
\delta\sigma
_{xx}^{\mathrm{MT(an)}}+\delta\sigma_{xx}^{\mathrm{MT(reg2)}}\right) \nonumber\\
&  =\frac{e^{2}}{\pi^{4}}\left(  \frac{h}{t}\right)
\sum_{m=0}^{M}\left\{
\sum_{k=-\infty}^{\infty}\frac{4\mathcal{E}_{m}^{\prime\prime
}\left(  t,h,|k|\right)  }{\mathcal{E}_{m}\left(  t,h,|k|\right)
}+\frac{{\pi^{3}}}{\gamma_{\phi}+\frac{2h}{t}\left(  m+1/2\right)  }%
\int_{-\infty}^{\infty}\frac{dz}{\sinh^{2}\pi z}\frac{\operatorname{Im}%
^{2}\mathcal{E}_{m}\left(  t,h,iz\right)  }{\operatorname{Re}^{2}%
\mathcal{E}_{m}\left(  t,h,iz\right)  +\operatorname{Im}^{2}\mathcal{E}%
_{m}\left(  t,h,iz\right)  }\right\}  . \label{MTgener}%
\end{align}
\end{widetext}where $M=(tT_{c0}\tau)^{-1}$ is the cut-off parameter.

\subsection{Asymptotic behavior}

\subsubsection{Contribution $\protect\delta\protect\sigma_{xx}^{\mathrm{%
MT(reg1)}}$}

Let us start with the evaluation of the contribution $\delta\sigma _{xx}^{%
\mathrm{MT(reg1)}}$ given by Eq. (\ref{MTreg1}).

\paragraph*{Vicinity of $T_{\mathrm{c0}},$ fields $h\ll1(H\ll H_{\mathrm{c2}%
}\left( 0\right) ).$}---
Here one can just use Eq. (\ref{e}) for integer argument $k$,
considering the smallness of $h/t$, expand the $\psi$-function, and perform the summation
exactly:

\begin{equation}
\delta\sigma_{xx}^{\mathrm{MT(reg1)}}\left( \epsilon\ll1,h\right) =-\frac{%
7\zeta\left( 3\right) e^{2}}{\pi^{4}}\left[ \psi\left( \frac {t}{2h}\right)
-\psi\left( \frac{1}{2}+\frac{\epsilon t}{2h}\right) \right] .  \label{regMT}
\end{equation}

\paragraph*{High temperatures $T\gg T_{\mathrm{c0}},$ weak fields $h\ll t$.}---
For the high-temperature asymptotic of Eq. (%
\ref{MTreg1}), we assume $\ln t\gg1$. The sum over $k$\ is determined
by $\mathcal{E}_{m}^{\prime\prime}\left( t,h,|k|\right) $ and converges
fast: it can be performed in first. The remaining sum over Landau
levels slowly diverges at large $m_{\max}$ and can be substituted by an
integral. The double logarithmic divergence of this integral at the upper
limit should be cut off at the limit corresponding $m_{\max}\sim\left( T_{%
\mathrm{c0}}\tau\right) ^{-1}$ which results in

\begin{equation}
\delta\sigma_{xx}^{\mathrm{MT(reg1)}}=-\frac{e^{2}}{\pi^{2}}\left[ \ln \ln%
\frac{1}{T_{\mathrm{c0}}\tau}-\ln\ln t\right] .  \label{hight}
\end{equation}
One can see that close to $T_{\mathrm{c0}}$ $\ln\left( \ln\ln t\right)
\rightarrow\ln\left( \frac{1}{\epsilon}\right) $, Eqs. (\ref{hight}) and (%
\ref{regMT}) therefore match each other.

\paragraph*{Fields close to the line $H_{\mathrm{c2}}\left( T\right).$}---
In this domain, as above, one can use the LLL approximation. Nevertheless,
since $t\ll h_{\mathrm{c2}}\left( t\right)$, many terms are involved in the sum over $k$\ in Eq. (%
\ref{MTreg1}). Therefore it is convenient to present  the function $\mathcal{E}%
_{0}\left( t,\widetilde{h},|k|\right) $ in this case in the form 
\begin{align*}
\mathcal{E}_{0}\left( t,\widetilde{h},|k|\right) & =\psi\left[ \frac{1+|k|}{2%
}+\frac{2}{\pi^{2}}\frac{h_{\mathrm{c2}}\left( t\right) }{t}\left( 1+%
\widetilde{h}\left( t\right) \right) \right] \\
& -\psi\left( \frac{1}{2}+\frac{2}{\pi^{2}}\frac{h_{\mathrm{c2}}\left(
t\right) }{t}\right)
\end{align*}
and taking into account the asymptotic expression (\ref{asspsi}) one finds 
\begin{align*}
\delta\sigma_{xx}^{\mathrm{MT(reg1)}} & =-\frac{e^{2}}{2}\frac{1}{\widetilde{%
h}\left( t\right) }\left( \frac{t}{2h}\right) \\
& -\frac{2e^{2}}{\pi^{2}}\left[ \ln\frac{2h}{\pi^{2}t}-\psi\left( 1+\frac{4}{%
\pi^{2}}\frac{h_{\mathrm{c2}}\left( t\right) }{t}\widetilde {h}\left(
t\right) \right) \right] .
\end{align*}
Along the line $H_{\mathrm{c2}}\left( T\right) ,$ in the region of \ classic
fluctuations $\widetilde{h}\lesssim t\ll h_{\mathrm{c2}}\left( t\right) $
Eq. (\ref{MTreg1}) contains two contributions: one, originating from the
term with $k=0$, is singular in $h\left( t\right) -h_{\mathrm{c2}}\left(
t\right) $, the second is logarithmic in reduced temperature: 
\begin{equation}
\delta\sigma_{xx}^{\mathrm{MT(reg1)}}=-\frac{e^{2}}{4}\frac{t}{h-h_{\mathrm{%
c2}}\left( t\right) }-\frac{2e^{2}}{\pi^{2}}\ln\frac {2h_{\mathrm{c2}}\left(
t\right) }{\pi^{2}t}.  \label{MT1reg}
\end{equation}
In the the regime of quantum fluctuations $t\lesssim\widetilde{h},$ we get
\begin{equation}
\delta\sigma_{xx}^{\mathrm{MT(reg1)}}=-\frac{2\gamma_{E}e^{2}}{\pi^{2}}\frac{%
t}{\widetilde{h}}-\frac{2e^{2}}{\pi^{2}}\ln\frac{1}{\widetilde{h}}.
\label{MTreg1quantum}
\end{equation}

\paragraph*{High fields $H\gg H_{\mathrm{c2}}\left( 0\right) ,$ temperatures $%
t\ll h.$}---
This domain is analogous to the previous one. As above,  we first perform the
summation over $k$ and integrate over Landau levels: 
\begin{equation}
\delta\sigma_{xx}^{\mathrm{MT(reg1)}}=-\frac{e^{2}}{\pi^{2}}\left( \ln \ln%
\frac{1}{T_{\mathrm{c0}}\tau}-\ln\ln\frac{2h}{\pi^{2}}\right) .
\label{highh}
\end{equation}
The only difference between Eqs. (\ref{hight}) and (\ref{highh}) consists in
the lower limit: in the former it is determined by the temperature while in
the latter its role is taken by the zero Landau level $\omega_{c}\gg T.$ Eq. (\ref%
{highh}) is valid for arbitrary temperatures smaller $\omega_{c}$ and it
obviously matches Eq. (\ref{MTreg1quantum}) along the axis of the magnetic field
($t=0$).

\subsubsection{Contribution $\protect\delta\protect\sigma_{xx}^{\mathrm{%
MT(an)}}+\protect\delta \protect\sigma_{xx}^{\mathrm{MT(reg2)}}$}

Now we consider the second part of the MT contribution Eq. (\ref{MTgener}),
namely $\delta\sigma_{xx}^{\mathrm{MT(an)}}+\delta\sigma_{xx}^{\mathrm{%
MT(reg2)}}$.

\paragraph*{Vicinity of $T_{\mathrm{c0}},$ fields $h\ll1(H\ll H_{\mathrm{c2}%
}\left( 0\right) ).$}---
In the vicinity of the critical temperature $T_{\mathrm{c0}}$ one should use
 the expansion Eqs. (\ref{e1})-(\ref{e2}) of $\mathcal{E}_{m}\left(
t,h,ix\right) $. For the second order correction it is sufficient to take on only
 the imaginary part of $\mathcal{E}_{m}\left( t,h,ix\right) $ into account.
Substituting correspondingly $\Re \mathcal{E}_{m}^{\left( 1\right) }\left(
\epsilon,h\ll1,ix\right) $ and $\Im \mathcal{E}_{m}^{\left( 2\right) }\left(
\epsilon,h\ll1,ix\right) $ to Eq. (\ref{MTgener}) and using the fact
that $\gamma_{\phi}\ll1$, the integral over $x$ can be easily performed [it
converges for  $x>x_{0}\sim\epsilon+\left( \frac{h}{t}\right) \left( 2m+1\right)
$]. 
The remaining summation is accomplished in terms of the $\psi$-functions and
its result consists of two terms: the first one corresponds to the anomalous MT term $%
\delta\sigma_{xx}^{\mathrm{MT(an)}}$,  while the second, $%
\delta\sigma_{xx}^{\mathrm{MT(reg2)}},$ exactly coincides in this region
with $\delta\sigma_{xx}^{\mathrm{MT(reg1)}}$ [Eq. (\ref{MT1reg})]. Therefore, we present the total $\delta\sigma _{xx}^{\mathrm{MT}%
}=\delta\sigma_{xx}^{\mathrm{MT(reg1)}}+\delta\sigma _{xx}^{\mathrm{MT(an)}%
}+\delta\sigma_{xx}^{\mathrm{MT(reg2)}}$, which takes the form 
\begin{widetext}%
\begin{equation}
\delta\sigma_{xx}^{\mathrm{MT}}=\frac{e^{2}}{8}\frac{1}{\epsilon-\gamma_{\phi}%
}\left[  \psi\left(  1/2+\frac{t\epsilon}{2h}\right)  -\psi\left(
1/2+\frac{t\gamma_{\phi}}{2h}\right)  \right]  -\frac{14\zeta\left(  3\right)
e^{2}}{\pi^{4}}\left[  \ln\left(  \frac{t}{2h}\right)  -\psi\left(
1/2+\frac{t\epsilon}{2h}\right)  \right]  . \label{anMTa}%
\end{equation}
\end{widetext}
This formula is valid in the vicinity of the critical temperature $T_{%
\mathrm{c0}},$ where we have three different regimes [weak fields $%
h\ll\epsilon,$ GL strong fields $\epsilon\ll h,$ and fields close to the
$h_{\mathrm{c2}}\left( \epsilon\right) $ line which is ''mirrored'' at $T_{c0}$].

\paragraph*{High temperatures $T\gg T_{\mathrm{c0}},$ weak fields $h\ll t$.}---
Here we discuss the high-temperature asymptotic. As it
was done before, we assume $\ln t\gg1$ and use Eqs. (\ref{e}) and (\ref{ehigh}%
). Integration over $x$ due to the factor $\cosh^{-2}\pi z$ involves
only the region $z\sim1$ and can be performed first. The sum over
Landau levels in this case converges at large $m_{\max}\sim t/h\gg1$ and it
can be substituted by an integral. The contributing part of the integration with logarithmic
accuracy turns out to be only the fraction containing $\gamma_{\phi}.$ As result we get:
\begin{equation}
\delta\sigma_{xx}^{\mathrm{MT(an)}}+\delta\sigma_{xx}^{\mathrm{MT(reg2)}}=%
\frac{\pi^{2}e^{2}}{192}\frac{\ln\frac{\pi^{2}}{2\gamma_{\phi}}}{\ln^{2}t}.
\label{MThigt}
\end{equation}
Despite the presence of the large logarithm $\ln\frac{\pi^{2}}{%
2\gamma_{\phi}}$ in this result in its numerator is relatively small with respect to Eq. (%
\ref{hight}) due to the large $\ln^{2}t$ in denominator of Eq. (\ref{highh}).

\paragraph*{Fields close to the line $H_{\mathrm{c2}}\left( T\right) $\emph{$%
. $}}---
As it was done in the case of the paraconductivity, let us use in Eq. (\ref%
{MTgener}) the asymptotic Eq. (\ref{emas}) and perform the calculations in the
LLL approximation. One easily finds that the result in this case is
also expressed in terms of the integral (\ref{JGL}): 
\begin{equation}
\delta\sigma_{xx}^{\mathrm{MT(an)}}+\delta\sigma_{xx}^{\mathrm{MT(reg2)}}=%
\frac{e^{2}}{\pi^{2}}\frac{{1}}{1+t\gamma_{\phi}/h}J_{GL}\left( \frac{4h_{%
\mathrm{c2}}\left( t\right) }{\pi^{2}t}\widetilde{h}\right) .
\label{MTanlow}
\end{equation}
We consider the case of low temperatures $t\ll h_{\mathrm{c2}}\left(
t\right) ,$ hence $\gamma_{\phi}\ll1\ll\frac{h}{t}\ $\ and we can omit it in
denominator. In result it turns out that in this region Eq. (\ref{MTanlow})
exactly coincides with the corresponding AL contribution Eq. (\ref%
{sigmaline1}). Therefore it is determined by Eq. (\ref%
{salgen}) along the whole line $H_{\mathrm{c2}}\left( T\right) ,$
i.e. for $t\ll h_{\mathrm{c2}}\left( t\right)$,  which was already analyzed in detail above. Looking at Eqs. (\ref%
{salgen}) one can see that in the region $h-h_{\mathrm{c2}}\left( t\right)
\ll t,$ strong cancellation takes place in the MT contribution $%
\delta\sigma_{xx}^{\mathrm{MT}}$ and only the logarithmic contribution
remains 
\begin{equation*}
\delta\sigma_{xx}^{\mathrm{MT}}=-\frac{2e^{2}}{\pi^{2}}\ln\frac {2h_{\mathrm{%
c2}}\left( t\right) }{\pi^{2}t}. 
\end{equation*}

In the regime of quantum fluctuations $t\lesssim h-h_{\mathrm{c2}}\left(
t\right) $ the contribution $\delta\sigma_{xx}^{\mathrm{MT(an)}}+\delta
\sigma_{xx}^{\mathrm{MT(reg2)}}\sim t^{2}$ and the linear (in $t$) part of $%
\delta\sigma_{xx}^{\mathrm{MT(reg1)}}$ are gradually frozen and  the MT contribution
 reaches the finite negative value at zero
temperature

\begin{equation*}
\delta\sigma_{xx}^{\mathrm{MT}}=-\frac{2e^{2}}{\pi^{2}}\ln\frac{1}{%
\widetilde{h}}-\frac{2\gamma_{E}e^{2}}{\pi^{2}}\frac{t}{\widetilde{h}\left(
t\right) }+O\left[ \left( \frac{t}{\widetilde{h}\left( t\right) }\right) ^{2}%
\right] . 
\end{equation*}

\begin{table*}[htb]
{\tiny 
\begin{tabular}{|c|c|c|c|}
\hline
domain & $\delta\sigma_{xx}^{\mathrm{MT(an)}}+\delta\sigma_{xx}^{\mathrm{%
MT(reg2)R}}$ & $\delta\sigma_{xx}^{\mathrm{MT(reg1)}}$ & $\delta\sigma_{xx}^{%
\mathrm{MT}}$ \\ \hline
\texttt{I} & $\frac{e^{2}}{8}\frac{1}{\epsilon-\gamma_{\phi}}\ln \frac{%
\epsilon}{\gamma_{\phi}}-\frac{7\zeta\left( 3\right) e^{2}}{\pi^{4}}%
\ln\left( \frac{1}{\epsilon}\right)$ & $-\frac{7\zeta\left( 3\right) e^{2}}{%
\pi^{4}}\ln\frac{1}{\epsilon}$ & $\frac{e^{2}}{8}\frac{1}{\epsilon
-\gamma_{\phi}}\ln\frac{\epsilon}{\gamma_{\phi}}-\frac{14\zeta\left(
3\right) e^{2}}{\pi^{4}}\ln\left( \frac{1}{\epsilon}\right)$ \\ \hline
\texttt{I} - \texttt{III} & $%
\begin{array}{c}
\frac{e^{2}}{8}\frac{1}{\epsilon-\gamma_{\phi}}\left[ \psi\left( 1/2+\frac{%
t\epsilon}{2h}\right) -\psi\left( 1/2+\frac{t\gamma_{\phi}}{2h}\right) %
\right] \\ 
-\frac{7\zeta\left( 3\right) e^{2}}{\pi^{4}}\left[ \ln\left( \frac{t}{2h}%
\right) -\psi\left( 1/2+\frac{t\epsilon}{2h}\right) \right]%
\end{array}%
$ & $-\frac{7\zeta\left( 3\right) e^{2}}{\pi^{4}}\left[ \psi\left( \frac {t}{%
2h}\right) -\psi\left( \frac{1}{2}+\frac{\epsilon t}{2h}\right) \right]$ & $%
\begin{array}{c}
\frac{e^{2}}{8}\frac{1}{\epsilon-\gamma_{\phi}}\left[ \psi\left( 1/2+\frac{%
t\epsilon}{2h}\right) -\psi\left( 1/2+\frac{t\gamma_{\phi}}{2h}\right) %
\right] \\ 
-\frac{14\zeta\left( 3\right) e^{2}}{\pi^{4}}\left[ \ln\left( \frac{t}{2h}%
\right) -\psi\left( 1/2+\frac{t\epsilon}{2h}\right) \right]%
\end{array}%
$ \\ \hline
\texttt{VII} & $\frac{e^{2}}{4}\frac{t}{h-h_{\mathrm{c2}}\left( t\right) }$
& $-\frac{e^{2}}{4}\frac{t}{h-h_{\mathrm{c2}}\left( t\right) }-\frac{2e^{2}}{%
\pi^{2}}\ln\frac{2h}{\pi^{2}t}$ & $-\frac{2e^{2}}{\pi^{2}}\ln\frac{2h}{\pi
^{2}t}$ \\ \hline
\texttt{VI} & $\frac{2\gamma_{E}e^{2}}{\pi^{2}}\frac {t}{\widetilde{h}}$ & $-%
\frac{2\gamma_{E}e^{2}}{\pi^{2}}\frac{t}{\widetilde{h}}-\frac{2e^{2}}{\pi^{2}%
}\ln\frac{2h}{\pi^{2}t}$ & $-\frac{2e^{2}}{\pi^{2}}\ln\frac{2h}{\pi^{2}t}$
\\ \hline
\texttt{IV} & $\frac{4e^{2}\gamma_{E}^{2}t^{2}}{3\pi^{2}\widetilde{h}^{2}}$
& $-\frac{2\gamma_{E}e^{2}}{\pi^{2}}\frac {t}{\widetilde{h}\left( t\right) }-%
\frac{2e^{2}}{\pi^{2}}\left[ \ln\frac {1}{2\widetilde{h}\left( t\right) }%
\right]$ & $-\frac{2e^{2}}{\pi^{2}}\left[ \ln\frac{1}{2\widetilde{h}}\right]
-\frac{2\gamma_{E}e^{2}}{\pi^{4}}\frac{t}{\widetilde{h}}$ \\ \hline
\texttt{IX} & $\frac{7\zeta\left( 3\right) \pi ^{2}e^{2}}{768}\frac{t^{2}}{%
h^{2}}\ln^{-2}\frac{2h}{\pi^{2}}$ & $-\frac{e^{2}}{\pi^{2}}\left( \ln\ln%
\frac{1}{T_{\mathrm{c0}}\tau}-\ln\ln\frac{2h}{\pi^{2}}\right)$ & $%
\begin{array}{c}
-\frac{e^{2}}{\pi^{2}}\left( \ln\ln\frac{1}{T_{\mathrm{c0}}\tau}-\ln\ln 
\frac{2h}{\pi^{2}}\right) \\ 
+\frac{7\zeta\left( 3\right) \pi^{2}e^{2}}{768}\frac{t^{2}}{h^{2}}\ln^{-2}%
\frac{2h}{\pi^{2}}%
\end{array}%
$ \\ \hline
\texttt{VIII} & $\frac{\pi^{2}e^{2}}{192}\frac {\ln\frac{\pi^{2}}{%
2\gamma_{\phi}}}{\ln^{2}t}$ & $-\frac{e^{2}}{\pi^{2}}\left[ \ln\ln\frac{1}{%
T_{\mathrm{c0}}\tau}-\ln\ln t\right]$ & $-\frac{e^{2}}{\pi^{2}}\left[ \ln\ln%
\frac{1}{T_{\mathrm{c0}}\tau}-\ln\ln t\right] +\frac{\pi ^{2}e^{2}}{192}%
\frac{\ln\frac{\pi^{2}}{2\gamma_{\phi}}}{\ln^{2}t}$ \\ \hline
\end{tabular}
}
\caption{Asymptotic behavior of the MT contributions in different domains,
see also Fig.~\protect\ref{fig.domains} and table~\protect\ref{tab.domains}}
\label{tab.MT}
\end{table*}

\paragraph*{High fields ($H\gg H_{\mathrm{c2}}\left( T\right) $).}---
In this domain we are far from the transition line $H_{\mathrm{c2}}\left(
T\right) $ and the LLL approximation is not applicable. Replacing the \ $\psi
- $function in the Eq. (\ref{e}) by the logarithm one can use the asymptotic
Eq. (\ref{en}) and gets%
\begin{equation*}
\delta\sigma_{xx}^{\mathrm{MT(an)}}+\delta\sigma_{xx}^{\mathrm{MT(reg2)}}=%
\frac{7\zeta\left( 3\right) \pi^{2}e^{2}}{768}\frac{t^{2}}{h^{2}}\frac {1}{%
\ln^{2}\frac{2h}{\pi^{2}}} 
\end{equation*}
which is beyond the accuracy of the large contribution $\delta\sigma_{xx}^{%
\mathrm{MT(reg1)}}$ [see Eq. (\ref{highh})] which, in result, determines the
value of $\delta\sigma_{xx}^{\mathrm{MT}}$ in strong fields.

Finally all asymptotic expressions for the MT diagram are summarized in
Table~\ref{tab.MT}.\\

\section{DOS renormalization: contribution of the diagrams 3-6}

\subsection{General expression}

We start with calculation of diagram 4. As above we use the
intermediate results of Ref.~\olcite{LV09} for the diagrams and then quantize the
motion of the center of mass of the Cooper pair in magnetic field. The
general expression for diagram 4 is given by
\begin{equation}
Q_{xx}^{(4)}(\omega_{\nu})=2e^{2}T^{2}\int\frac{d^{2}q}{(2\pi)^{2}}\sum
_{k,n}L\left( q,\Omega_{k}\right) \lambda^{2}\left( q,\varepsilon
_{n},\Omega_{k-n}\right) I_{xx}^{(4)}  \label{q5}
\end{equation}
with the integral $I_{xx}^{(4)}$ of the four electron Green functions calculated
exactly in Ref.~\olcite{LV09} in the same spirit as it was demonstrated above: 
\begin{align*}
& I_{xx}^{(4)}=\int\frac{d^{2}p}{(2\pi)^{2}}v_{x}^{2}~G^{2}\left(
p,\varepsilon_{n}\right) G\left( p,\varepsilon_{n+\nu}\right) G\left(
p,\Omega_{k-n}\right) \\
& =-2\pi\nu_{0}\mathcal{D}\tau^{2}\left[ \Theta\left( \varepsilon
_{n}\varepsilon_{n+\nu}\right) \Theta\left( \varepsilon_{n}\varepsilon
_{n-k}\right) \right. \\
& \left. +\Theta\left( -\varepsilon_{n}\varepsilon_{n+\nu}\right)
\Theta\left( -\varepsilon_{n}\varepsilon_{n-k}\right) \right] .
\end{align*}
Substituting this expression to Eq. (\ref{q5}) one finds 
\begin{widetext}
\begin{equation}
Q_{xx}^{(4)}(\omega_{\nu})=-4\pi\nu_{0}\mathcal{D}e^{2}T^{2}\int\frac{d^{2}%
q}{(2\pi)^{2}}\sum_{k}L\left(  q,|\Omega_{k}|\right)  \left[  2\sum
_{n=0}^{\infty}\frac{\Theta\left(  \varepsilon_{n}+\Omega_{k}\right)
}{\left[  2\varepsilon_{n}+\Omega_{k}+\mathcal{D}\mathbf{q}^{2}\right]  ^{2}%
}-\sum_{n=0}^{\nu-1}\frac{\Theta\left(  \varepsilon_{n}+\Omega_{k}\right)
}{\left[  2\varepsilon_{n}+\Omega_{k}+\mathcal{D}\mathbf{q}^{2}\right]  ^{2}%
}\right]  .
\end{equation}
\end{widetext}

The first term in this expression does not depend on external frequency and the
corresponding part of the electro-magnetic response operator does not
contribute to conductivity. In the remaining part $\widetilde{Q}%
_{xx}^{(4)}(\omega_{\nu})$ one can perform the summation over fermionic
frequency and obtain it in the form of a sum of two terms:
\begin{align*}
& \widetilde{Q}_{xx}^{(4,1)}(\omega_{\nu}) =\frac{\nu_{0}\mathcal{D}e^{2}}{%
4\pi}\int\frac{d^{2}q}{(2\pi)^{2}}\sum_{k=0}^{\infty}L\left( q,|\Omega
_{k}|\right) \\
& \left[ \psi^{\prime}\left( \frac{1}{2}+\frac{\Omega_{k}+\mathcal{D}\mathbf{%
q}^{2}}{4\pi T}\right) -\psi^{\prime}\left( \frac{1}{2}+\frac{%
2\omega_{\nu}+\Omega_{k}+\mathcal{D}\mathbf{q}^{2}}{4\pi T}\right) \right] ,\\
& \widetilde{Q}_{xx}^{(4,2)}(\omega_{\nu}) =\frac{\nu_{0}\mathcal{D}e^{2}}{%
4\pi}\int\frac{d^{2}q}{(2\pi)^{2}}\sum_{k=1}^{\nu-1}L\left( q,|\Omega
_{k}|\right) \\
& \left[ \psi^{\prime}\left( \frac{1}{2}+\frac{\Omega_{k}+\mathcal{D}\mathbf{%
q}^{2}}{4\pi T}\right) -\psi^{\prime}\left( \frac{1}{2}+\frac{%
2\omega_{\nu}-\Omega_{k}+\mathcal{D}\mathbf{q}^{2}}{4\pi T}\right) \right] .
\end{align*}
The analytical continuation of the first one is trivial and it gives the
first contribution to the conductivity, which in Landau representation takes the
form

\begin{equation*}
\delta\sigma_{xx}^{(4,1)}=\left( \frac{2e^{2}}{\pi^{4}}\right) \left( \frac{h%
}{t}\right) \sum_{m=0}^{M}\sum_{k=0}^{\infty}\frac{\mathcal{E}%
_{m}^{\prime\prime}\left( t,h,k\right) }{\mathcal{E}_{m}\left( t,h,k\right) }
\end{equation*}

The analytical continuation of $\widetilde{Q}_{xx}^{(4,2)}(\omega_{\nu})$ is
completely analogous to that one performed above in the case of the anomalous
MT part. As a result the total contribution of diagrams 3 and 4 can be presented
as a sum of two very different terms 
\begin{widetext}
\begin{equation}
\delta\sigma_{xx}^{(3+4)}=\frac{4e^{2}}{\pi^{4}}\left(  \frac{h}{t}\right)
\sum_{m=0}^{M}\left[  \sum_{k=0}^{\infty}\frac{\mathcal{E}_{m}^{\prime\prime
}\left(  t,h,k\right)  }{\mathcal{E}_{m}\left(  t,h,k\right)  }+\frac{\pi}{2}
\int_{-\infty}^{\infty}\frac{dx}{\sinh^{2}\left(  \pi x\right)  }%
\frac{\operatorname{Im}\mathcal{E}_{0}\left(  t,h,ix\right)  \operatorname{Im}%
\mathcal{E}_{0}^{\prime}\left(  t,h,ix\right)  }{\operatorname{Re}%
^{2}\mathcal{E}_{0}\left(  t,h,ix\right)  +\operatorname{Im}^{2}%
\mathcal{E}_{0}\left(  t,h,ix\right)  }\right]  . \label{s56}%
\end{equation}
\end{widetext}

Next, we discuss diagram 5. Its contribution can be
written in the same way as above:

\begin{align*}
Q_{xx}^{\left( 5\right) }(\omega_{\nu}) & =\frac{e^{2}T^{2}v_{F}^{2}}{%
2\pi\nu_{0}\tau}\int\frac{d^{2}q}{(2\pi)^{2}}\sum_{n,k}L\left( q,\Omega
_{k}\right) \lambda^{2}\left( q,\varepsilon_{n},\Omega_{k-n}\right) \\
& \times I^{\left( 5\right) }\left( \varepsilon_{n},\varepsilon_{n+\nu
}\right) I^{\left( 5\right) }\left( \varepsilon_{n},-\varepsilon
_{n-k}\right) ,
\end{align*}
where the integral%
\begin{align}
I^{\left( 5\right) }\left( \varepsilon_{n},\varepsilon_{n+\nu}\right) & =\int%
\frac{d^{2}p}{(2\pi)^{2}}~G^{2}\left( p,\varepsilon_{n}\right) G\left(
p,\varepsilon_{n+\nu}\right)  \notag \\
& =2\pi i\nu_{0}\tau^{2}\mathrm{sgn}\varepsilon_{n+\nu}\Theta\left(
-\varepsilon_{n+\nu}\varepsilon_{n}\right) .  \label{iprod}
\end{align}
As result:%
\begin{align}
Q_{xx}^{\left( 5\right) }(\omega_{\nu}) & =-4\pi\nu_{0}\mathcal{D}%
e^{2}T^{2}\int\frac{d^{2}q}{(2\pi)^{2}}\sum_{k=-\infty}^{\infty}L\left(
q,\Omega_{k}\right)  \notag \\
& \times\sum_{n=-\nu}^{-1}\frac{\Theta\left( \Omega_{k}-\varepsilon
_{n}\right) }{\left[ |2\varepsilon_{n}-\Omega_{k}|+\mathcal{D}\mathbf{q}^{2}%
\right] ^{2}}.  \label{q7h}
\end{align}
Further evaluation of this expression is very similar to that one of \ $%
\widetilde{Q}_{xx}^{\left( 5\right) }(\omega_{\nu})$. In particular, after
summation over fermionic frequencies, $Q_{xx}^{\left( 7\right) }(\omega_{\nu
}) $ is presented in the form of two sums over bosonic frequencies, one in
the limits $k\in\lbrack0,\infty),$ the other $k\in\lbrack1,\nu-1]$ and
following step-by-step the same procedure of the analytical continuation as
before, one finds that $\delta\sigma_{xx}^{\left( 5,1\right)
}=-\delta\sigma_{xx}^{(4,1)}, $ and $\delta\sigma_{xx}^{\left( 5,2\right)
}=\delta\sigma_{xx}^{(4,2)}$. Therefore we get
\begin{align}
& \delta\sigma_{xx}^{(5+6)}=\frac{4e^{2}}{\pi^{4}}\left( \frac{h}{t}\right)
\sum_{m=0}^{M}\left[ -\sum_{k=0}^{\infty}\frac{\mathcal{E}_{m}^{\prime\prime
}\left( t,h,k\right) }{\mathcal{E}_{m}\left( t,h,k\right) }\right.
\label{s78} \\
& +\left. \frac{\pi}{2}\int_{-\infty}^{\infty}\frac{dx}{\sinh^{2}\left( \pi
x\right) }\frac{\Im \mathcal{E}_{0}\left( t,h,ix\right) \Im \mathcal{E}%
_{0}^{\prime}\left( t,h,ix\right) }{\Re ^{2}\mathcal{E}_{0}\left(
t,h,ix\right) +\Im ^{2}\mathcal{E}_{0}\left( t,h,ix\right) }\right] .  \notag
\end{align}
Evaluating the sum and integral close to $T_{\mathrm{c0}}$ one can see that the
first term in Eq. (\ref{s56}) is twice larger than the second one. Comparing
Eq. (\ref{s56}) to Eq. (\ref{s78}) we obtain the old result: $3\delta
\sigma_{xx}^{(5+6)}=-\delta\sigma_{xx}^{(3+4)},$ \olcite{ARV83}, used later in 
Refs.~\olcite{ILVY93,BDKLV93,LV09,GL01} . But it is necessary to stress, that the
last statement {\it is not universal:} far from the critical temperature, or
at low temperatures, close to $H_{\mathrm{c2}}\left( 0\right) ,$ the
integrals in Eqs. (\ref{s56}) - (\ref{s78}) are small with respect to the
contribution of the sums. Regarding the latter, they enter in Eqs. (\ref%
{s56}) -(\ref{s78}) with the opposite sign. After the summation in $%
\delta\sigma_{xx}^{\mathrm{(5-8)}}$ these just cancel each other (in this region
of temperatures $\delta\sigma_{xx}^{(3+4)}\approx-\delta%
\sigma_{xx}^{(5+6)}). $ 
To avoid misunderstanding~\cite{GL01}, it is more
convenient to use the total contribution of the DOS-like diagrams 3-6 in the
form: 
\begin{widetext}
\begin{equation}
\delta\sigma_{xx}^{\mathrm{(DOS)}}=\frac{4e^{2}}{\pi^{3}}\frac{h}{t}\sum_{m=0}%
^{M}\int_{-\infty}^{\infty}\frac{dx}{\sinh^{2}\pi x}\frac{\operatorname{Im}%
\mathcal{E}_{m}\left(  t,h,ix\right)  \operatorname{Im}\mathcal{E}_{m}%
^{\prime}\left(  t,h,ix\right)  }{\operatorname{Re}^{2}\mathcal{E}_{m}\left(
t,h,ix\right)  +\operatorname{Im}^{2}\mathcal{E}_{m}\left(  t,h,ix\right)  }.
\label{sigdos}%
\end{equation}
\end{widetext}

\subsection{Asymptotic behavior}

\subsubsection{Vicinity of $T_{\mathrm{c0}},$ fields $h\ll1(H\ll H_{\mathrm{%
c2}}\left( 0\right) )$}

 In this case $\ln t=\epsilon\ll1,$ the $\psi$-function in Eq.(\ref{e})
can be expanded. The function $\mathcal{E}_{m}\left( t,h,ix\right) $ is
determined by Eq.(\ref{e1}). Its substitution to Eq.(\ref{sigdos})
results in%
\begin{align}
\delta\sigma_{xx}^{\mathrm{DOS}} & =-\frac{14\zeta\left( 3\right) e^{2}}{%
\pi^{4}}\left[ \ln\left( 1/2h\right) -\psi\left( 1/2+\frac{\epsilon}{2h}%
\right) \right]  \notag \\
& =-\frac{14\zeta\left( 3\right) e^{2}}{\pi^{4}}\left\{ 
\begin{array}{c}
\ln\left( 1/\epsilon\right) ,\;h\ll\epsilon \\ 
\ln\left( \frac{1}{2h}\right) ,\;\epsilon\ll h\ll1%
\end{array}
\right.  \label{sigGL}
\end{align}
This expression is valid in the vicinity of the critical temperature $T_{%
\mathrm{c0}}$ and exactly reproduces existing results~\cite%
{ILVY93,BDKLV93}.

\subsubsection{High temperatures, high fields}

Next, we discuss the high-temperature asymptotic. As it
was done above, we assume $\ln t\gg1$ and use Eqs. (\ref{e})-(\ref{ehigh}).
The sum over Landau levels in this case converges at large $m_{\max}\sim
t/h\gg1$ and can be substituted by an integral. The main integral contribution
comes only from the region up to $x\sim1$ and can be performed first. One gets%
\begin{equation*}
\delta\sigma_{xx}^{\mathrm{DOS}}=-\frac{\pi^{2}e^{2}}{192\ln^{2}t}. 
\end{equation*}
We see that this result differs from that one of Ref.~\olcite{ARV83}. The cancellation
of the sums of Eqs. (\ref{s56}) - (\ref{s78} in $\delta\sigma_{xx}^{\mathrm{%
DOS}}$ removes\ the double logarithmic term $\ln\ln t$ from it.
Nevertheless, such terms in $\delta\sigma_{xx}^{(tot)}$ still appear from
the regular MT term and, as we will see below, from diagrams 9 and 10.

In the limit of high fields $h\gg t$ the summation over Landau levels gives:%
\begin{equation*}
\delta\sigma_{xx}^{\mathrm{DOS}}=-\frac{7\zeta\left( 3\right) \pi^{2}e^{2}}{%
384}\left( \frac{t}{h}\right) ^{2}\frac{1}{\ln^{2}\frac{2h}{\pi^{2}}}. 
\end{equation*}

\subsubsection{Above the line $H_{\mathrm{c2}}\left( T\right) $ but $t\ll h_{%
\mathrm{c2}}\left( t\right) $}

Using the asymptotic Eq. (\ref{emas}) one can perform the integration in Eq.
(\ref{sigdos}) and express $\delta \sigma _{xx}^{\mathrm{DOS}}$ in terms of
the integral $J_{GL}:$ 
\begin{equation*}
\delta \sigma _{xx}^{\mathrm{DOS}}=-\frac{e^{2}}{\pi ^{2}}\sum_{m=0}^{M}%
\frac{J_{GL}\left[ \frac{4h\left( 2m+1\right) }{\pi ^{2}t}\ln \frac{4h}{\pi
^{2}}\left( m+\frac{1}{2}\right) \right] }{\left( 2m+1\right) }.
\end{equation*}%
Close to the line $H_{\mathrm{c2}}\left( T\right) $ we can restrict ourselves
to the LLL and immediately get%
\begin{equation*}
\delta \sigma _{xx}^{\mathrm{(DOS)}}=-\frac{e^{2}}{\pi ^{2}}J_{GL}\left( 
\frac{4\widetilde{h}}{\pi ^{2}t}\right) .
\end{equation*}%
Looking on the asymptotic behavior of $J\left( r\right) $ at low
temperatures, one notices that in contrast to the statement of Ref~\olcite{GL01} \
the group of diagrams 5-8 does not give any contribution to $\delta \sigma
_{xx}^{(tot)}$ when temperature tends to zero. Nevertheless, a non-trivial
contribution of quantum fluctuations $\sim \ln \widetilde{h}$, found in Ref.~\olcite%
{GL01}, exists due to the regular MT term and diagrams 9 and 10.\\

\section{Renormalization of the diffusion coefficient: contribution of
diagrams 7-10}

\subsection{General expression}

We start with the calculation of diagram 7:
\begin{align}
& Q_{xx}^{(7)}(\omega_{\nu})=2e^{2}T^{2}\sum_{k,n}\int\frac{d^{2}q}{%
(2\pi)^{2}}L\left( q,\Omega_{k}\right) \lambda\left( \mathbf{q}%
,\varepsilon_{n},\Omega_{k}-\varepsilon_{n}\right)  \notag \\
& \lambda\left( \mathbf{q},\varepsilon_{n+\nu},\Omega_{k}-\varepsilon
_{n+\nu}\right) C\left( q,\varepsilon_{n+\nu},\Omega_{k}-\varepsilon
_{n}\right) I_{1}^{\left( 7\right) }I_{2}^{\left( 7\right) },  \label{q3}
\end{align}
where the integrals of the Green's function products can be calculated in
the standard way:%
\begin{align}
& I_{\left( 1\right) }^{(7)}\left( \varepsilon_{n},\varepsilon_{n+\nu
},\Omega_{k-n}\right) =\int\frac{d^{D}p}{(2\pi)^{D}}v_{x}\left( p\right)
G\left( p,\varepsilon_{n}\right) G\left( p,\varepsilon_{n+\nu}\right)  \notag
\\
& \times G\left( q-p,\Omega_{k-n}\right) =4\pi\nu_{0}\mathcal{D}\mathbf{q}%
_{x}\tau^{2}\theta\left( \varepsilon_{n}\varepsilon_{n+\nu}\right)
\theta\left( -\varepsilon_{n}\Omega_{k-n}\right)  \label{I3}
\end{align}%
\begin{equation*}
I_{\left( 2\right) }^{(7)}=I_{\left( 1\right) }^{(7)}\left(
\Omega_{k-n-\nu},\Omega_{k-n},\varepsilon_{n+\nu}\right) . 
\end{equation*}
Substitution of these expressions to Eq. (\ref{q3}) and accounting for the fact
that $\overline{\mathcal{D}\mathbf{q}_{x}^{2}}=\mathcal{D}\mathbf{q}^{2}/2$
results in%
\begin{widetext}
\begin{align}
Q_{xx}^{(7)}=8\pi\nu_{0}\mathcal{D}e^{2}T^{2}\int\frac
{\mathcal{D}\mathbf{q}^{2}d^{2}q}{(2\pi)^{2}}\sum_{k=-\infty}^{\infty}L\left(
q,\Omega_{k}\right)  \sum_{n=-\infty}^{\infty}\frac{\theta\left(
-\varepsilon_{n}\left(  \Omega_{k}-\varepsilon_{n}\right)  \right)
}{\left\vert 2\varepsilon_{n}-\Omega_{k}\right\vert +\mathcal{D}\mathbf{q}%
^{2}}\frac{\theta\left(  -\varepsilon_{n+\nu}\left(
\Omega_{k}-\varepsilon _{n+\nu}\right)  \right)  }{\left\vert
2\varepsilon_{n+\nu}-\Omega _{k}\right\vert
+\mathcal{D}\mathbf{q}^{2}}\frac{\theta\left(  \varepsilon
_{n}\varepsilon_{n+\nu}\right)  }{\left\vert
2\varepsilon_{n}+\omega_{\nu }-\Omega_{k}\right\vert
+\mathcal{D}\mathbf{q}^{2}}. \nonumber
\end{align}
The $\theta-$function $\theta\left(
\varepsilon_{n}\varepsilon_{n+\nu}\right) $\ defines the limits of summation
over fermionic frequencies $n$ as $(-\infty,-\nu-1]$ and $[0,\infty).$
Changing the sign of summation in the first interval and then shifting the
variable of summation $\varepsilon_{n}+\omega_{\nu}\rightarrow%
\varepsilon_{n^{\prime}}$, one finds that the expression is even in $\Omega_{k}$,
which allows to present $Q_{xx}^{(7)}(\omega_{\nu})$ in the form of an
analytical function of $\omega_{\nu },$ to perform the analytical
continuation $\omega_{\nu}\rightarrow-i\omega$, and to expend it over small $%
\omega:$

\begin{equation}
Q_{xx}^{(7)R}(\omega)=-8\pi\nu_{0}\mathcal{D}e^{2}T^{2}\sum_{k=-\infty
}^{\infty}\int\mathcal{D}\mathbf{q}^{2}L\left(  q,\Omega_{k}\right)
\frac{d^{2}q}{(2\pi)^{2}}\sum_{n=0}^{\infty}\left\{  \frac{1}{\left[
2\varepsilon_{n}+|\Omega_{k}|+\mathcal{D}\mathbf{q}^{2}\right]  ^{3}}%
+\frac{3i\omega}{\left[  2\varepsilon_{n}+|\Omega_{k}|+\mathcal{D}%
\mathbf{q}^{2}\right]  ^{4}}\right\}  .\label{q3a}
\end{equation}
\end{widetext}
The corresponding contribution to the conductivity is determined
by the imaginary part of Eq. (\ref{q3a}). Quantizing the motion of Cooper
pairs and going over to the Landau representation, one finds 
\begin{equation}
\delta\sigma_{xx}^{(7+8)}=\frac{2e^{2}}{\pi^{6}}\left( \frac{h}{t}\right)
^{2}\sum_{m=0}^{M}\left( m+\frac{1}{2}\right) \sum_{k=-\infty}^{\infty}\frac{%
8\mathcal{E}_{m}^{\prime\prime\prime}\left( t,h,|k|\right) }{\mathcal{E}%
_{m}\left( t,h,|k|\right) }.  \label{s34}
\end{equation}
Comparing this formula with Eq. (\ref{MTgener}) one can see that beyond the
vicinity of $T_{\mathrm{c0}}$ the contribution of diagrams 7 and 8 given
by the Eq.(\ref{s34}) cancels the regular MT contribution [given by the
first term of Eq. (\ref{MTgener})].

Finally we proceed with the calculation of diagram 9. Two
integrals of the three Green function blocks in it are equal and coincide with $%
I_{1}^{\left( 7\right) }.$ Substituting Eq. (\ref{I3}) to the general
expression for $Q_{xx}^{(9)}(\omega_{\nu})$ and performing the summation over
fermionic frequencies in the spirit of the above calculations, one finds 
\begin{align}
& Q_{xx}^{(9)}(\omega_{\nu}) =-\frac{e^{2}T}{4\pi^{2}\nu_{0}\tau^{4}\omega_{%
\nu}^{2}}\sum_{k=-\infty}^{\infty}\int\frac{d^{2}q}{(2\pi)^{2}}\mathcal{D}%
\mathbf{q}^{2}L\left( q,\Omega_{k}\right)  \notag \\
& \times\left[ \Psi_{1}\left( \left\vert \Omega_{k}\right\vert ,\omega
_{\nu}\right) -\Psi_{2}\left( \left\vert \Omega_{k+\nu}\right\vert
,\omega_{\nu}\right) \right] =Q_{\left( 1\right) }^{(9)}+Q_{\left( 2\right)
}^{(9)},  \label{q9}
\end{align}
where%
\begin{align}
\Psi_{\gamma}\left( x,\omega_{\nu}\right) & =\left[ \psi\left( \frac {1}{2}+%
\frac{\omega_{\nu}+x+\mathcal{D}\mathbf{q}^{2}}{4\pi T}\right) -\psi\left( 
\frac{1}{2}+\frac{x+\mathcal{D}\mathbf{q}^{2}}{4\pi T}\right) \right.  \notag
\\
& \left. -\frac{\omega_{\nu}}{\left( 4\pi T\right) }\psi^{\prime}\left( 
\frac{1}{2}+\frac{\omega_{\nu}}{4\pi T}\delta_{\gamma2}+\frac{x+\mathcal{D}%
\mathbf{q}^{2}}{4\pi T}\right) \right]  \label{psi1}
\end{align}
with $\gamma=1,2.$

There is no problem to perform analytical continuation of the first term of
Eq. (\ref{q9}): the function $\Psi_{1}\left( \left\vert
\Omega_{k}\right\vert ,\omega_{\nu}\right) $ is analytical in its argument $%
\omega_{\nu}$, and the corresponding contribution to Eq. (\ref{q9}) can be
continued in the standard way $\omega_{\nu}\rightarrow-i\omega\rightarrow0.$
Expanding Eq. (\ref{psi1}) with $\gamma=1$ over $%
\omega$ one finds the essential contribution to the electromagnetic response
operator: 
\begin{align}
Q_{\left( 1\right) }^{(9)R}(\omega) & =-i\omega\frac{\mathcal{D}\nu
_{0}e^{2}T}{3\left( 4\pi T\right) ^{3}}\sum_{k=-\infty}^{\infty}\int\mathcal{%
D}\mathbf{q}^{2}\frac{d^{2}q}{(2\pi)^{2}}  \notag \\
& \times L\left( q,\Omega_{k}\right) \psi^{\prime\prime\prime}\left( \frac{1%
}{2}+\frac{\left\vert \Omega_{k}\right\vert +\mathcal{D}\mathbf{q}^{2}}{4\pi
T}\right) .  \label{q91}
\end{align}

The evaluation of the second term of Eq. (\ref{q9}) turns out to be much
more sophisticated, since $\omega_{\nu}$ appears in $\Psi_{2}\left(
\left\vert \Omega_{k+\nu}\right\vert ,\omega_{\nu}\right) $ not only as
parameter but also in the argument $\left\vert \Omega_{k+\nu}\right\vert $\
of this non-analytical function. The situation is analogous to the AL
contribution and the same method of analytical continuation has to be
applied. The corresponding sum over bosonic frequencies is transformed in an
integral over the contour ${\cal C}$ shown in Fig. \ref{fig.AL} with three regions of
different analytic behavior: 
\begin{align*}
Q_{\left( 2\right) }^{(9)}(\omega_{\nu}) & =\frac{1}{2\pi i}\frac{\mathcal{D}%
\nu_{0}e^{2}}{\omega_{\nu}^{2}}\int\mathcal{D}\mathbf{q}^{2}\frac{d^{2}q}{%
(2\pi)^{2}} \\
& \times\oint _{C}\coth\frac{z}{2T}L\left( q,-iz\right) \psi_{1}\left(
\left\vert \Omega_{k+\nu}\right\vert ,\omega_{\nu}\right) .
\end{align*}
After shifting of the variable $z$ of the integral over the line $\Im
z=-\omega_{\nu}$ as $-iz+\omega_{\nu}\rightarrow-iz^{\prime}$, one
gets $Q_{\left( 2\right) }^{(9)}(\omega_{\nu})$ already as an analytical
function of $\omega_{\nu}:$ 
\begin{align}
Q_{\left( 2\right) }^{(9)}(\omega_{\nu}) & =\frac{\mathcal{D}\nu_{0}e^{2}}{%
\pi\omega_{\nu}^{2}}\int\mathcal{D}\mathbf{q}^{2}\frac{d^{2}q}{(2\pi)^{2}}%
\int_{-\infty}^{\infty}dz\coth\left( \frac{z}{2T}\right)  \notag \\
& \times\left[ \Psi_{2}^{R}(-iz+\omega_{\nu},\omega_{\nu})\Im L^{R}\left(
q,-iz\right) \right.  \notag \\
& \left. +L^{A}\left( q,-iz-\omega_{\nu}\right) \Im \Psi
_{2}^{R}(-iz,\omega_{\nu})\right] .  \label{q92}
\end{align}
Obviously, this expression can be continued in $\omega_{\nu}$ in the
standard way $\omega_{\nu}\rightarrow-i\omega$.

We are interested in the imaginary part of $Q_{\left( 2\right)
}^{(9)R}(\omega )$, i.e. only $\Im \Psi _{2}^{R}(-iz-i\omega ,-i\omega )$
and $\Im \Psi _{2}^{R}(-iz,-i\omega )$ are essential. They can be can be
written explicitly from Eq. (\ref{psi1}): 
\begin{equation}
\Im \Psi _{2}^{R}(-iz,-i\omega )=-\frac{\omega ^{3}}{3\left( 4\pi T\right)
^{4}}\Re \psi ^{\prime \prime \prime }\left( \frac{1}{2}+\frac{-iz+\mathcal{D%
}\mathbf{q}^{2}}{4\pi T}\right)   \label{psire}
\end{equation}%
with $\Im \Psi _{2}^{R}(-iz-i\omega ,-i\omega )=5\Im \Psi
_{2}^{R}(-iz,-i\omega ).$ Since we are interested\ only in the linear 
$\omega$-part of $\Im Q_{\left( 2\right) }^{(9)R}(\omega )$ in the
analytically continued Eq. (\ref{q92}), one can omit $i\omega $ in the
argument of $L^{A}\left( q,-iz+i\omega \right) $ and recall that $\Im
L^{A}\left( q,-iz\right) =-\Im L^{R}\left( q,-iz\right) $. One gets:
\begin{align*}
\Im Q_{\left( 2\right) }^{(9)}(\omega )& =\frac{4\omega \mathcal{D}\nu
_{0}e^{2}}{3\pi \left( 4\pi T\right) ^{4}}\int \mathcal{D}\mathbf{q}^{2}%
\frac{d^{2}q}{(2\pi )^{2}}\int_{-\infty }^{\infty }dz\coth \left( \frac{z}{2T%
}\right)  \\
& \times \Re \psi ^{\prime \prime \prime }\left( \frac{1}{2}+\frac{-iz+%
\mathcal{D}\mathbf{q}^{2}}{4\pi T}\right) \Im L^{R}\left( q,-iz\right) .
\end{align*}%
Now one can see that the integrand function is odd in $z$ and its
integration with symmetric limits gives zero. Hence, in linear
approximation $\Im Q_{\left( 2\right) }^{(9)}(\omega )=0$ and the second
term of \ Eq. (\ref{q9}) does not contribute to conductivity. Going over
to the dimensionless variables in Eq. (\ref{q91}) and to the Landau
representation, one finds that $\delta \sigma _{xx}^{(9)}=-\delta \sigma
_{xx}^{(7)}/3.$ Finally, the total contribution of diagrams 7-10,
determining the renormalization of the one-particle diffusion coefficient in
the presence of superconducting fluctuations, is 
\begin{equation}
\delta \sigma _{xx}^{\mathrm{7-10}}=\frac{4e^{2}}{3\pi ^{6}}\left( \frac{h}{t%
}\right) ^{2}\sum_{m=0}^{M}\sum_{k=-\infty }^{\infty }(m+\frac{1}{2})\frac{8%
\mathcal{E}_{m}^{\prime \prime \prime }\left( t,h,|k|\right) }{\mathcal{E}%
_{m}\left( t,h,|k|\right) }.  \label{DCRf}
\end{equation}

\subsection{\protect\bigskip Asymptotic behavior}

\subsubsection{Vicinity of $T_{\mathrm{c0}},$ fields $h\ll1(H\ll H_{\mathrm{%
c2}}\left( 0\right) )$}

In contrast to the AL, MT and DOS contributions, due to presence of \ the
multiplier $\mathcal{D}\mathbf{q}^{2}$\ in the numerator of Eq. (\ref{q9})
[corresponding to $(m+\frac{1}{2})$ in Eq. (\ref{DCRf}) close to the
critical temperature $T_{\mathrm{c0}}$], the value $\delta \sigma _{\left(
2\right) }^{(DCR)}$ turns out to be not singular in $\epsilon $ at all.
Substituting the summations in Eq. (\ref{DCRf}) by integrals, one finds 
\begin{equation}
\delta \sigma _{xx}^{\mathrm{7-10}}\left( \epsilon \ll 1\right) =\frac{e^{2}%
}{3\pi ^{2}}\ln \ln \frac{1}{T_{\mathrm{c0}}\tau }+O\left( \epsilon \right) ,
\label{DCR1}
\end{equation}%
which just gives a temperature independent constant. Let us stress that
this constant is necessary for matching of the results in domain \rI and
\rVII of Fig.~\ref{fig.domains}.

\subsubsection{High temperatures, high fields}

In this domain of the phase diagram, we cannot omit  $\ln t\gg 1$ in the denominator of Eq.
(\ref{DCRf}), but the above consideration still is applicable.
As a result we get
\begin{equation}
\delta \sigma _{xx}^{\mathrm{7-10}}\left( t\gg \max \{1,h\}\right) =\frac{%
e^{2}}{3\pi ^{2}}\left( \ln \ln \frac{1}{T_{\mathrm{c0}}\tau }-\ln \ln
t\right) .  \label{DCR2}
\end{equation}%
In the limit of high fields $h\gg t$%
\begin{equation}
\delta \sigma _{xx}^{\mathrm{7-10}}\left( h\gg \max \{1,t\}\right) =\frac{%
e^{2}}{3\pi ^{2}}\left( \ln \ln \frac{1}{T_{\mathrm{c0}}\tau }-\ln \ln \frac{%
2h}{\pi ^{2}}\right) .  \label{DCR3}
\end{equation}

\subsubsection{Above the line $H_{\mathrm{c2}}\left( T\right) $ but $t\ll h_{%
\mathrm{c2}}\left( t\right) $}

In this region one can restrict the consideration to the LLL approximation and
use the asymptotic expression (\ref{emas}). In complete analogy with the
case of the regular part of the MT contribution one finds%
\begin{equation}
\delta \sigma _{xx}^{\mathrm{7-10}}\left( t\ll 1,\widetilde{h}\right) =\frac{%
4e^{2}}{3\pi ^{2}}\ln \frac{1}{\widetilde{h}}.  \label{DCR4}
\end{equation}%
One can notice the close connection between the $\delta \sigma _{xx}^{%
\mathrm{7-10}}$ and the $\delta \sigma _{xx}^{\mathrm{MT(reg1)}}$ contributions,
which is why the Eqs. (\ref{DCR1})-(\ref{DCR4}) should be considered side by
side with Eqs. (\ref{regMT})-(\ref{highh}).


\begin{thebibliography}{99}
\bibitem{footnote1} It is worth mentioning that the contribution of diagrams
7-10, which represents the renormalization of the diffusion coefficient due
to the presence of fluctuations (we will call this group the \textit{DCR}
diagrams) was never distinguished from the DOS contributions before. It was
believed that these diagrams are not singular at all close to the critical
temperature, but far from the critical temperature these form together with
diagrams 3-6 long, double logarithmic tails in temperature in the
fluctuation conductivity.
\end{thebibliography}
\end{document}